\pdfoutput=1

\documentclass[11pt,twoside,a4paper,cmspaper,final,collab]{cms-tdr}
\def\svnVersion{457143P}\def\cmsCernNoTag{CERN-EP-2017-121}\def\cmsCernDate{\today}\def\cmsMessage{Published in the Journal of High Energy Physics as \href{http://dx.doi.org/10.1007/JHEP03(2018)166}{\doi{10.1007/JHEP03(2018)166.}}}

\begin{document}\cmsNoteHeader{SUS-16-039}

\hyphenation{had-ron-i-za-tion}
\hyphenation{cal-or-i-me-ter}
\hyphenation{de-vices}
\RCS$Revision: 447574 $
\RCS$HeadURL: svn+ssh://svn.cern.ch/reps/tdr2/papers/SUS-16-039/trunk/SUS-16-039.tex $
\RCS$Id: SUS-16-039.tex 447574 2018-02-23 08:48:56Z folguera $
\newlength\cmsFigWidth
\ifthenelse{\boolean{cms@external}}{\setlength\cmsFigWidth{0.85\columnwidth}}{\setlength\cmsFigWidth{0.4\textwidth}}
\ifthenelse{\boolean{cms@external}}{\providecommand{\cmsLeft}{top\xspace}}{\providecommand{\cmsLeft}{left\xspace}}
\ifthenelse{\boolean{cms@external}}{\providecommand{\cmsRight}{bottom\xspace}}{\providecommand{\cmsRight}{right\xspace}}

\newcommand{\fulllumi}{35.9\fbinv}
\newcommand{\ewklumi}{35.9\fbinv}
\newcommand{\MT}{\ensuremath{M_\mathrm{T}}\xspace}
\newcommand{\MTT}{\ensuremath{M_{\mathrm{T}2}}\xspace}
\newcommand{\Njets}{\ensuremath{N_\text{jets}}\xspace}
\newcommand{\Mll}{\ensuremath{M_{\ell\ell}}\xspace}
\newcommand{\WJ}{\ensuremath{\PW{}\text{+jets}}\xspace}
\newcommand{\ZJ}{\ensuremath{\cPZ{}\text{+jets}}\xspace}
\newcommand{\WZ}{\ensuremath{\PW\cPZ}\xspace}
\newcommand{\ttZ}{\ensuremath{\ttbar\cPZ}\xspace}
\newcommand{\ttW}{\ensuremath{\ttbar\PW}\xspace}
\newcommand{\slep}{\ensuremath{\widetilde{\ell}}\xspace}
\newcommand{\ptRatio}{\ensuremath{\pt^\text{ratio}}\xspace}
\newcommand{\ptRel}{\ensuremath{\pt^\text{rel}}\xspace}
\newcommand{\miniIso}{\ensuremath{I_\text{mini}}\xspace}
\providecommand{\NA}{\ensuremath{\text{---}}}

\cmsNoteHeader{SUS-16-039}
\title{Search for electroweak production of charginos and neutralinos in multilepton final states in proton-proton collisions at $\sqrt{s}=13\TeV$ }

\date{\today}

\abstract{ Results are presented from a search for the direct electroweak production of charginos and neutralinos in signatures with either two or more leptons (electrons or muons) of the same electric charge, or with three or more leptons, which can include up to two hadronically decaying tau leptons. The results are based on a sample of proton-proton collision data collected at $\sqrt{s}=13\TeV$, recorded with the CMS detector at the LHC, corresponding to an integrated luminosity of 35.9\fbinv. The observed event yields are consistent with the expectations based on the standard model. The results are interpreted in simplified models of supersymmetry describing various scenarios for the production and decay of charginos and neutralinos. Depending on the model parameters chosen, mass values between 180\GeV and 1150\GeV are excluded at 95\% CL. These results significantly extend the parameter space probed for these particles in searches at the LHC. In addition, results are presented in a form suitable for alternative theoretical interpretations.}

\hypersetup{%
pdfauthor={CMS Collaboration},%
pdftitle={Search for electroweak production of charginos and neutralinos in multilepton final states in proton-proton collisions at sqrt(s) =  13 TeV},%
pdfsubject={CMS},%
pdfkeywords={CMS, physics, supersymmetry}}

\maketitle
\section{Introduction}
\label{sec:intro}
The standard model (SM) describes the vast majority of particle physics phenomena.
So far, it has withstood a multitude of challenges from precision measurements. Searches for physics beyond the SM carried out by
various experiments also have not revealed convincing evidence for the existence of such phenomena. A recent triumph of the SM is the 2012
discovery of a Higgs boson (\PH) by  the ATLAS and CMS Collaborations at the CERN LHC~\cite{Higgs1,Higgs2,Higgs3}.
However, there are several open challenges that cannot be explained by the SM, such as the  hierarchy problem~\cite{BARBIERI198863,WITTEN1981513,DIMOPOULOS1981150} (and fine tuning), and the absence of a dark matter candidate.
Supersymmetry (SUSY)~\cite{Ramond:1971gb,Golfand:1971iw,Neveu:1971rx,Volkov:1972jx,Wess:1973kz,Wess:1974tw,Fayet:1974pd,Nilles:1983ge,Martin:1997ns}
is an extension of the SM that introduces an additional symmetry between bosons and fermions, and predicts superpartners, or ``sparticles'', to the SM particles.
This extension offers a solution to several limitations of the SM, including those cited above. In particular, in the case of conserved $R$-parity~\cite{Wess:1974tw},
SUSY particles are created in pairs, and  the lightest SUSY particle (LSP) is stable, making it a possible dark matter candidate.
Furthermore, the existence of relatively light superpartners can lead to the cancellation of the large quantum corrections to the Higgs boson mass, addressing the hierarchy problem.

Thus far, no evidence for such new particles has been found. Constraints have been placed on the masses of the colored superpartners (squarks and gluinos) ranging from several hundred \GeV to about 2\TeV, depending on the assumptions entering into the models used for the interpretation of the results~\cite{Sirunyan:2017uyt,Sirunyan:2017kqq,Sirunyan:2017cwe,Sirunyan:2017uyt,Aaboud:2017faq,Aaboud:2016ejt}. The cross sections associated with electroweak production of SUSY particles are far lower than those for strong production. This directly translates into significantly lower exclusion limits, ranging from about 100 to 700\GeV~\cite{SUS-13-006,SUS-14-002,Aad:2015eda,Chatrchyan:2013sza} on the masses of sparticles produced exclusively via the electroweak interaction. This would be the dominant production mechanism of sparticles if the colored superpartners are too heavy to be produced.

This paper describes a search for direct production of charginos and neutralinos, mixtures of the SUSY partners of the electroweak gauge and Higgs bosons, decaying to two, three, or more charged leptons, and significant missing transverse momentum (\ptmiss). In events with two light leptons (electrons or muons), the leptons are required to have the same charge; in events with three or more leptons, up to two may be hadronically decaying tau leptons ($\tauh$). We use a data sample of pp collisions recorded during 2016 with the CMS detector corresponding to an integrated luminosity of \fulllumi. Similar searches have been reported by the CMS and ATLAS collaborations for the lower-energy LHC Run~1 ~\cite{SUS-13-006,SUS-14-002,Aad:2015eda}.

\section{Supersymmetric models}
\label{sec:models}
This search targets scenarios of direct electroweak production of charginos $\PSGcpmDo$ and neutralinos $\PSGczDt$,
which decay into final states containing two, three, or four charged leptons ($\Pe^{\pm},\mu^{\pm},\tau^\pm$).
The results are interpreted using simplified models~\cite{Chatrchyan:2013sza,Alves:2011wf}.
In such models, the masses and the decay modes of the relevant particles are the only free parameters.

In the case of $\PSGcpmDo\PSGczDt$ production, the $\PSGcpmDo$ and $\PSGczDt$ are assumed
to be mass-degenerate and wino-like, i.e. superpartners of the SU(2)$_L$ gauge fields,
and the \PSGczDo is set to be bino-like, i.e. a superpartner of the U(1)$_Y$ gauge field~\cite{PDG2016}.
The masses of the pure wino-like and bino-like gauginos are governed by two complex gaugino
Majorana mass parameters, $M_2$ and $M_1$, and can assume any values.
In this scenario, the \PSGczDo is the lightest SUSY particle (LSP).

For the effective $\PSGczDo\PSGczDo$ production, the $\PSGczDt$, $\PSGcpmDo$, and $\PSGczDo$ are assumed to be
higgsino-like, i.e. are superpartners of the Higgs doublets. In this case,
scenarios in which the $\PSGczDo$ is next-to-LSP (NLSP) are considered.

\subsection*{Production of $\PSGcpmDo\PSGczDt$ with 2-body decays through sleptons}

\begin{figure}[htbp]
\centering
\includegraphics[width=0.32\linewidth]{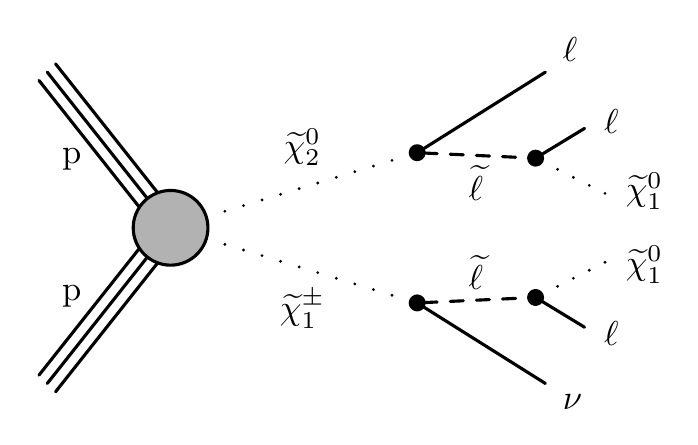}
\includegraphics[width=0.32\linewidth]{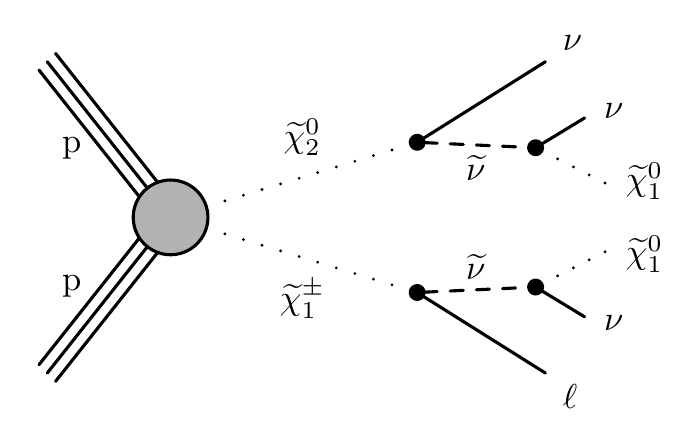}
\caption{Chargino and neutralino pair production with decays mediated by sleptons and sneutrinos.}
\label{fig:TChiNeuSlepSneu}
\end{figure}

In the first scenario considered, the charginos and neutralinos decay to leptons via intermediate sleptons or sneutrinos, the SUSY partners of charged leptons and neutrinos, as shown in Fig.~\ref{fig:TChiNeuSlepSneu}. The combination of gauge eigenstates that make up the neutralinos, charginos, and their masses, will determine whether their decays through sleptons and sneutrinos (which are assumed to be mass-degenerate) lead to all three lepton flavors with equal probability, or if they prefer to decay to $\tau$ leptons.  Three different scenarios for the decays are considered:

\begin{itemize}
  \item {$\PSGcpmDo\PSGczDt$ production with $\slep_L$-mediated decays:} The chargino and neutralino decay via sleptons or sneutrinos to all lepton flavors with the same branching fraction (``flavor-democratic'' scenario). As the decay through sleptons or sneutrinos happens with equal probability, only 50\% of the decays will lead to three-lepton final states.
  \item {$\PSGcpmDo\PSGczDt$ production with $\slep_R$-mediated decays:} In this case, because the $\slep_R$ couples to the chargino via its higgsino component, chargino decays to $\slep_R$ strongly favor the production of a $\tau$ lepton (``$\tau$-enriched'' scenario).
  The neutralino still decays to all three flavors. In this model, both left-handed sleptons and sneutrinos are considered to be heavy and decoupled; they do not participate in this process.
  \item {$\PSGcpmDo\PSGczDt$ production with $\sTau$-mediated decays:} The first- and second-generation sleptons and sneutrinos are decoupled and the chargino and neutralino only decay via a $\sTau_R$.  We will refer to this model as the ``$\tau$-dominated'' scenario. Left-handed sleptons and sneutrinos are considered to be heavy and decoupled; they do not participate in this process.
\end{itemize}

In these simplified models, the slepton mass is assumed to lie between the common $\PSGcpmDo$ and $\PSGczDt$ masses, and that of the $\PSGczDo$. In addition, the branching fraction to leptons is taken to be 100\%. Three different mass assumptions are considered: $m_{\slep} = m_{\sNu} = m_{\PSGczDo} + x\, (m_{\PSGczDt} - m_{\PSGczDo})$,  with $x =0.05$, 0.5 and 0.95. When $x = 0.05$ or 0.95, one of the three leptons is very soft and may escape detection. The same-sign (SS) final state is used in these cases to recover some of these events without the penalty of increasing the SM background.

\subsection*{Production of $\PSGcpmDo\PSGczDt$ with 2-body decays to $\PW$, $\PZ$, and Higgs bosons}

\begin{figure}[htbp]
\begin{center}
\includegraphics[width=0.32\linewidth]{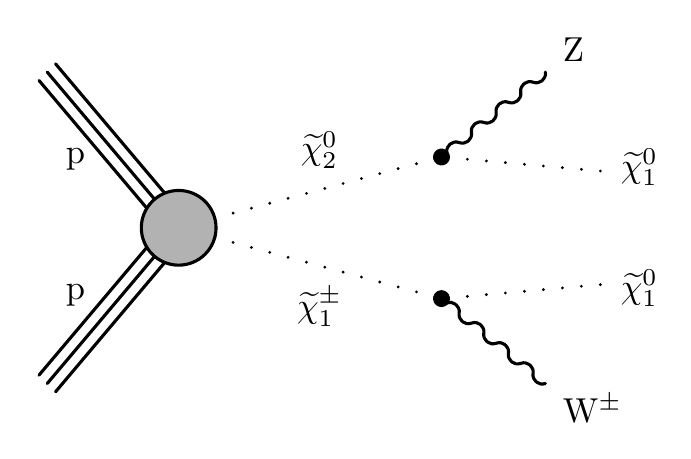}
\includegraphics[width=0.32\textwidth]{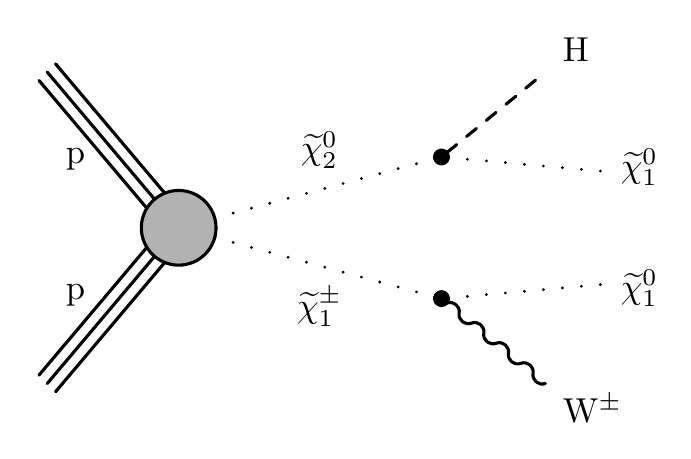}
\caption{Chargino and neutralino pair production with the chargino decaying to a $\PW$ boson and the LSP and the neutralino decaying to (left) a Z boson and the LSP or (right) a Higgs boson and the LSP. }
\label{fig:TChiNeuWZ}
\end{center}
\end{figure}

In the second scenario, we assume that the sleptons are too heavy and that the $\PSGcpmDo$ and $\PSGczDt$ undergo direct decay to the LSP via the emission of a $\PW$ boson, $\PZ$ boson, or Higgs boson as depicted in Fig.~\ref{fig:TChiNeuWZ}. The chargino decays to a $\PW$ and the $\PSGczDo$, while the neutralino can decay either to a $\PZ$ or a Higgs boson and the $\PSGczDo$.
The Higgs boson is expected to have SM-like properties and branching fraction if all the other Higgs bosons are much heavier ~\cite{Martin:1997ns}. If the Higgs boson decays to $\PW\PW$, $\PZ\PZ$, or $\tau\tau$, and each $\PW$ or $\PZ$ decays leptonically, one can expect multiple leptons in the final state. However, compared to the other models included in this analysis, the leptonic branching fractions are rather small, namely 3.3\% for $\PW\PZ$ to three leptons and 2.9\% for $\PW\PH$ to three leptons, including taus.

 \subsection*{Production of $\PSGczDo\PSGczDo$ in models with four higgsinos}

Finally we consider the production of pairs of neutralinos that decay via a \PZ or a Higgs boson.
As for the pair of neutralinos ($\PSGczDt\PSGczDt$ or $\PSGczDo\PSGczDo$) the production cross section is vanishingly small~\cite{Beenakker:1999xh,Fuks:2012qx,Fuks:2013vua},
we consider a specific gauge-mediated SUSY breaking (GMSB) model with four higgsinos ($\PSGczDt$, $\PSGcpmDo$, $\PSGczDo$)
and an effectively massless gravitino $\PXXSG$ as the LSP \cite{Matchev:1999ft,Ruderman:2011vv,Meade:2009qv}.

The cross sections for higgsino pair production are computed at NLO plus next-to-leading-log (NLL) precision in a limit of mass-degenerate higgsino states $\PSGczDt$,  $\PSGcpmDo$, and $\PSGczDo$, with all the other sparticles assumed to be heavy and decoupled.
Following the convention of real mixing matrices and signed neutralino masses~\cite{Skands:2003cj}, we set the sign of the mass of $\PSGczDo$ ($\PSGczDt$) to $+1$ ($-1$).
The lightest two neutralino states are defined as symmetric (anti-symmetric) combinations of higgsino states by setting the product of the elements $N_{i3}$ and $N_{i4}$ of the neutralino mixing matrix $N$ to $+0.5$ ($-0.5$) for $i = 1$ (2). The elements $U_{12}$ and $V_{12}$ of the chargino mixing matrices $U$ and $V$ are set to 1.

\begin{figure}[htbp]
 \centering
 \includegraphics[width=0.32\linewidth]{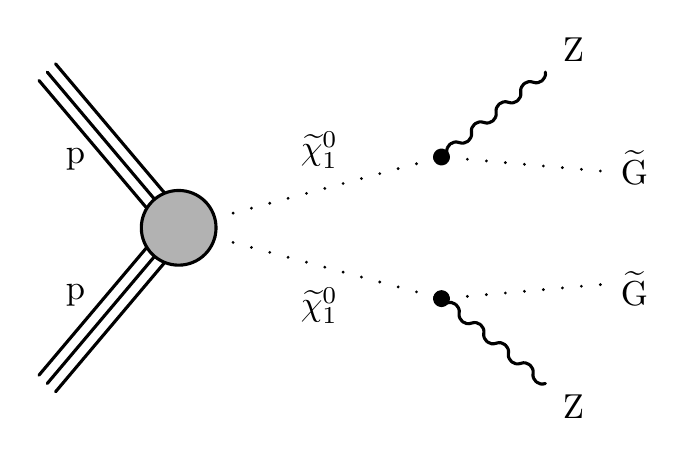}
 \includegraphics[width=0.32\linewidth]{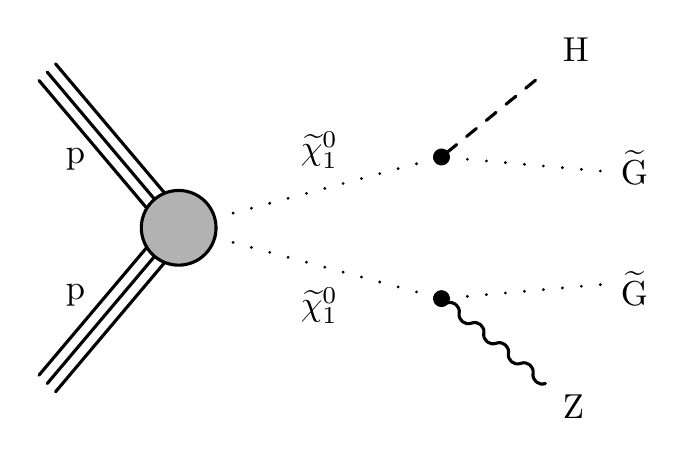}
 \includegraphics[width=0.32\linewidth]{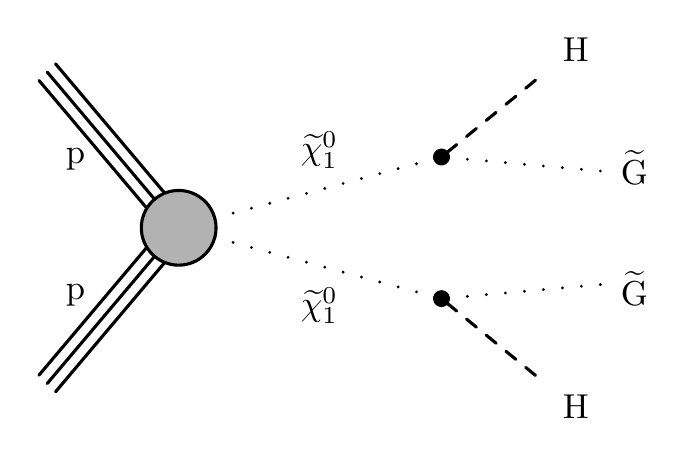}
 \caption{A GMSB model with higgsino pair production. The $\PSGczDt$, $\PSGcpmDo$, and $\PSGczDo$ are nearly mass-degenerate with $\PSGczDo$ decaying to \PZ or Higgs bosons and $\PXXSG$ LSP.}

 \label{fig:TNeuNeu}
 \end{figure}

Since the \PSGczDt, \PSGcpmDo, and \PSGczDo are nearly mass degenerate, the heavier higgsinos  \PSGcpmDo and \PSGczDt decay to the \PSGczDo via soft particles
which escape  detection.
Therefore the sum of the various possible production processes of higgsinos
($\PSGcpmDo\PSGczDt$, $\PSGczDo\PSGczDt$, $\PSGcpmDo\widetilde{\chi}^\mp_1$, $\PSGcpmDo\PSGczDo$)
describes an effective $\PSGczDo\PSGczDo$ production mechanism at the LHC.
In the considered scenario, each $\PSGczDo$ promptly decays to a \PZ or a Higgs boson and the gravitino LSP.
Different final states depending on the assumptions on the NLSP are possible, as shown in Fig.~\ref{fig:TNeuNeu}.

\section{The CMS detector}
\label{sec:cmsdectector}
The central feature of the CMS apparatus is a superconducting solenoid of 6\unit{m} internal diameter,
providing a magnetic field of 3.8\unit{T}. Within the solenoid volume are a silicon pixel
and strip tracker, a lead tungstate crystal electromagnetic calorimeter, and a brass scintillator
hadron calorimeter, each composed of a barrel and two endcap sections. Forward calorimeters
extend the pseudorapidity ($\eta$)~\cite{Chatrchyan:2008zzk} coverage provided by the barrel and endcap detectors.
Muons are measured in gas-ionization detectors embedded in the steel flux-return yoke outside the solenoid.
The first level of the CMS trigger system, composed of custom hardware processors, uses information from the calorimeters and muon detectors to select events of interest in a fixed time interval of less than 4\mus. The high-level trigger processor farm further decreases the event rate from around 100\unit{kHz} to around 1\unit{kHz}, before data storage. A more detailed description of the CMS detector, together with a definition of the coordinate system used and the relevant kinematic variables, can be found in Ref.~\cite{Chatrchyan:2008zzk}.

\label{sec:section_name}

\section{Event selection and Monte Carlo simulation}
\label{sec:eventsel}

The events are recorded if they satisfy the requirements of the CMS two-level trigger system. As we consider different lepton multiplicities in the final state, a combination of several trigger algorithms is required to cover all possible cases and to maximize the acceptance. Events with at least two light-flavor leptons ($\Pe$ or $\mu$) rely mostly on dilepton triggers with transverse momentum ($\pt$) $>$23\,(17)\GeV for the leading electron (muon) and $\pt > 12\,(8)\GeV$ for the subleading electron (muon). For events with at least two leptons, the double lepton triggers are not fully efficient. Single lepton triggers are used to increase the acceptance for these events.

For the final state with two $\tauh$s and one electron or muon, we use single lepton triggers requiring an isolated $\Pe$ ($\mu$) with $\pt > 27\,(24)\GeV$. Typical trigger efficiencies for leptons satisfying the offline selection criteria described below are 98\%\,(92\%) per electron (muon). In final states with three or more leptons, the total trigger efficiency is close to 100\%.

In the offline analysis, the information from all subdetectors is combined by the CMS particle-flow algorithm~\cite{Sirunyan:2017ulk}. The particle flow algorithm provides a global interpretation of the event and reconstructs and identifies individual particles. The particles are classified into charged hadrons, neutral hadrons, photons, electrons, and muons.

We require electrons to have $\abs{\eta} < 2.5$ to ensure that they are within the tracking volume, and $\pt >$ 10$\GeV$. The particle flow algorithm for electrons uses a multivariate discriminant, built from variables that characterize the shower shape and track quality~\cite{Khachatryan:2015hwa}. To exclude electrons from photon conversions, we reject candidates that have missing measurements in the innermost layers of the tracking system or are matched to a secondary conversion vertex candidate~\cite{Khachatryan:2015hwa}.

Muon candidates are reconstructed by combining the information from the silicon tracker and the muon spectrometer in a global fit~\cite{Chatrchyan:2012xi}. An identification selection is performed using the quality of the geometrical matching between the tracker and the muon system measurements. Only muons within the muon system acceptance $\abs{\eta}<2.4$ having a minimum $\pt$ of 10$\GeV$ are considered.

Light lepton candidates are required to be consistent with originating from the primary vertex, defined as the one with the largest $\pt^2$ sum of the tracks pointing to that vertex \cite{Chatrchyan:2014fea}. The transverse impact parameter $d_0$ of a lepton track with respect to this vertex must not exceed 0.5 mm, and the longitudinal displacement  $d_z$ of that impact point must not exceed 1.0 mm. Additionally they must satisfy a requirement on the impact parameter significance $S_{\mathrm{IP3D}} \equiv |d_{\rm 3D}|/\sigma(d_{\rm 3D}) < 8$, where $d_{\rm 3D}$ is the three-dimensional displacement with respect to the primary vertex and $\sigma(d_{\rm 3D})$ is its uncertainty as estimated from the track fit. Furthermore, leptons are required to be isolated. An isolation variable \miniIso~\cite{Rehermann:2010vq,Khachatryan:2016kod} is computed as the ratio of the scalar \pt sum of charged hadrons, neutral hadrons, and photons within a cone around the lepton candidate direction at the vertex to the transverse momentum $\pt(\ell)$ of the lepton candidate. The cone radius $\Delta R=\sqrt{(\Delta\eta)^2+(\Delta\phi)^2}$  (where $\phi$ is the azimuthal angle in radians) depends on $\pt(\ell)$ as:

\begin{equation}
 \Delta R \left(\pt(\ell)\right) = \frac{10\GeV}{\min\left[\max\left(\pt(\ell), 50\GeV\right), 200\GeV\right]}.
  \end{equation}

The varying isolation cone definition takes into account the increased collimation of the decay products
of a particle as its $\pt$ increases, and it reduces the inefficiency from accidental overlap between the lepton and other objects in an event. Loosely isolated leptons are required to have $\miniIso < 0.4$. Electrons and muons that pass all the aforementioned requirements are referred to as \textit{loose} in this analysis.

In order to discriminate between leptons (``prompt'' leptons) originating from decays of heavy particles, such as $\PW$ and $\cPZ$ bosons, or SUSY particles, and those produced in hadron decays or in photon conversions, as well as misidentified hadrons (``nonprompt'' leptons), we use a multivariate discriminator based on a boosted decision tree (BDT)~\cite{Roe:2004na,Khachatryan2014}. This BDT takes the following variables as inputs:  $d_\mathrm{0}$, $d_\mathrm{z}$, $S_{\mathrm{IP3D}}$, and $\miniIso$; variables related to the jet closest to the lepton, such as the ratio between the $\pt$ of the lepton and the $\pt$ of the jet ($\ptRatio$), the b tagging discriminator value of the jet, the number of charged particles in the jet, and the \ptRel variable:
 \begin{equation}
    \ptRel=\frac{\left|\left(\vec{p}(\text{jet})-\vec{p}(\ell)\right) \times \vec{p}(\ell)\right| }{|\vec{p}(\text{jet})-\vec{p}(\ell)|}.
  \end{equation}

Other identification variables, such as the muon segment compatibility and the electron identification multivariate discriminant are also included.
The BDT is trained using simulation with prompt leptons from $\ttbar\cPZ$ and with nonprompt leptons from $\ttbar$ processes. Leptons satisfying a requirement on this discriminant in addition to the loose requirements are referred to as \textit{tight} leptons. Two working points are defined, one with higher efficiency for the three or more lepton channel and one with high nonprompt background rejection for the SS dilepton channel. The identification efficiency measured in data for electrons passing tight criteria varies between 40~(20)\% for tracks with $\pt<20\GeV$  in the barrel (endcap) region and the plateau efficiency of 90~(80)\% for those with $\pt>50\GeV$ in the barrel (endcap) region, while the misidentification rate for non-prompt electrons is between 3\% and 7\% depending on the \pt. For muons, the efficiency is between 82\% for $\pt<20\GeV$  and 100\% for $\pt>40\GeV$, the misidentification rate for non-prompt muons goes from 2\% up to 10\% depending on \pt.

The $\tauh$ candidates are reconstructed with the hadron-plus-strips  algorithm~\cite{Khachatryan:2015dfa}. They  are   required   to  pass   the   ``decay  mode   finding''
discriminator~\cite{Khachatryan:2015dfa}, selecting one-  or three-prong decay modes, with or  without additional $\pi^0$ particles. In  addition, they  must  fulfill  $\pt >  20$\GeV,  $\abs{\eta}  < 2.3$,  and  isolation  requirements computed in  a cone defined by
$\Delta{R}  =   0.5$ centered on the $\tauh$ direction.
   The  typical   $\tauh$  identification
efficiency of  these selection requirements is  50\%, while the  jet misidentification rate is  well below
0.1\%~\cite{CMS-PAS-TAU-16-002}.

Particle-flow candidates are clustered into jets using the anti-$\kt$ algorithm~\cite{Cacciari:2008gp} with a distance parameter of 0.4, as implemented in the \FASTJET package~\cite{Cacciari:fastjet1,Cacciari:fastjet2}. Jets are required to satisfy quality requirements~\cite{Chatrchyan:2011ds} to remove those likely arising from anomalous energy deposits. Charged hadrons are not considered if they do not originate from the selected primary vertex. After the estimated contribution of neutral particles from additional $\Pp\Pp$ interactions in the same beam crossing (pileup) is subtracted by using the average \pt in the event per unit area~\cite{Cacciari:2007fd}, jet energies are corrected for residual nonuniformity and nonlinearity of the detector response using simulation and data~\cite{Chatrchyan:2011ds,Khachatryan:2016kdb}. Only jets with $\pt > 25\GeV$, $\abs{\eta} < 2.4$, and separated from any lepton candidate by $\Delta{R} > 0.4$ are retained.

To identify jets originating from b quarks, the combined secondary vertex algorithm~\cite{Chatrchyan:2012jua,BTV-16-002} is used. Jets with $\pt > 25\GeV$ and $\abs{\eta}<2.4$ are considered b quark jets (``b jets") if they satisfy the requirements of the medium working point of the algorithm. These requirements result in an efficiency of approximately 63\% for tagging a bottom quark jet, and a mistagging rate of 1.5\% for light-flavor jets, as measured in \ttbar events in data. Simulated events are corrected for the differences in the performance of the algorithm between data and simulation. Events with at least one identified b jet are rejected in the analysis to reduce the \ttbar background. 

The missing transverse momentum $\ptmiss$ is obtained as the magnitude of the negative vector sum $\ptvecmiss$ of the transverse momenta of all reconstructed particle-flow candidates
consistent with originating from the primary vertex and is further adjusted to account for jet energy corrections applied to the event~\cite{JME-13-003}.
This quantity is used in the definitions of the search regions presented in the following sections.

The Monte Carlo (MC) simulated samples, which include the effects of pileup, are used to estimate the  background from SM processes with prompt leptons (see Section~\ref{sec:backgrounds}) and to calculate the selection efficiency for various new-physics scenarios. The samples are reweighted to match the pileup profile in the data. The SM background samples are  produced with the \MGvATNLO v2.2.2 or v2.3.3 generator~\cite{MADGRAPH5} at leading-order (LO)~\cite{Alwall:2007fs} or next-to-leading-order (NLO)~\cite{Frederix:2012ps} accuracy in perturbative quantum chromodynamics, including up to one or two additional partons in the matrix element calculations.
The exceptions are the diboson samples, which are  produced with the \POWHEG~v2.0~\cite{Melia:2011tj,Nason:2013ydw} generator without
additional partons in the matrix element calculations. The NNPDF3.0LO~\cite{Ball:2014uwa}  parton distribution functions (PDFs) are used for the simulated samples generated at LO and the NNPDF3.0NLO~\cite{Ball:2014uwa} PDFs for those generated at NLO. Parton showering and hadronization  are described using the \PYTHIA~8.212 generator~\cite{Sjostrand:2007gs} with the CUETP8M1  tune~\cite{Skands:2014pea,CMS-PAS-GEN-14-001}.
A double counting of the partons generated with \MGvATNLO and those with \PYTHIA is removed using
the MLM~\cite{Alwall:2007fs} and the FXFX~\cite{Frederix:2012ps} matching schemes, in the LO and NLO samples, respectively.
The CMS detector response for the background samples is  modelled with the \GEANTfour package~\cite{Geant}.

Signal samples are generated with \MGvATNLO at LO precision, including up to two additional partons
in the matrix element calculations. The NNPDF3.0LO~\cite{Ball:2014uwa}  parton distribution functions (PDFs) are used. Parton showering and hadronization are modelled with \PYTHIA as described above. SUSY particles, are also modelled with \PYTHIA, while the detector simulation is performed with the CMS fast simulation package~\cite{Abdullin:2011zz}. Any residual differences in the detector response description between the \GEANTfour and fast simulation are corrected for, with corresponding uncertainties in the signal acceptance taken into account.

\section{Search strategy}
\label{sec:strategy}
This search is designed to target the scenarios described in Section~\ref{sec:models} of direct electroweak production of charginos $\PSGcpmDo$ and neutralinos $\PSGczDt$ leading to final states with either two, three, or four leptons and little hadronic activity. The specific strategy of the analysis is guided by the assumption that $R$-parity is conserved, hence leading to the presence of particles in the final states that evade detection, yielding a sizable \ptmiss.

The small cross section of the electroweak production drives the analysis design, which includes all the possible final states to enhance the discovery potential. Therefore, the analysis is subdivided into several categories defined by the number of leptons in the event, their flavors, and their charges. Each of these categories is further subdivided into bins defined by the kinematic variables that enhance the discrimination against SM backgrounds and the sensitivity to possible mass hierarchies of new particles.

Among the SM processes yielding the same final states as those targeted in this search there are WZ production, nonprompt leptons, external and internal (where the emitted photon is virtual) conversions, rare SM processes (\ie, multiboson production or single-boson production in association with a \ttbar pair), and charge misidentification. The dominant source of background varies depending on the considered category and thus the search strategy is tailored accordingly.

\subsection{Two same-sign dilepton category} \label{sec:dilepton}

Although most of the targeted models naturally yield three charged lepton final states, for compressed-spectrum scenarios, e.g. when the mass splitting between the next-to-lightest SUSY particle and the LSP is small, one of the leptons from the decay chain of a neutralino can be very soft, such that it may not fulfill the selection requirements. By accepting events with two SS leptons, we recover some of these missing events while keeping the SM background under control.

We require two SS leptons with $\pt>25\,(20)\GeV$ for the leading and $\pt>15\,(10)\GeV$ for the trailing electron (muon), no third lepton with $\pt>20,10,20$\GeV (\Pe,$\mu$,$\tau$) passing the tight identification criteria, and $\ptmiss > 60\GeV$. To suppress the dominant $\PW\PZ$ background, events are vetoed if they contain an opposite-sign same-flavor (OSSF) pair formed from loose electrons or muons within a $\pm$15\GeV window around the $\PZ$ boson mass, taken to be 91\GeV. To reduce the contribution from the processes with low-mass resonances, events are vetoed if they contain an OSSF pair with an invariant mass below 12\GeV.

The events are first divided into two categories: with and without a jet of $\pt > 40\GeV$. Signal processes would populate the one-jet category when accompanied by initial-state radiation (ISR). In compressed scenarios where the electroweak production and decay of sparticles produces limited \ptmiss, the compensating boost to the sparticle system may raise the laboratory \ptmiss above the selection threshold. Further binning is done in \ptmiss, the minimum transverse mass ($\MT = \sqrt{\smash[b]{2\ptmiss \pt^\ell[1-\cos(\Delta\phi)]}}$) computed for each lepton, and the $\pt$ of the dilepton system ($\pt^{\ell\ell}$). The bins with enough events are also split by charge to help constrain charge-asymmetric backgrounds. This categorization is summarized in Table~\ref{tab:regionsSS}.

\begin{table}[tbh]
\centering
\topcaption{Search regions for events with two SS light-flavor leptons.}
\label{tab:regionsSS}
\resizebox{\textwidth}{!}{
\begin{tabular}{|c|c|c|c|c|c|c|}
\hline
 $\Njets$ & 	$\MT$ (\GeVns{}) & 	$\pt^{\ell\ell}$ (\GeVns{})  &  $\ptmiss < 100\GeV $ & $100 \le \ptmiss < 150\GeV $ & $150 \le \ptmiss < 200\GeV$ & $\ptmiss \ge 200\GeV$\\
\hline\hline
\multirow{6}{*}{0} & \multirow{4}{*}{$<$100 } & \multirow{2}{*}{$<$50 } & \multirow{2}{*}{SS01} & SS02 ($++$) & \multirow{2}{*}{SS04} & \multirow{2}{*}{SS05}\\
  &     & & & SS03 ($--$) & & \\ \cline{3-7}
  & &  \multirow{2}{*}{$>$50 } & \multirow{2}{*}{SS06} & SS07 ($++$) & \multirow{2}{*}{SS09}&  \multirow{2}{*}{SS10} \\
  &     & & & SS08 ($--$) & & \\ \cline{2-7}
  & \multirow{2}{*}{$>$100 } &  & \multirow{2}{*}{SS11} & SS12 ($++$) & \multirow{2}{*}{SS14} & \multirow{2}{*}{SS15}\\
  &     & & & SS13 ($--$) & & \\ \hline

\multirow{6}{*}{1} & \multirow{4}{*}{$<$100 } & \multirow{2}{*}{$<$50 } & \multirow{2}{*}{SS16} & SS17 ($++$) & \multirow{2}{*}{SS19} & \multirow{2}{*}{SS20}\\
  &     & & & SS18 ($--$) & & \\ \cline{3-7}
  & &  \multirow{2}{*}{$>$50 } & \multirow{2}{*}{SS21} & SS22 ($++$) & \multirow{2}{*}{SS24}&  \multirow{2}{*}{SS25} \\
  &     & & & SS23 ($--$) & & \\ \cline{2-7}
  & \multirow{2}{*}{$>$100 } &  & \multirow{2}{*}{SS26} & SS27 ($++$) & \multirow{2}{*}{SS29} & \multirow{2}{*}{SS30}\\
  &     & & & SS28 ($--$) & & \\ \hline
							
\end{tabular}}
\end{table}

\subsection{Three or more leptons} \label{sec:trilepton}
Most of the targeted models described in Section~\ref{sec:models} and depicted in Figs.~\ref{fig:TChiNeuSlepSneu} and \ref{fig:TChiNeuWZ} yield three isolated leptons and significant \ptmiss in the final state.

Events are selected on the condition that they have $\ptmiss > 50\GeV$ and contain at least three leptons, of which at most two are $\tauh$ candidates. The leading electron (muon) must satisfy $\pt > 25\,(20)\GeV$, while the subleading electron (muon), if present, must satisfy $\pt > 15\,(10)\GeV$. These criteria are driven by the $\pt$ thresholds of the dilepton triggers used in the analysis. If the leading lepton is a muon and the other leptons are electrons or $\tauh$ candidates, the muon threshold is increased to $\pt > 25\GeV$. For events with one $\Pe$ or $\mu$ and two $\tauh$, all leptons are additionally constrained to have $\abs{\eta} < 2.1$, and the electron (muon) must have $\pt > 30\,(25)\GeV$. These requirements are imposed to ensure that the selected events have a high efficiency with respect to the required triggers. To reduce the contribution from processes with low-mass resonances, events are vetoed if they contain an OSSF loose-lepton pair with an invariant mass below 12\GeV. Additionally, in events with exactly three leptons containing an OSSF pair of two $\Pe$ or $\mu$, the invariant mass of three leptons is required not to be consistent with the mass of a $\PZ$ boson ($\abs{M_{3\ell}-M_{\PZ}} > 15\GeV$) in order to suppress contributions from asymmetric photon conversions.

These events are then classified according to the number of identified leptons and their flavor. We distinguish between final states with three and more than three leptons and among final states with differing \tauh content as follows:

\begin{itemize}
  \item Events with three light-flavor leptons (electrons or muons).
  \item Events with two light-flavor leptons and a $\tauh$.
  \item Events with one light-flavor lepton and two $\tauh$s.
  \item Events with at least four light-flavor leptons and no $\tauh$.
  \item Events with at least three light-flavor leptons and one $\tauh$.
  \item Events with at least two light-flavor leptons and two $\tauh$.
\end{itemize}

These categories are then further subdivided according to their kinematic properties to define the different signal regions. Further binning of the events in the aforementioned categories is described in detail in the remainder of this Section.

\subsubsection{Three light leptons (signal regions A and B)} \label{sec:3light}
In most of the cases two out of the three leptons ($\Pe$ or $\mu$) will form an OSSF pair. This is signal region { ``A"}. We further divide the events into three bins of invariant mass of the dilepton pair, $\Mll$, in order to separate processes that include a Z boson in the decay chain from those that do not.
One of the $\Mll$ bins is defined to be below the $\PZ$ mass ($\Mll < 75\GeV$) and the second one contains the events with $\Mll$ above the $\PZ$ mass ($\Mll > 105\GeV$),  which enhances the sensitivity to the scenarios with various mass splittings between the $\PSGczDt$ and $\PSGczDo$.
The third one is defined as the Z mass window ($75 \le \Mll < 105\GeV $), and it is expected to contain the bulk of the standard model background events. In the case of three same-flavor leptons, the OSSF pair with the invariant mass closest to the mass of the $\PZ$ boson is used.
The transverse mass $\MT$ of the third lepton in the event is computed with respect to \ptmiss. For the SM $\PW\PZ\to3\ell\nu$ process,
the $\MT$ variable computed in this way is steeply falling around the $\PW$ mass, and facilitates discrimination against the dominant background from $\PW\PZ$ production in the search.
Both variables, $\MT$ and \ptmiss, are used to further categorize the events with most of the standard model background expected in low $\MT$ and \ptmiss bins.
These signal regions are summarized in Table~\ref{tab:regionsA}.

\begin{table}[tbh]
\centering
\topcaption{Search regions corresponding to category A, events with three electrons or muons that form at least one opposite-sign same-flavor (OSSF) pair. Search region A15$^\dagger$
overlaps with the $\PW\PZ$ control region of the analysis, and is not used in the interpretation.}
\label{tab:regionsA}
\resizebox{0.8\textwidth}{!}{
\begin{tabular}{|c|c|c|c|c|}
\hline
$\MT$ (\GeVns{}) & \ptmiss (\GeVns{}) & $\Mll < 75\GeV$ & $75 \le \Mll < 105\GeV $ & $\Mll\ge105\GeV $\\
\hline\hline
\multirow{7}{*}{$ 0 - 100$}   & $50-100  $ & A01 & A15$^\dagger$ & A32 \\ \cline{2-5}
                              & $100-150 $ & A02 & A16 & A33 \\ \cline{2-5}
                              & $150-200 $ & A03 & A17 & A34 \\ \cline{2-5}
                              & $200-250 $ & A04 & A18 & A35 \\ \cline{2-5}
                              & $250-400 $ & \multirow{3}{*}{  A05} & A19 & \multirow{3}{*}{  A36} \\ \cline{2-2}\cline{4-4}
                              & $400-550 $ &     & A20 &     \\ \cline{2-2}\cline{4-4}
                              & $\geq$550  &     & A21 &     \\ \hline
\multirow{4}{*}{$ 100 - 160$} & $50-100  $ & A06 & A22 & A37 \\ \cline{2-5}
                              & $100-150 $ & A07 & A23 & A38 \\ \cline{2-5}
                              & $150-200 $ & A08 & A24 & A39 \\ \cline{2-5}
                              & $\geq$200  & A09 & A25 & A40 \\ \hline
\multirow{6}{*}{$\geq$160 }   & $50-100  $ & A10 & A26 & A41 \\ \cline{2-5}
                              & $100-150 $ & A11 & A27 & A42 \\ \cline{2-5}
                              & $150-200 $ & A12 & A28 & A43 \\ \cline{2-5}
                              & $200-250 $ & A13 & A29 & \multirow{3}{*}{  A44} \\ \cline{2-4}
                              & $250-400 $ & \multirow{2}{*}{  A14} & A30 &   \\ \cline{2-2}\cline{4-4}
                              & $\geq$400  &     & A31 &   \\ \hline
\end{tabular}}
\end{table}

Signal region { ``B"} contains the events in which no OSSF pair is found, further sorted into two bins each for $\Mll$ and $\MT$. The low $\MT$ bins are then further subdivided into two \ptmiss bins. Most of these events arise from a leptonic decay of $\PZ\to\tau\tau$; therefore, the $\Mll$ is calculated from the opposite-sign (OS) dilepton pair whose invariant mass is closest to the mean dilepton mass determined from $\PZ\to\tau\tau$ simulation, which is 50\GeV. An event lacking an OS pair (all three leptons have the same sign) most likely arises from $\PZ\to\tau\tau$  with one lepton charge misidentified. In this case the event is assigned to the lowest $\Mll$ bin, and the $\MT$ is taken to be the minimum $\MT$ calculated from any of the three leptons and the \ptmiss. These signal regions are summarized in Table~\ref{tab:regionsB}. For this category, the dominant background arises from nonprompt leptons.

\begin{table}[tbh]
\centering	
\topcaption{Search regions corresponding to category B, events with three $\Pe$ or $\mu$ that do not form an opposite-sign same-flavor (OSSF) pair.}
\label{tab:regionsB}
\resizebox{0.6\textwidth}{!}{
\begin{tabular}{|c|c|c|c|}
\hline
$\MT$ (\GeVns{}) & \ptmiss (\GeVns{}) & $\Mll < 100\GeV$ & $\Mll\ge100\GeV $\\
\hline\hline
\multirow{2}{*}{$0-120$}   & $50-100$  & B01  & B04 \\ \cline{2-4}
                              & $>$100     & B02  & B05 \\ \hline
$>$120                       & $>$50      & B03  & B06 \\ \hline
\end{tabular}}
\end{table}

\subsubsection{Three leptons with at least one $\tauh$ (signal regions C to F)}
\label{sec:3l1tau}

A third category { ``C''} is built from events with two  $\Pe$ or $\mu$ forming an OSSF pair and a $\tauh$; it uses the same three $\Mll$ bins as in category { ``A''}, again in order to separate off-Z and on-Z regions. For these events, the two-lepton ``stransverse mass'', $\MTT$~\cite{MT2variable,MT2variable2}, replaces $\MT$ for the further subdivision of the bins, as $\MTT$ is a more powerful discriminator with respect to the dominant $\ttbar$ background containing two leptons and two neutrinos in the final state. $\MTT$ is computed with the leptons that are most likely to be the prompt ones from a $\PW$ boson decay, and in case of $\ttbar$ processes it has an endpoint at the $\PW$ boson mass; it is computed as: 

\begin{equation}
\MTT = {\underset{\vec{p}_\text{T1}^\text{miss}+\vec{p}_\text{T2}^\text{miss} = \ptvecmiss}{\min}}\Bigl[\max\bigl\{\MT(\ptvec^{\ell_1}, \vec{p}_\text{T1}^\text{miss}),\MT(\ptvec^{\ell_2}, \vec{p}_\text{T2}^\text{miss}) \bigr\} \Bigr],
\end{equation}

where the minimization is done over all possible momenta $\vec{p}_\text{T1}^\text{miss}$ and $\vec{p}_\text{T2}^\text{miss}$ summing to the observed $\ptvecmiss$.
The probability to misidentify $\tauh$ is significantly larger than that to misidentify an electron or muon; hence $\MTT$ is computed with a pair of light leptons in this category. The $\MTT$ bins are defined so that the vast majority of the SM backgrounds are at low $\MTT$, especially the $\ttbar$ contribution. For the signal regions containing a $\PZ$ boson candidate, the categorization in terms of $\MTT$ is not performed. The complete set of requirements defining the signal regions for events in this category  is given in Table~\ref{tab:regionsC}.

For events with a $\tauh$ and two light leptons that do not form an OSSF pair (\ie, $\Pe^{\pm}\Pe^{\pm}$, $\mu^{\pm}\mu^{\pm}$, $\mu^{\pm}\Pe^{\mp}$, $\mu^{\pm}\Pe^{\pm}$), the OS pair, if present, with the invariant mass closest to the corresponding dilepton mass expected from a $\PZ\to\tau\tau$ decay (50\GeV for $\Pe\mu$ and 60\GeV for $\Pe\tauh$ or $\mu\tauh$) is used for the event categorization. If no OS pair is present, the event is counted in the lowest $\Mll$ bin.
An additional splitting of the regions high in SM background
with $\Mll < 100\GeV$ is introduced to enhance the sensitivity to various new-particle spectra. Further categorization is performed depending on whether the $\Pe$ or $\mu$ form an OS (category { ``D''}) or SS (category { ``E''}) pair. The final search region binning is shown in Tables~\ref{tab:regionsD} and~\ref{tab:regionsE}. The $\MTT$ variable is computed with a pair of the OS light leptons if it is present, otherwise a light lepton leading in $\pt$ and a $\tauh$ is used.

The last category (category { ``F''}) includes events with two $\tauh$s and an $\Pe$ or $\mu$, for which the binning is shown in Table~\ref{tab:regionsF}. The $\MTT$ variable is computed with the light lepton and the leading $\tauh$. For all these categories ( {``C''} to {``F''}), the dominant source of background arises from nonprompt leptons.

\begin{table}[tbh]
\centering
\caption{Search region definition corresponding to category C, events with two $\Pe$ or $\mu$ forming an opposite-sign same-flavor (OSSF) pair and one $\tauh$ candidate.
Regions where there is a $\PZ$ boson candidate are not split into $\MTT$ categories.}
\label{tab:regionsC}
\resizebox{0.8\textwidth}{!}{
\begin{tabular}{|c|c|c|c|c|}
\hline
\ptmiss (\GeVns{}) & $75 \le \Mll < 105\GeV $ & $\MTT(\ell_{1},\ell_{2})$ (\GeVns{}) & $\Mll < 75\GeV$ & $\Mll\ge105\GeV$\\
\hline\hline
$50-100  $ & C06                  & \multirow{7}{*}{$ 0 - 100$} & C01 & C12 \\ \cline{1-2}\cline{4-5}
$100-150 $ & C07                  &                             & C02 & C13 \\ \cline{1-2}\cline{4-5}
$150-200 $ & C08                  &                             & C03 & C14 \\ \cline{1-2}\cline{4-5}
$200-250 $ & \multirow{2}{*}{  C09} &                             & C04 & C15 \\ \cline{1-1}\cline{4-5}
$250-300 $ &                         &                             & \multirow{3}{*}{  C05} & \multirow{3}{*}{  C16} \\ \cline{1-2}
$300-400 $ & C10                  &                             &        &  \\ \cline{1-2}
$\geq$400  & C11                  &                             &        &  \\ \hline
$50-200 $  &                         & \multirow{2}{*}{$\geq$100 } & \multicolumn{2}{c|}{  C17} \\ \cline{1-1}\cline{4-5}
$\geq$200  &                         &                             & \multicolumn{2}{c|}{  C18} \\ \hline
\end{tabular}}
\end{table}

\begin{table}[tbh]
\centering
\topcaption{Search region definition corresponding to category D, events with one $\Pe$ and one $\mu$ of OS and one $\tauh$ candidate.}
\label{tab:regionsD}
\resizebox{0.8\textwidth}{!}{
\begin{tabular}{|c|c|c|c|c|}
\hline
$\MTT(\ell_{1},\ell_{2})$ (\GeVns{}) & \ptmiss (\GeVns{}) & $\Mll < 60\GeV$ & $60 \le \Mll < 100\GeV$ & $\Mll\ge100\GeV $\\
\hline\hline
\multirow{5}{*}{$ 0 - 100$}   & $50-100  $ & D01 & D06 & D11 \\ \cline{2-5}
                              & $100-150 $ & D02 & D07 & D12 \\ \cline{2-5}
                              & $150-200 $ & D03 & D08 & D13 \\ \cline{2-5}
                              & $200-250 $ & D04 & D09 & \multirow{2}{*}{  D14} \\ \cline{2-4}
                              & $\geq250 $ & D05 & D10 &  \\ \hline
\multirow{2}{*}{$\geq$100 }   & $50-200  $ & \multicolumn{3}{c|}{D15$\ \  $} \\ \cline{2-5}
                              & $\geq$200  & \multicolumn{3}{c|}{D16$\ \  $} \\ \hline
\end{tabular}}
\end{table}

\begin{table}[tbh]
\centering
\topcaption{Search region definition corresponding to category E, events with two $\Pe$ or $\mu$ of same sign and one $\tauh$ candidate.}
\label{tab:regionsE}
\resizebox{0.8\textwidth}{!}{
\begin{tabular}{|c|c|c|c|c|}
\hline
$\MTT(\ell_{1},\tau)$ (\GeVns{}) & \ptmiss (\GeVns{}) & $\Mll < 60\GeV$ & $60 \le \Mll < 100\GeV$ & $\Mll\ge100\GeV $\\
\hline\hline
\multirow{5}{*}{$ 0 - 100$}   & $50-100  $ & E01 & E06 & \multirow{5}{*}{  E11} \\ \cline{2-4}
                              & $100-150 $ & E02 & E07 &  \\ \cline{2-4}
                              & $150-200 $ & E03 & E08 &  \\ \cline{2-4}
                              & $200-250 $ & E04 & E09 &  \\ \cline{2-4}
                              & $\geq$250  & E05 & E10 &  \\ \hline
$\geq$100                     & $\geq 50 $ & \multicolumn{3}{c|}{E12$\ \  $} \\ \hline
\end{tabular}}
\end{table}

\begin{table}[tbh]
\centering
\topcaption{Search region definition corresponding to category F, events with one electron or muon and two $\tauh$ candidates.}
\label{tab:regionsF}
\resizebox{0.6\textwidth}{!}{
\begin{tabular}{|c|c|c|c|}
\hline
$\MTT(\ell,\tau_{1})$ (\GeVns{}) & \ptmiss (\GeVns{}) & $\Mll < 100\GeV$ & $\Mll\ge100\GeV $\\
\hline\hline
\multirow{6}{*}{$ 0 - 100$}   & $50-100  $ & F01 & F07 \\ \cline{2-4}
                              & $100-150 $ & F02 & F08 \\ \cline{2-4}
                              & $150-200 $ & F03 & F09 \\ \cline{2-4}
                              & $200-250 $ & F04 & \multirow{3}{*}{  F10} \\ \cline{2-3}
                              & $250-300 $ & F05 &  \\ \cline{2-3}
                              & $\geq$300  & F06 &  \\ \hline
\multirow{2}{*}{$\geq$100 }   & $50-200  $ & \multicolumn{2}{c|}{  F11} \\ \cline{2-4}
                              & $\geq$200  & \multicolumn{2}{c|}{  F12} \\ \hline
\end{tabular}}
\end{table}

\subsubsection{More than three leptons (signal regions G to K)} \label{sec:4l}
The remaining signal regions comprise events with at least four leptons. This category benefits from much lower SM backgrounds compared to the three-lepton category, but suffers from low branching fractions for the signals considered.

The signal regions are formed according to the number of OSSF pairs (with any lepton entering at most one pair) and the number of $\tauh$s in the event.
This separation is motivated by the production of a $\PZ$ or Higgs boson in the decay chain that would then decay into two light-flavor leptons or two $\tauh$ candidates.

The data are further subdivided in intervals of \ptmiss so as to more efficiently discriminate between signal and background. The search region definitions and their notations are summarized in Table~\ref{tab:regions4L}.

\begin{table}[tbh]
\centering
\topcaption{Search region definition corresponding to categories G--K, events with four or more leptons. Categorization is made based on the number of OSSF pairs (nOSSF).}
\label{tab:regions4L}
\resizebox{0.7\textwidth}{!}{
\begin{tabular}{|c|c|c|c|c|c|}
\hline
\multirow{2}{*}{\ptmiss (\GeVns{})} & \multicolumn{2}{c|}{$0\tauh$} & $1\tauh$ & \multicolumn{2}{c|}{$2\tauh$} \\ \cline{2-6}
                                 & $\mathrm{nOSSF}\geq 2$ & $\mathrm{nOSSF}<2$ & $\mathrm{nOSSF}\geq 0$&$\mathrm{nOSSF}\geq 2$&$\mathrm{nOSSF}<2$\\
\hline\hline
$0-50    $ & G01 & H01 & I01 & J01 & K01 \\ \hline
$50-100  $ & G02 & H02 & I02 & J02 & K02 \\ \hline
$100-150 $ & G03 & H03 & I03 & J03 & \multirow{3}{*}{K03} \\ \cline{1-5}
$150-200 $ & G04 & \multirow{2}{*}{H04} & \multirow{2}{*}{I04} & \multirow{2}{*}{J04} & \\\cline{1-2}
$\geq$200  & G05 &     &     &     & \\ \hline
\end{tabular}}
\end{table}

\subsection{Aggregated signal regions} \label{sec:regionsSSR}
To facilitate the use of these results to test models not included in Section~\ref{sec:models} of this paper, we provide a set of ``aggregated signal regions'' defined by much simpler selections. Typically, the sensitivity to the signal models obtained by using the aggregate regions is weaker by a factor of two compared with that of the full analysis. The definitions of all aggregated regions are summarized in Table~\ref{tab:regionsSSRmulti}.

\begin{table}[tbh]
\centering
\topcaption{Definition of the aggregated regions for multilepton and two SS dilepton final states.}
\label{tab:regionsSSRmulti}
\resizebox{0.95\textwidth}{!}{
\begin{tabular}{|c|l|l|}
\hline
Bin & Final state & Definition \\
\hline\hline
1 & 2 SS leptons &  0 jets, $\MT>100\GeV$ and $\ptmiss>140\GeV$  \\
2 & 2 SS leptons &  1 jet , $\MT<100\GeV$, $\pt^{\ell\ell}<100\GeV$ and $\ptmiss>200\GeV$ \\
3 & 3 light leptons & $\MT>120\GeV$ and $\ptmiss>200\GeV$ \\
4 & 3 light leptons & $\ptmiss>250\GeV$  \\
5 & 2 light leptons and 1 tau & $\MTT(\ell_1,\tau)>50\GeV$ and $\ptmiss>200\GeV$ \\
6 & 1 light lepton and 2 taus & $\MTT(\ell,\tau_1)>50\GeV$ and $\ptmiss>200\GeV$ \\
7 & 1 light lepton and 2 taus & $\ptmiss>75\GeV$ \\
8 & more than 3 leptons & $\ptmiss>200\GeV$ \\ \hline
\end{tabular}
}
\end{table}

\section{Backgrounds}
\label{sec:backgrounds}
The SM backgrounds leading to the final states under consideration can be divided into the following categories:
\begin{itemize}
\item { WZ or W$\gamma^*$ production:} When both $\PW$ and $\PZ$ or $\gamma^*$ bosons decay leptonically, these events produce the same signature as the new physics scenarios targeted by this analysis: three energetic and isolated leptons and a sizable $\ptmiss$ due to a neutrino from the $\PW$ boson decay. This source is by far the dominant background in the searches with three $\Pe$ or $\mu$, including an OSSF dilepton pair. A SS dilepton signature may also be produced when the W boson is accompanied by a $\gamma^{*}$ or off-shell Z boson, when one of the leptons from the Z or $\gamma^{*}$ decay fails the applied selection criteria (such as a $\PZ$ boson mass veto or a minimum $\pt$ requirement on a vetoed lepton), or when the $\PZ$ boson decays to $\tau$ leptons yielding a semileptonic (one $\tau$ decays hadronically and one decays to leptons) final state.

\item { Nonprompt e, $\boldsymbol\mu$,  and $\boldsymbol\tau_\text{ h}$:} Depending on the lepton multiplicity, this background is dominated by $\WJ$ (especially in the SS dilepton regions), \ttbar, or Drell--Yan processes. This category contributes the largest background contribution in the trilepton search regions, either that contain a $\tauh$ candidate, or that do not contain an OSSF pair.

\item { External and internal conversions:} These processes contribute to the SS dilepton or trilepton final state when the production of a $\PW$ or a $\PZ$ boson is accompanied by radiation of an initial- or final-state photon and this photon undergoes an asymmetric internal or external conversion in which one of the leptons has very low $\pt$. This soft lepton has a high probability of failing the selection criteria of the analysis, leading to a reconstructed two- (in case of a $\PW$ boson) or three-lepton (in case of a $\PZ$ boson) final state. This background mostly contributes to categories with an OSSF pair and to final states
with two SS leptons.

\item { Rare SM processes with multiple prompt leptons:} Rare SM processes that yield a SS lepton pair or three or more leptons include multiboson production (two or more bosons, including any combination of \PW, \PZ, \PH, or a prompt $\gamma$), single-boson production in association with a \ttbar pair, and double parton scattering. Some of these processes have a very small production rate, and are in some cases further suppressed by the b jet veto. The contribution of such processes is estimated from MC simulation.

\item { Electron charge misidentification:} A background from charge misidentification arises from events  with an OS pair of isolated $\Pe\mu$ or $\Pe\Pe$ in which the charge of one of the electrons is misreconstructed. In most cases, this arises from severe bremsstrahlung in the tracker material. This is a small background, manifesting itself in the SS dilepton category or in the category with a SS dilepton pair and a $\tauh$ candidate.
\end{itemize}

The \WZ background contribution is normalized to data in a dedicated control region containing events with three light leptons: only events with an OSSF pair with an invariant mass of   $75 <\Mll<105\GeV$  are selected. Additional requirements on these events are: $\MT < 100\GeV$ and $35<\ptmiss<100\GeV$. The purity of this WZ selection is approximately 86\%. This definition overlaps with the search region SR~A15 of the trilepton search category. As a consequence, the latter region is not used for the interpretation of the results in terms of new-physics models.

A good description of the $\MT$ distribution in our \WZ simulation is crucial in this search, especially in the tail where new physics may appear. The tail of the $\MT$ distribution is a result of, in order of importance, the accidental usage of a wrong pair of leptons to compute the mass of the Z candidate and the $\MT$ of the W candidate (``mispairing'' of the leptons), the $\ptmiss$ resolution, and the W boson width. The prediction of lepton mispairing from simulation is confirmed in a control sample in the data similar to the one described above but only allowing events with an OSSF pair of different flavor than the third lepton and using the OS pair of leptons of different flavor in the $\Mll$ computation. More details on these checks can be found in Section~\ref{sec:systematics}. 

The background from  nonprompt light leptons is estimated using  the ``tight-to-loose'' ratio method, which is described in detail in Ref.~\cite{Khachatryan:2016kod}.  The probability for a loosely defined light lepton to pass the full set of selection criteria  is measured in a multijet data sample enriched in nonprompt leptons, called  the measurement region. Once measured, this  probability is applied in a sample of  events that pass the  full kinematic selection, but  where at least one  of the leptons fails the  nominal selection while passing the loose requirements, in  order to predict the  number of events from nonprompt  leptons entering each search region. The  contribution from nonprompt $\tauh$ leptons is  estimated in a  similar way. This  time, the ``tight-to-loose''  ratio is measured  in a $\ZJ$ enriched  control sample in data  in which a $\tauh$  candidate is required to  be present in addition to  an OSSF  pair consistent  with the $\PZ$  boson decay.  The residual  contribution from prompt leptons in the  measurement and application regions is subtracted using  MC simulation. It is verified in  both MC simulation  and low-$\ptmiss$ data control  regions that this  method describes the background from the nonprompt leptons entering  the different search regions within a systematic uncertainty of 30\%.

The modelling of the conversion background is verified in a data control region enriched in both
external and internal conversions from the $\PZ$+jets process with $\Z \to\ell\ell\gamma^{(*)}$ and
$\gamma^{(*)}\to\ell\ell$, where one of the leptons is out of acceptance.
This control region is defined by $\abs{\Mll - M_\PZ} > 15\GeV$, $\abs{M_{3\ell} - M_\PZ} < 15\GeV$,
and $\ptmiss < 50\GeV$.
The expected background yields are found to agree with the observed counts in data within the statistical uncertainties.
Scale factors are derived for the modelling of the asymmetric conversions to electrons or muons
in the $\PZ\gamma^{(*)}$ process after subtraction of the residual nonprompt lepton and $\PW\Z$ backgrounds.
The scale factors are found to be $1.04\pm0.11$ and $1.25\pm0.24$, respectively, and are used to derive the systematic uncertainty on this process.

The charge misidentification background in the SS dilepton channel is estimated by reweighting the events with OS lepton pairs by the charge misidentification probability. For electrons, this probability is obtained from simulation and cross-checked on an on-Z $\Pe^{\pm}\Pe^{\pm}$ control region in data, and is in the range $10^{-5}--10^{-3}$ depending on the electron's \pt and $\eta$. Studies of simulated events indicate that the muon charge misidentification probability is negligible.

\section{Systematic uncertainties}
\label{sec:systematics}
The systematic uncertainties in the background estimates and signal acceptance affect both
the overall normalization of the yields and the relative populations of processes in the search regions.
The systematic uncertainties considered in this analysis are summarized in Table~\ref{tab:systSummary}.

Experimental uncertainties include those in the lepton selection efficiency, the trigger efficiency, the jet
energy scale, and the $\cPqb$ tag veto efficiency.

Lepton identification efficiencies are computed with the tag-and-probe technique~\cite{Chatrchyan:2012xi,Khachatryan:2015hwa}, with an uncertainty of 3\% per lepton. The $\tauh$ identification efficiency is similarly determined with
an uncertainty of 5\%~\cite{CMS-PAS-TAU-16-002}.

The trigger efficiency uncertainty is obtained from measuring efficiencies of all trigger components
with the tag-and-probe technique, and is estimated to be less than ~3\%.

\begin{table}[h!]
\centering
    \topcaption{Summary of systematic uncertainties in the event yields in the search regions and their treatment. Uncertainties are allowed to vary only the normalization of all the bins at once, or both the shape and the normalization (allowing for different correlations across the bins).
The upper group lists uncertainties related to experimental effects for all processes whose yield is estimated from simulation;
the middle group lists uncertainties in these yields related to the event simulation process itself.
The third group lists uncertainties for background processes whose yield is estimated from data. Finally, the last group describes
uncertainties related to the extraction of the signal acceptance in MC simulation.}
    \label{tab:systSummary}
\small
\begin{tabular}{lcc}
\hline
Source          & Estimated uncertainty (\%) & Treatment \\
\hline
$\Pe/\mu$ selection &  3  & normalization \\
$\tauh$ selection &  5  & normalization \\
Trigger efficiency &  3 & normalization \\
Jet energy scale & 2--10 & shape \\
$\cPqb$ tag veto & 1--2 & shape \\
Pileup & 1--5 & shape\\
Integrated luminosity & 2.5 & normalization \\
\hline
Scale variations and PDF (\ttZ and \ttW) & 15 & normalization \\
Theoretical (ZZ) & 25 & normalization \\
Conversions & 15 & normalization \\
Other backgrounds & 50 & normalization \\
Monte Carlo statistical precision & 1--30 & normalization \\
\hline
Nonprompt leptons (closure) & 30 & normalization \\
Nonprompt leptons ($\PW/\cPZ$ bkg. subtraction) & 5--20 & shape \\
Charge misidentification & 20 & normalization \\
\WZ normalization & 10 & normalization\\
\WZ shape & 5--50 & shape\\
\hline
ISR uncertainty                           & 1--5 & shape  \\
Scale variations for signal processes     & 1--2 & shape  \\
Lepton efficiencies				& 2 & normalization \\
Signal acceptance (\ptmiss modelling) & 1--5 & shape \\
\hline
\end{tabular}
\end{table}

The jet energy scale uncertainty varies between 2 and 10\%, depending on the $\pt$ and $\eta$ of the jet. This uncertainty affects other event quantities such as the $\cPqb$ tag veto, $\ptmiss$, $\MT$, and $\MTT$, and is computed by shifting the energies of all jets coherently and propagating the variation to all of these kinematic variables. Correlation effects due to the migration of events from one search region to another are taken into account. These variations yield estimated uncertainties ranging from 2 to 10\% in the simulated signal and background yields in the different search regions. Similarly, the $\cPqb$ jet veto efficiency is corrected for the differences between data and simulation, and an associated uncertainty with this correction is 1--2\%, which takes into account the kinematic difference between signal events and those used to measure such efficiency. The uncertainty in the modelling of pileup is computed by modifying slightly the pileup reweighting profile and  is measured to be 1--5\%, depending on the search region. The uncertainty in the integrated luminosity is 2.5\%~\cite{CMS-PAS-LUM-17-001}.

The dominant source of background for most of the event categories described in Section \ref{sec:strategy} is $\WZ$ production. Systematic uncertainties are derived from detailed evaluation of the modelling of this process. The uncertainty in the normalization of the $\WZ$ background is 10\%. This includes statistical uncertainties in the yields in the control sample used for normalization, and in the subtraction of the non-$\WZ$ contributions to the sample. An additional uncertainty stems from the modelling of the $\MT$ shape in the simulation of the $\WZ$ process. To estimate the effect of a potentially different $\ptmiss$ resolution and W width in data and simulation on the W boson $\MT$ distribution shape, we verify the $\MT$ shape prediction from simulation in $\PW\gamma$ and $\WJ$ control samples in data. After applying a threshold $\pt > 40 \GeV$ on the photon to suppress the contribution of $\PW\gamma$ events produced by final-state radiation (FSR) which influences $\MT$, we find the $\PW$ boson $\MT$ distribution shapes in simulated $\PW \gamma$, $\WJ$, and WZ processes to be the same. We thus proceed to measure the W$\gamma$ and $\WJ$ $\MT$ shape in a dedicated control sample in which an energetic, well-identified, and isolated photon passing the aforementioned $\pt$ threshold is required, together with a lepton passing the same criteria as those selected in the trilepton search regions, and $\ptmiss > 50\GeV$. A minimum separation of $\Delta R > 0.3$ is required between the lepton and the photon to further reduce the FSR contribution. The residual $\WJ$ contribution in this control sample is about 20\%, and consists of events where a jet has very high electromagnetic fraction, and thus is not subject to large mismeasurements that might influence the $\MT$ tail. Residual contamination from processes other than $\PW \gamma$ or $\WJ$ is subtracted using MC, and the $\MT$ shape measured in this control region is compared to the one predicted by the $\WZ$ simulation. The measured shape is found to agree well with the prediction from simulation within the statistical uncertainties, and the precision of this comparison is used to derive systematic uncertainties on the high-$\MT$ bins of the trilepton search. This uncertainty is between 5 and 40\%, from the comparison of the $\MT$ shape in the $\WZ$ simulation with the one measured in the $\PW\gamma$ control region. The size of this uncertainty increases for higher $\MT$ values and is driven by the statistical uncertainty of the $\PW\gamma$ control sample. Systematic uncertainties on the modelling of $\Mll$ are neglected as lepton momentum scale and resolution effects are too small to have significant effects in this analysis~\cite{Chatrchyan:2012jua,Chatrchyan:2014mua}. 

Further uncertainties in background yields estimated from simulations arise from the unknown higher-order effects in the theoretical calculations of the cross sections, and from uncertainties in the knowledge of the proton PDFs. The uncertainties from the PDFs are estimated by using the envelope of several PDF sets~\cite{Butterworth:2015oua}.
The effect of these theoretical uncertainties is $15\%$ for \ttW and \ttZ, and $25\%$ for ZZ backgrounds. Theoretical uncertainties are also considered for the remaining minor backgrounds estimated purely from simulation, in which 15\% uncertainty is assigned to processes with a prompt $\gamma$ modelled with NLO accuracy corresponding to the precision of the scale factor measured in a dedicated control region, and 50\% to other rare processes.

Other sources of uncertainties are associated with the backgrounds that are derived from, or normalized in, data control samples. The nonprompt background prediction has an uncertainty of 30\% assigned to both light-lepton and $\tauh$ cases. This uncertainty arises from the performance of the method in the simulation (closure) in various regions of parameter space, and is given by the observed deviations between the estimated and observed yields in the control sample.

The uncertainty in the measurement of the charge misidentification background is derived from the difference between the yields of on-Z $\Pe^{\pm}\Pe^{\pm}$ events in data and simulation. This uncertainty is found to be 20\%.

The \MADGRAPH modelling of ISR, which affects the total transverse
momentum ($\pt^\text{ISR}$) of the system of SUSY particles, is improved by reweighting the $\pt^\text{ISR}$
distribution in MC SUSY events. This reweighting procedure is based on studies of the transverse momentum
of $\PZ$ events in data~\cite{Chatrchyan:2013xna}.
The reweighting factors range between 1.18 at $\pt^\text{ISR}\approx125\GeV$ and
0.78 for $\pt^\text{ISR}>600\GeV$.
The deviation of the reweighting factors from 1.0 is taken as the systematic uncertainty
on the reweighting procedure.

Additional uncertainties in the signal acceptance extraction are considered for the renormalization and factorization scale variations by simultaneously varying them by a factor of 2 up and down~\cite{Beenakker:1999xh,Fuks:2012qx,Fuks:2013vua}, \ptmiss modelling and lepton efficiencies due to the differences in simulations between signal and background samples.
\section{Results}
\label{sec:results}
\begin{figure}[!hbtp]
\centering
\includegraphics[width=0.45\textwidth]{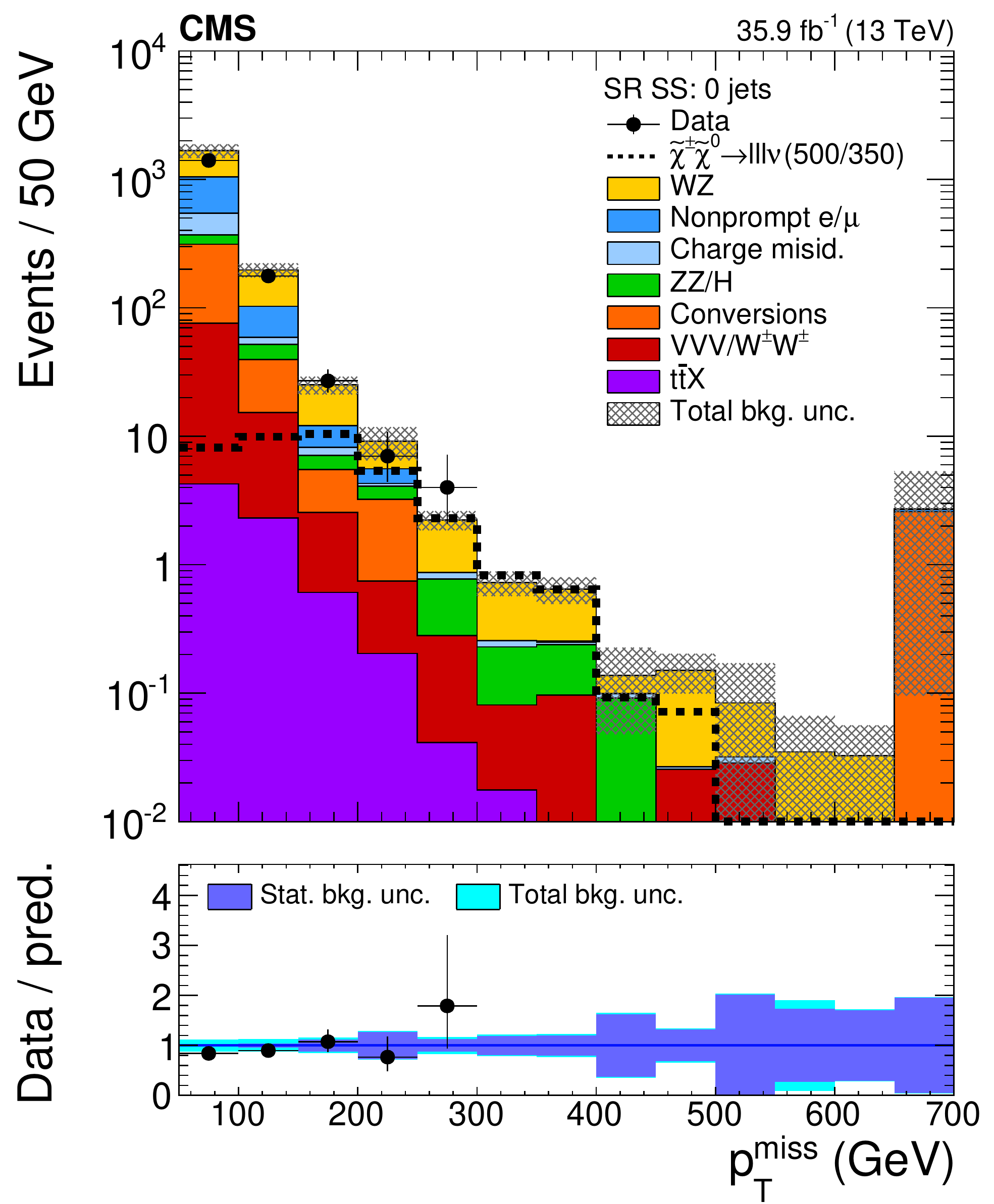}
\includegraphics[width=0.45\textwidth]{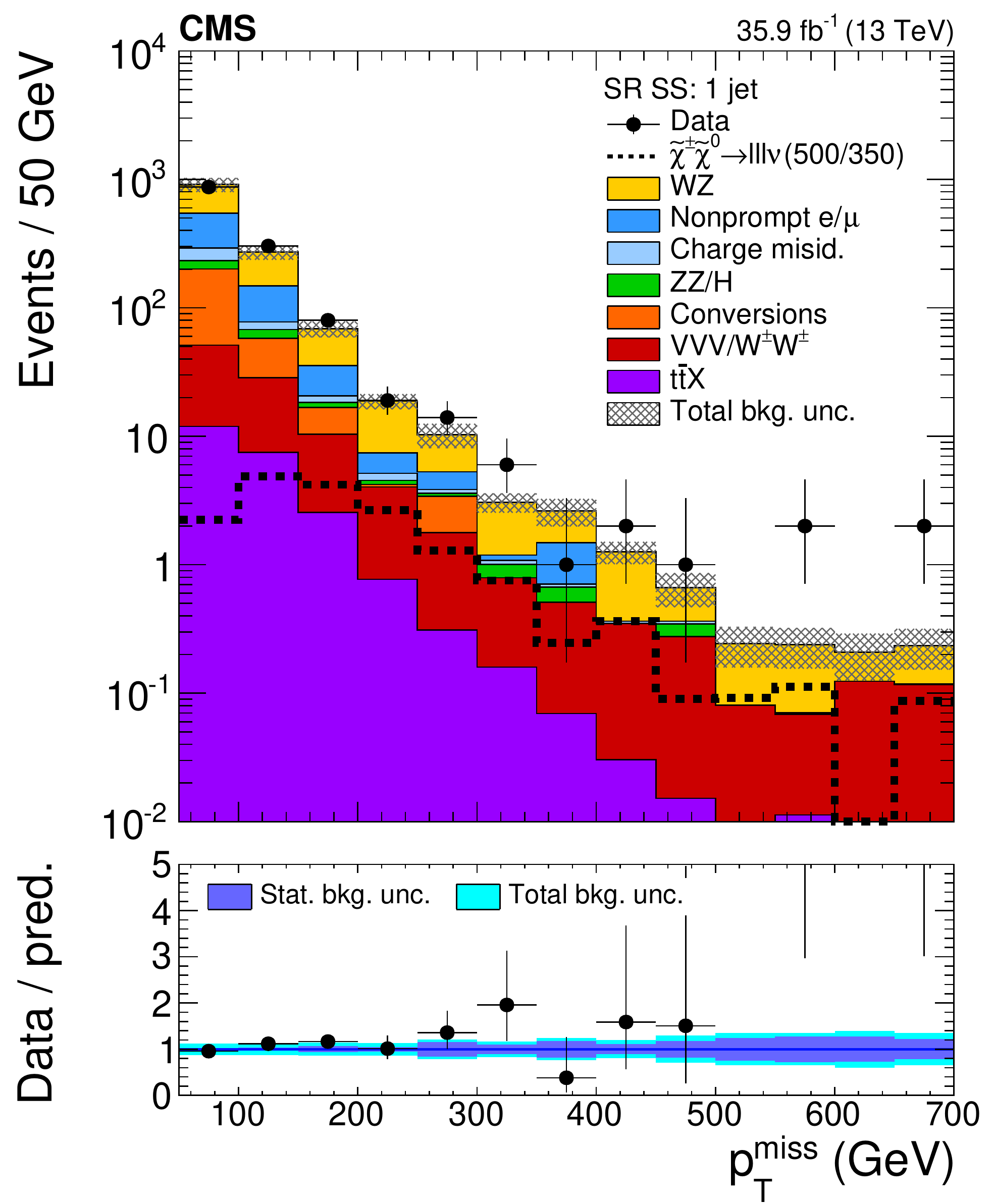}
\caption{Distribution of $\ptmiss$ for events with 2 SS leptons and
0 jets (left) or 1 jet (right). An example signal mass point in the flavor-democratic
model with mass parameter $x=0.05$ is displayed for illustration. The numbers in the
parentheses denote the $\PSGcpmDo$ and $\PSGczDo$ masses, namely $m_{\PSGcpmDo}=m_{\PSGczDt}=500\GeV$
and $m_{\PSGczDo}=350\GeV$. The last bin contains the overflow events.}
\label{fig:yields:2lss:met}
\end{figure}

\begin{figure}[!hbtp]
\centering
\includegraphics[width=0.45\textwidth]{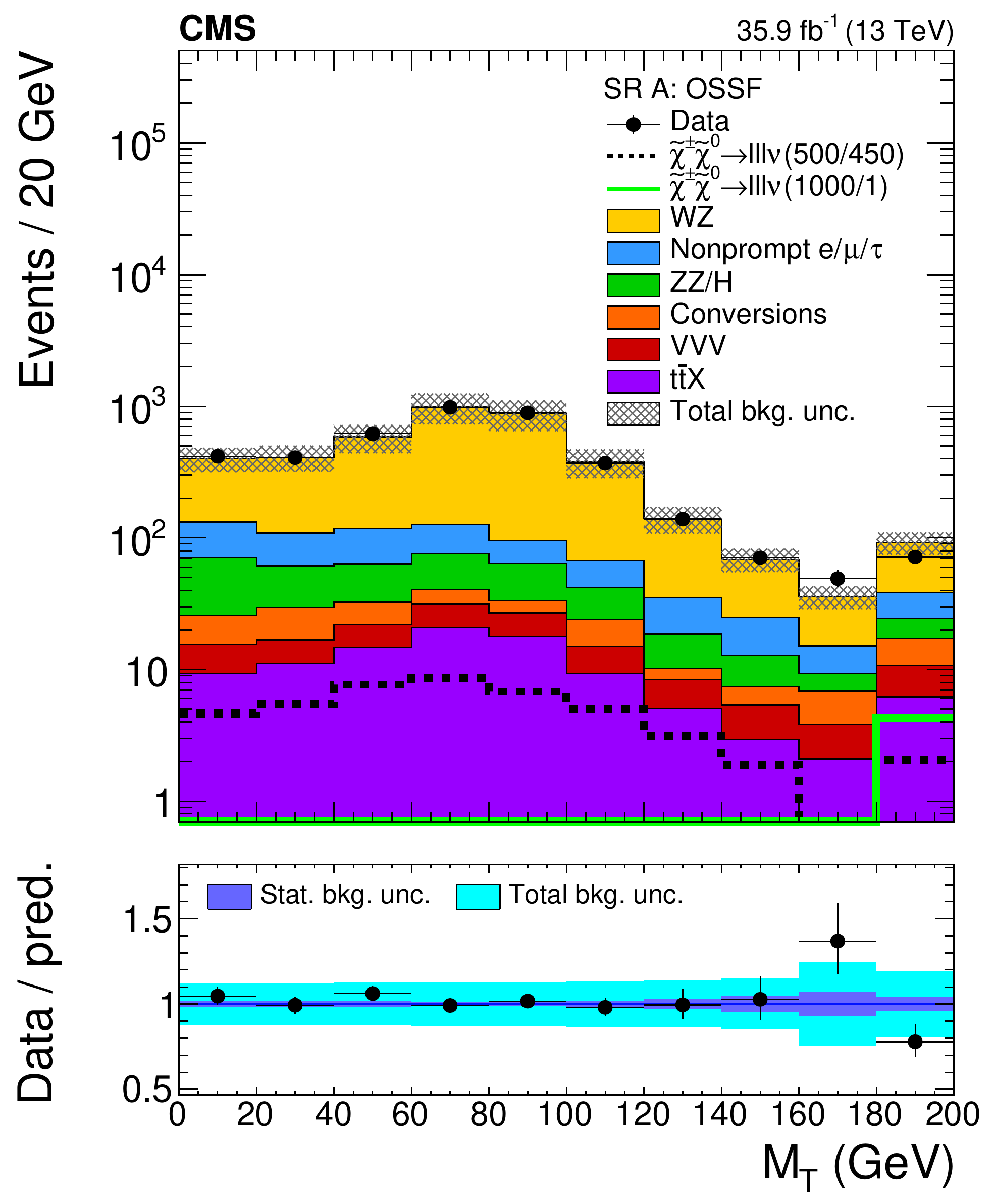}
\includegraphics[width=0.45\textwidth]{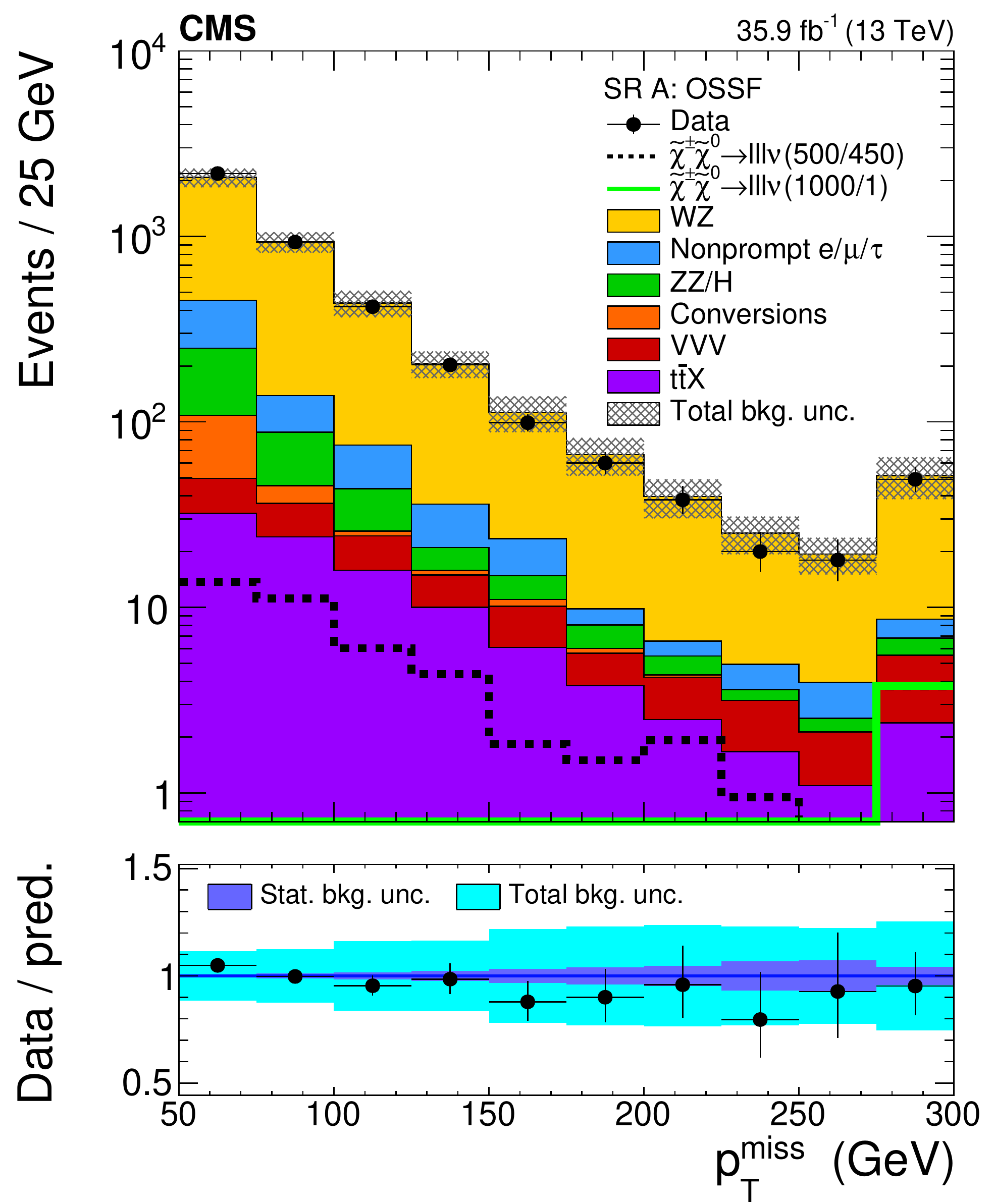} \\
\includegraphics[width=0.45\textwidth]{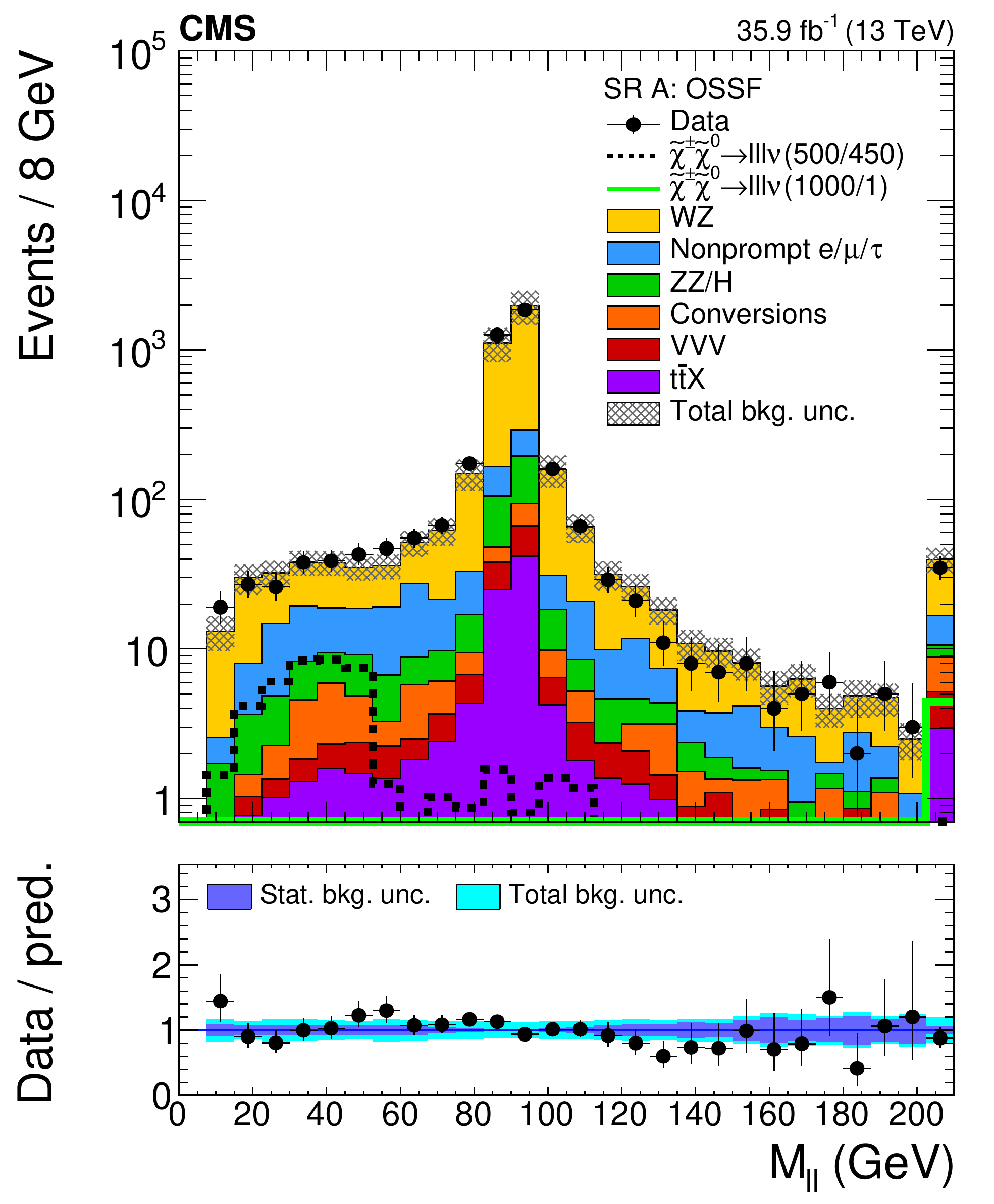}
\caption{Distribution of kinematical observables used in the event selection
for events entering signal regions A: the transverse mass of the third lepton
(upper left), the $\ptmiss$ (upper right) and the $\Mll$ of the OSSF pair (lower).
Two signal mass points in the flavor-democratic model with mass parameter
$x=0.5$ are displayed for illustration. The notation is analogous to that
used in Fig.~\ref{fig:yields:2lss:met}.}
\label{fig:yields:BRA:evt}
\end{figure}

\begin{figure}[!hbtp]
\centering
\includegraphics[width=0.45\textwidth]{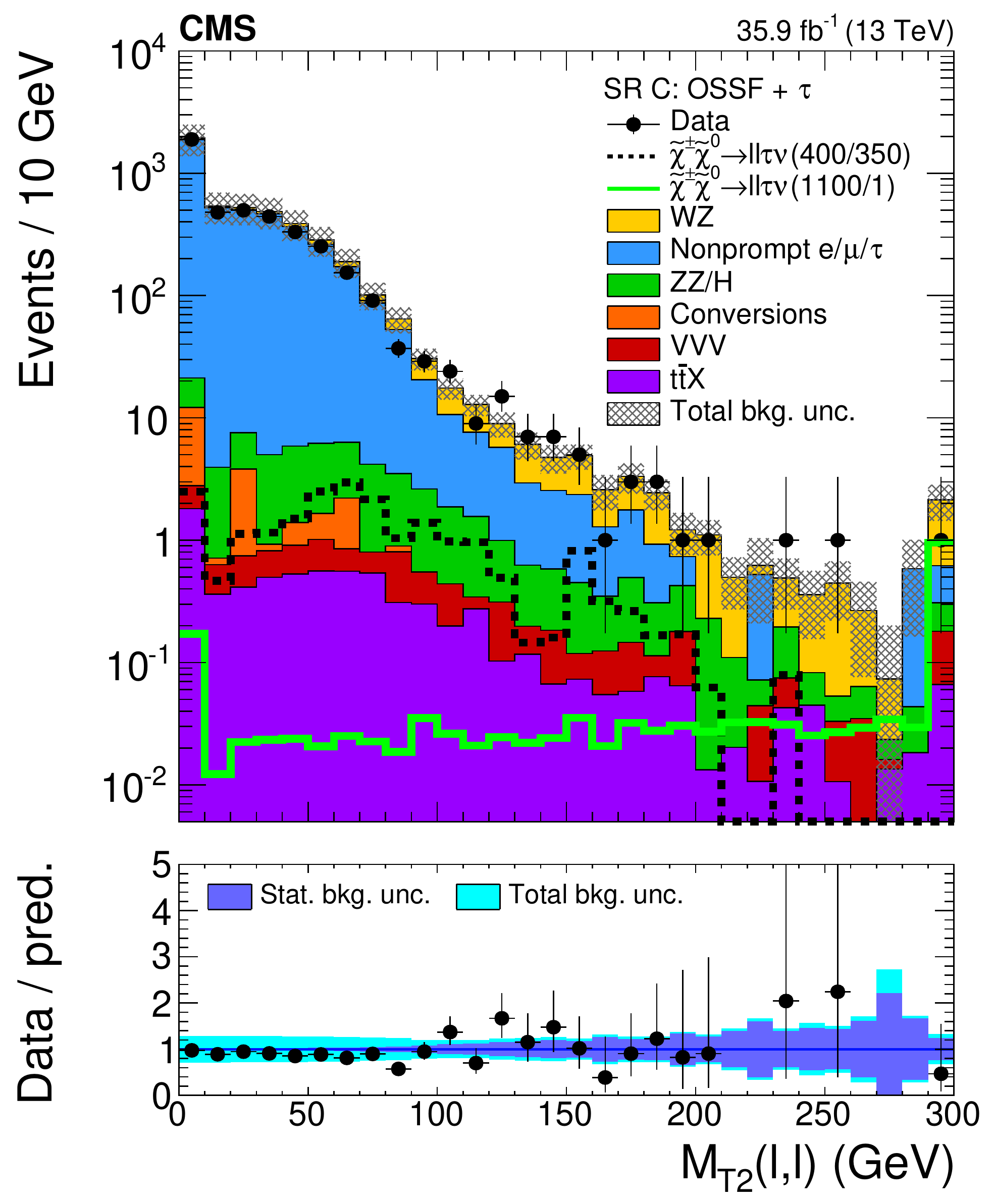}
\includegraphics[width=0.45\textwidth]{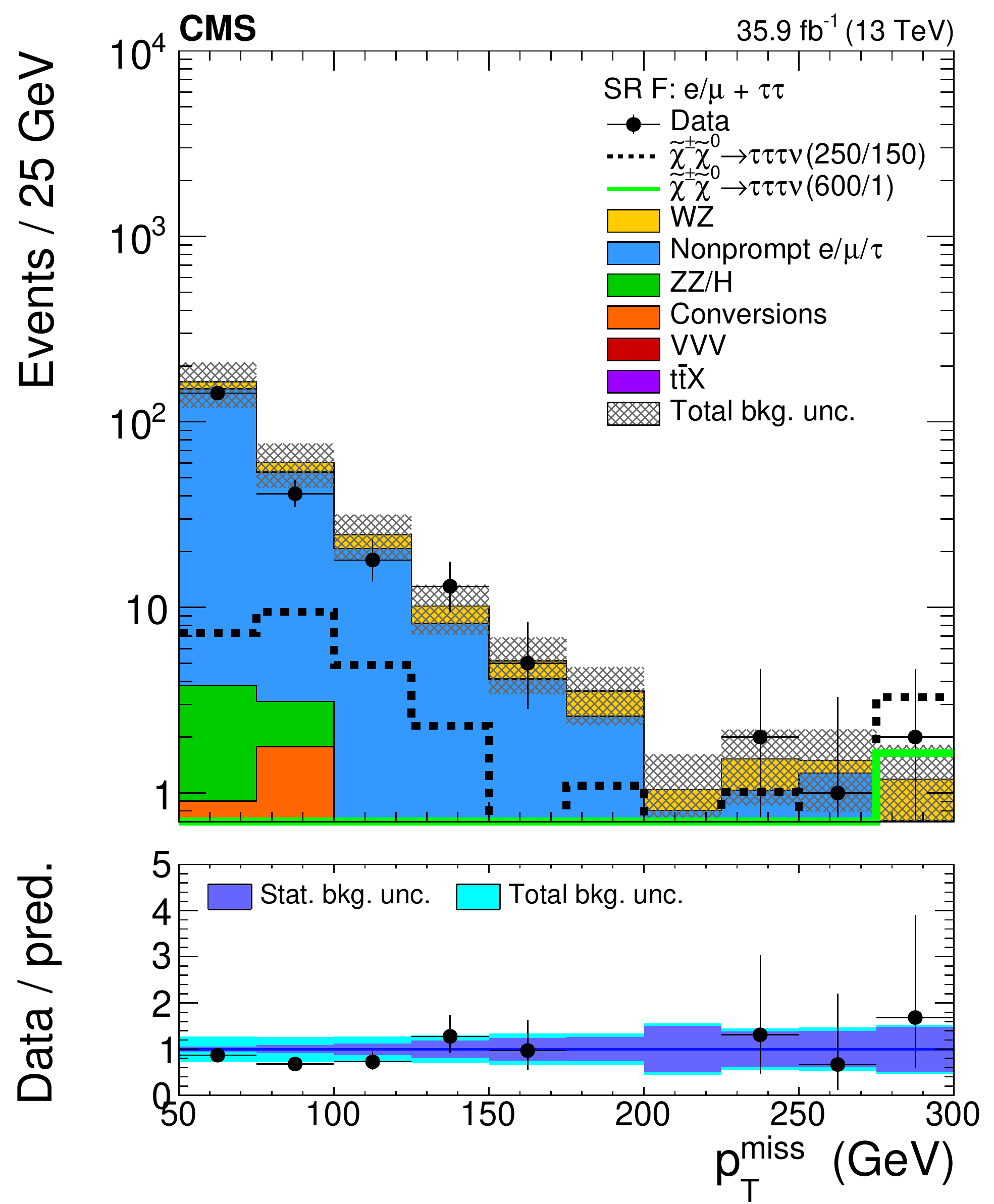}
\caption{Distributions in the stransverse mass for events with two OSSF light leptons and one $\tauh$ (left) and in $\ptmiss$ for events with one light-flavor lepton and two $\tauh$s. Two signal mass points in the  $\tau$-enriched (left) and the $\tau$-dominated (right) scenarios with mass parameter $x=0.5$ are displayed for illustration. The notation is analogous to that used in Fig.~\ref{fig:yields:2lss:met}. }
\label{fig:yields:BRCF:evt}
\end{figure}

\begin{figure}[!hbtp]
\centering
\includegraphics[width=0.45\textwidth]{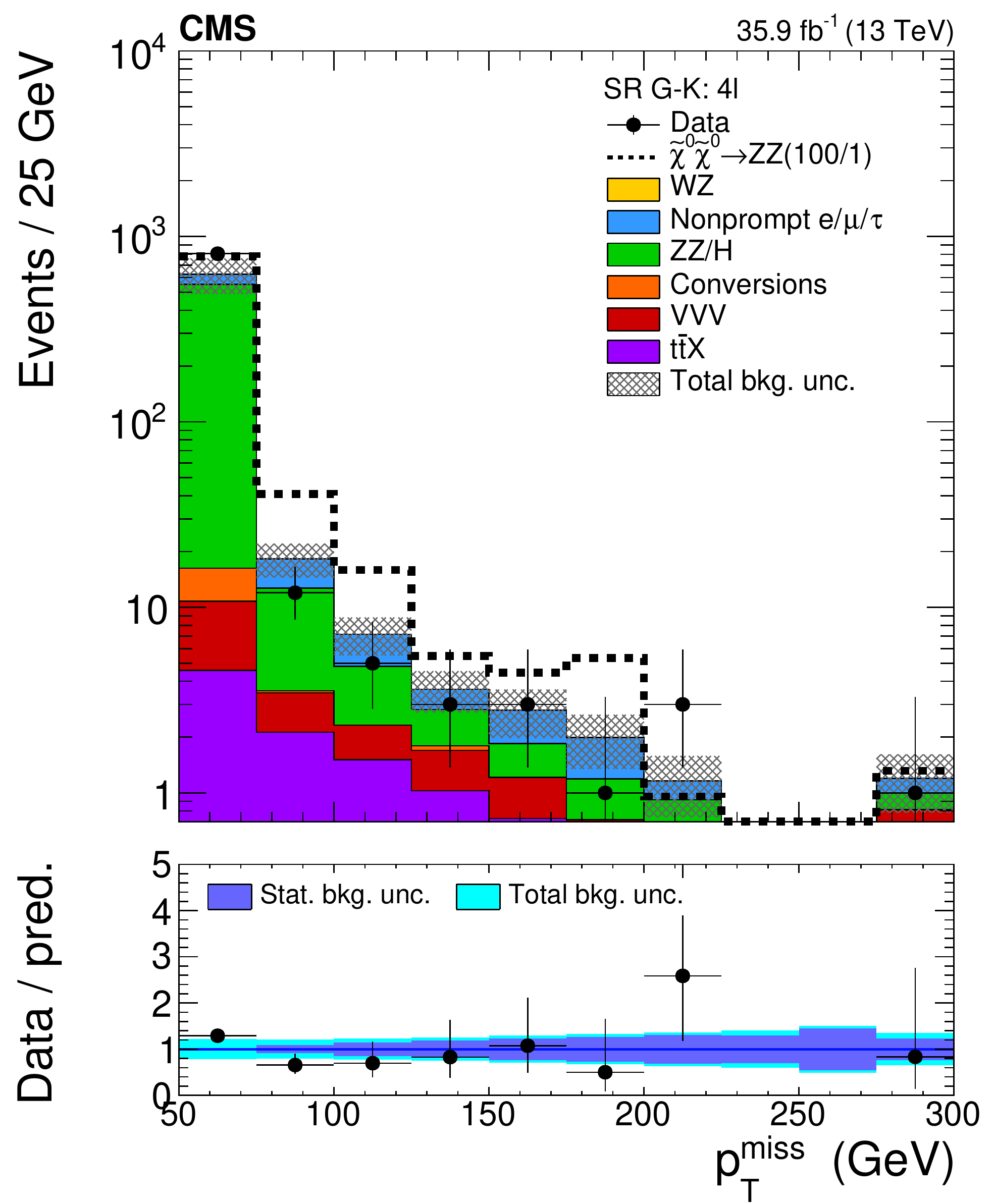}
\caption{Distribution in $\ptmiss$ for events with 4 or more leptons entering search categories G--K. An example signal mass point in the $\PSGczDo\PSGczDo$ production GMSB model is displayed for illustration. The numbers in the parenthesis denote the $\PSGczDo$ and $\PXXSG$ masses, namely $m_{\PSGczDo}=100\GeV$  and $m_{\PXXSG}=1\GeV$. }
\label{fig:yields:BRGK:met}
\end{figure}

The estimation methods for the dominant background sources, $\WZ$ production and nonprompt leptons, have been extensively validated. Such checks are based on both simulation and data control regions as described in Sections \ref{sec:backgrounds} and \ref{sec:systematics}. In particular, the modelling of the most relevant kinematic distributions used to define the signal regions is validated for each category. Distributions in \ptmiss are shown in Fig.~\ref{fig:yields:2lss:met} for the SS dilepton channel for events with 0 or 1 jet. Key kinematic distributions for the three-lepton channel are displayed in  Fig.~\ref{fig:yields:BRA:evt} and for three lepton events with at least one \tauh in Fig.~\ref{fig:yields:BRCF:evt}. Fig.~\ref{fig:yields:BRGK:met} displays the \ptmiss for events with four leptons. An example signal mass point in the $\PSGcpmDo$-$\PSGczDo$ mass plane for which each category has sensitivity, is also shown in these figures.

The expected and observed yields are summarized in Table~\ref{tab:resultsSS} for the SS dilepton channel, in Tables~\ref{tab:resultsA}--\ref{tab:resultsF} for the trilepton channel, and in Table~\ref{tab:yields:SR4l} for the four-lepton channel. The observed event counts are consistent with those expected from the SM backgrounds.

\begin{figure}[!hbtp]
\centering
\includegraphics[width=\textwidth]{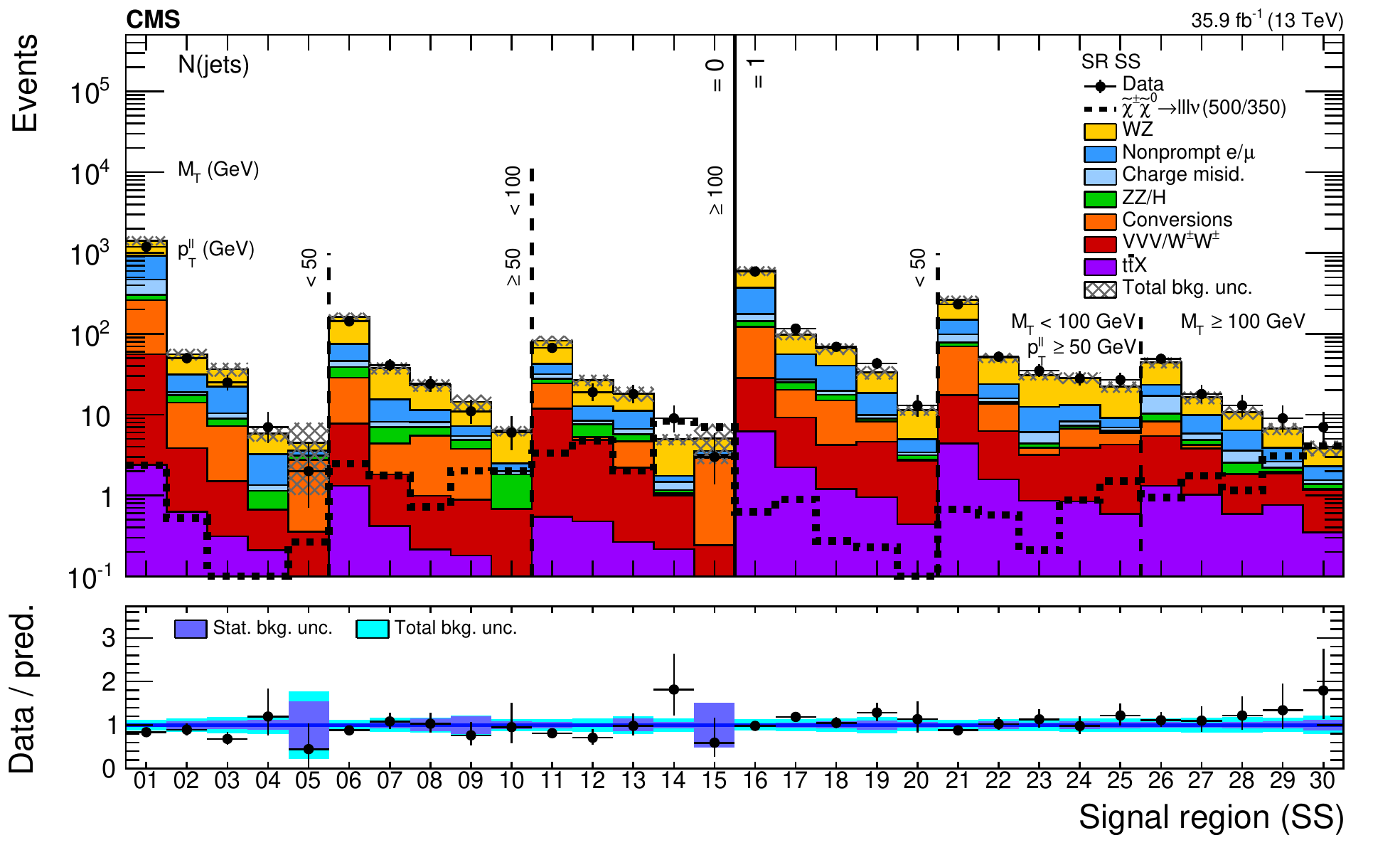}
\caption{Expected and observed yields comparison in the SS dilepton category. An example signal  mass point in the flavor-democratic model with mass parameter $x=0.05$ is  displayed for illustration. The notation is analogous to that used in  Fig.~\ref{fig:yields:2lss:met}.}
\label{fig:results_SR_ss}
\end{figure}

\begin{figure}[!hbtp]
\centering
\includegraphics[width=0.95\textwidth]{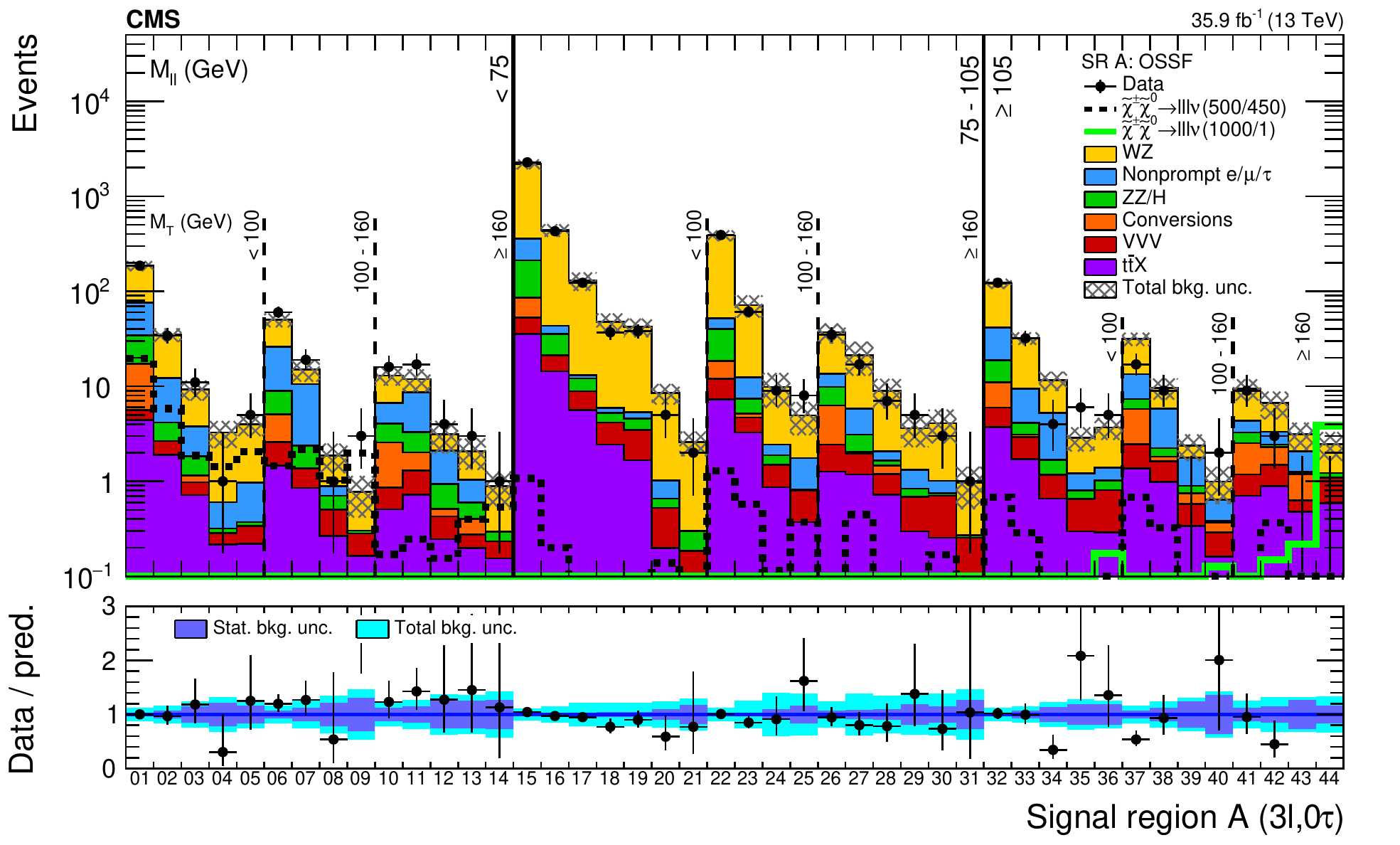}
\includegraphics[width=0.95\textwidth]{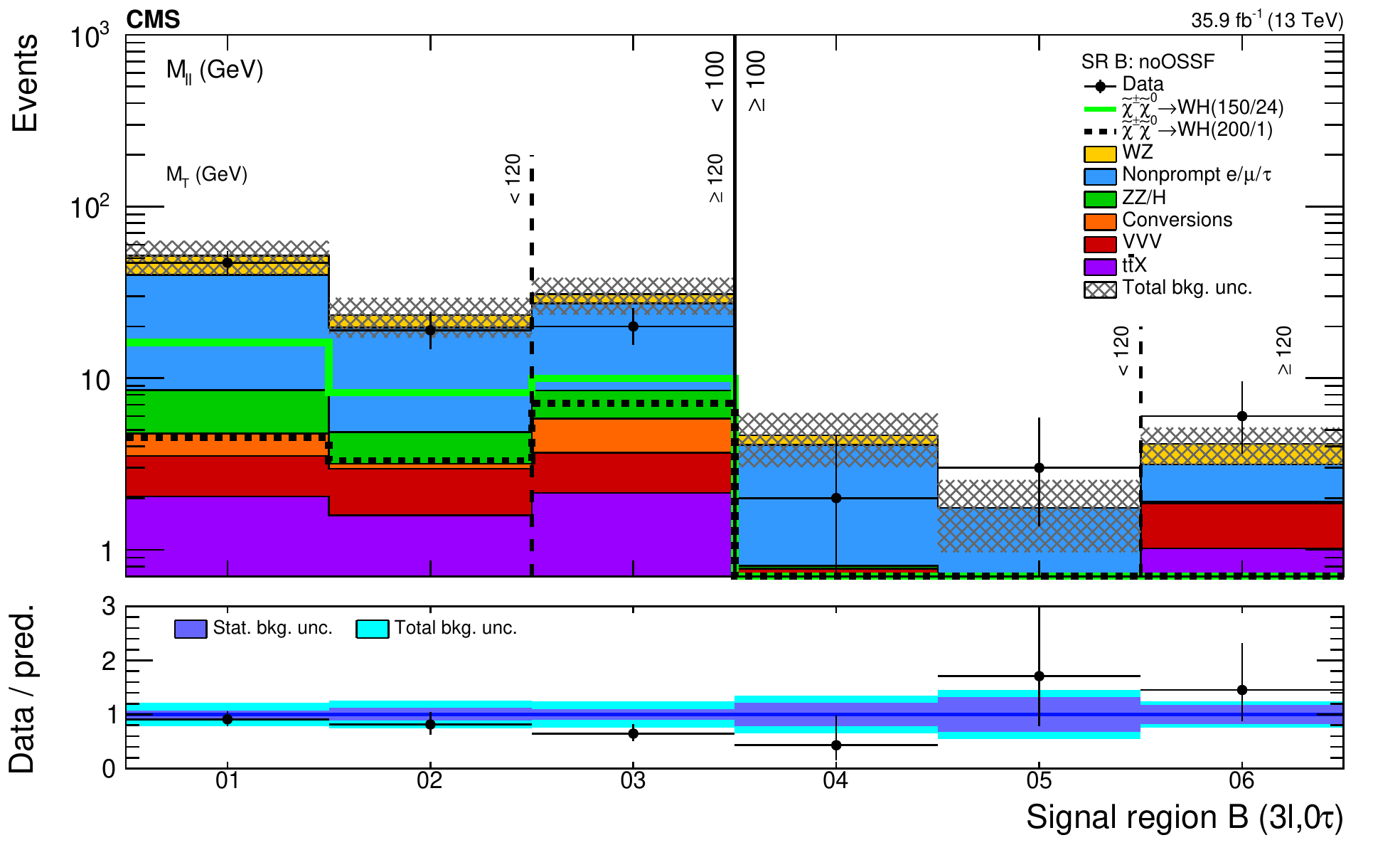}
\caption{Expected and observed yields comparison in category A (upper) and
category B (lower) signal regions, i.e. 3 light flavor leptons including at
least one OSSF pair (A) or no OSSF pair (B). SR~A15 is replaced by the
WZ control region in interpretations of the results. Two signal mass points
in the flavor-democratic model with mass parameter $x=0.5$ are displayed
for illustration. The notation is analogous to that used in
Fig.~\ref{fig:yields:2lss:met}.}
\label{fig:yields:SR3lAB}
\end{figure}

\begin{figure}[!hbtp]
\centering
\includegraphics[width=0.95\textwidth]{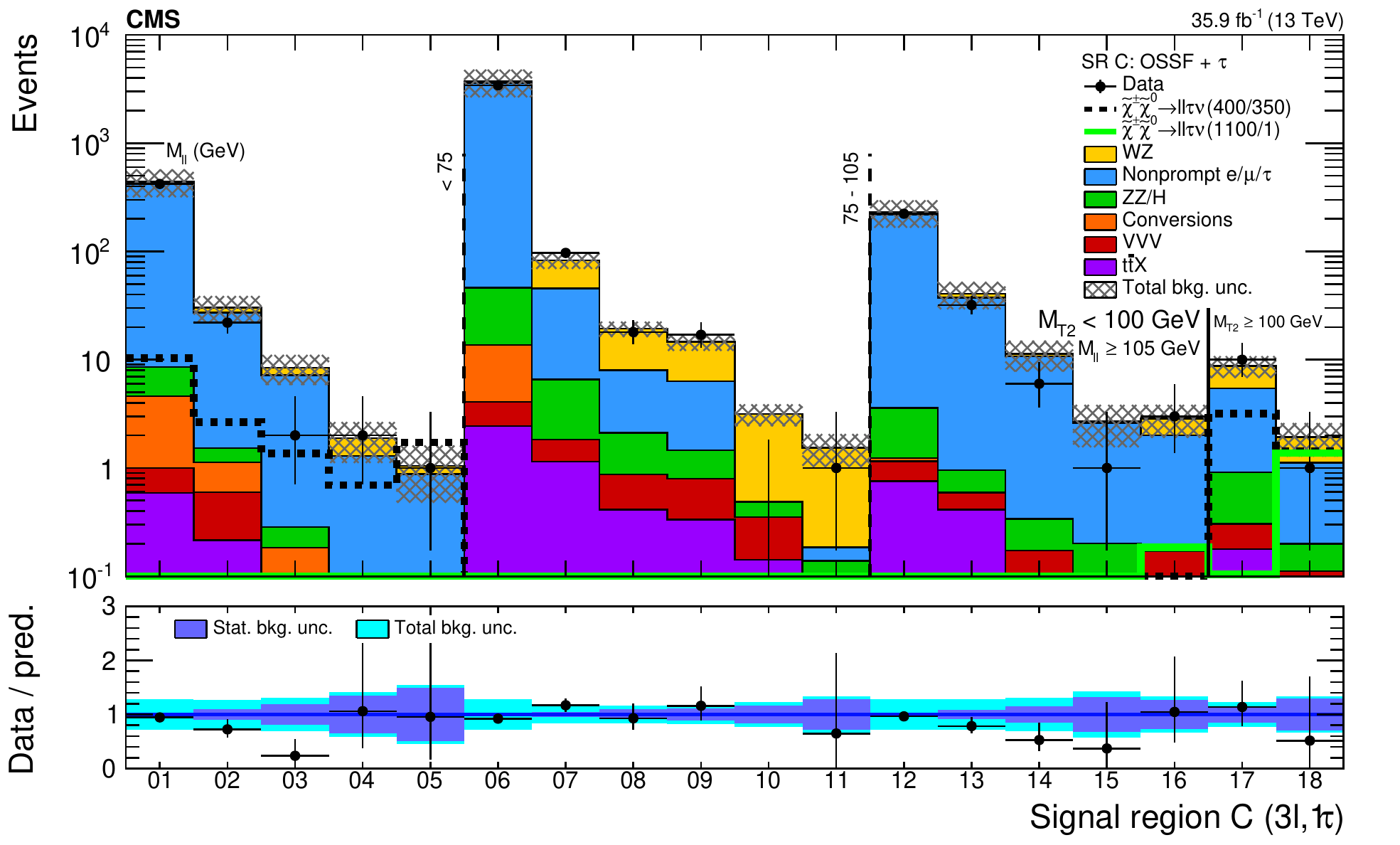}
\includegraphics[width=0.95\textwidth]{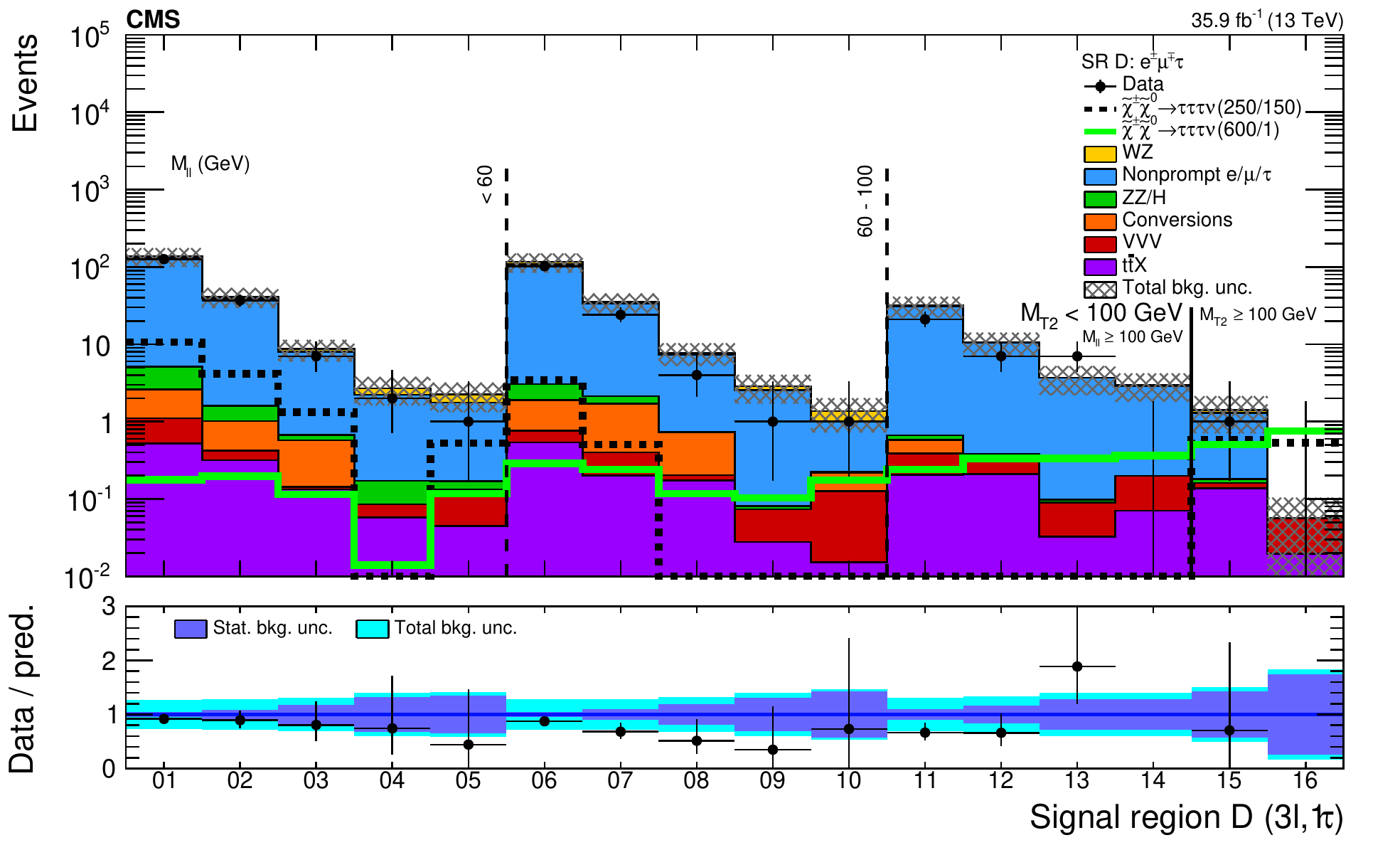}
\caption{Expected and observed yields comparison in events with one $\tauh$: categories C (upper)
and D (lower). Two signal mass points in the $\tau$-enriched (upper) and the $\tau$-dominated (lower) scenarios
with mass parameter $x=0.5$ are displayed for illustration. The notation is analogous to that
used in Fig.~\ref{fig:yields:2lss:met}.}
\label{fig:yields:SR3lCD}
\end{figure}

\begin{figure}[!hbtp]
\centering
\includegraphics[width=0.95\textwidth]{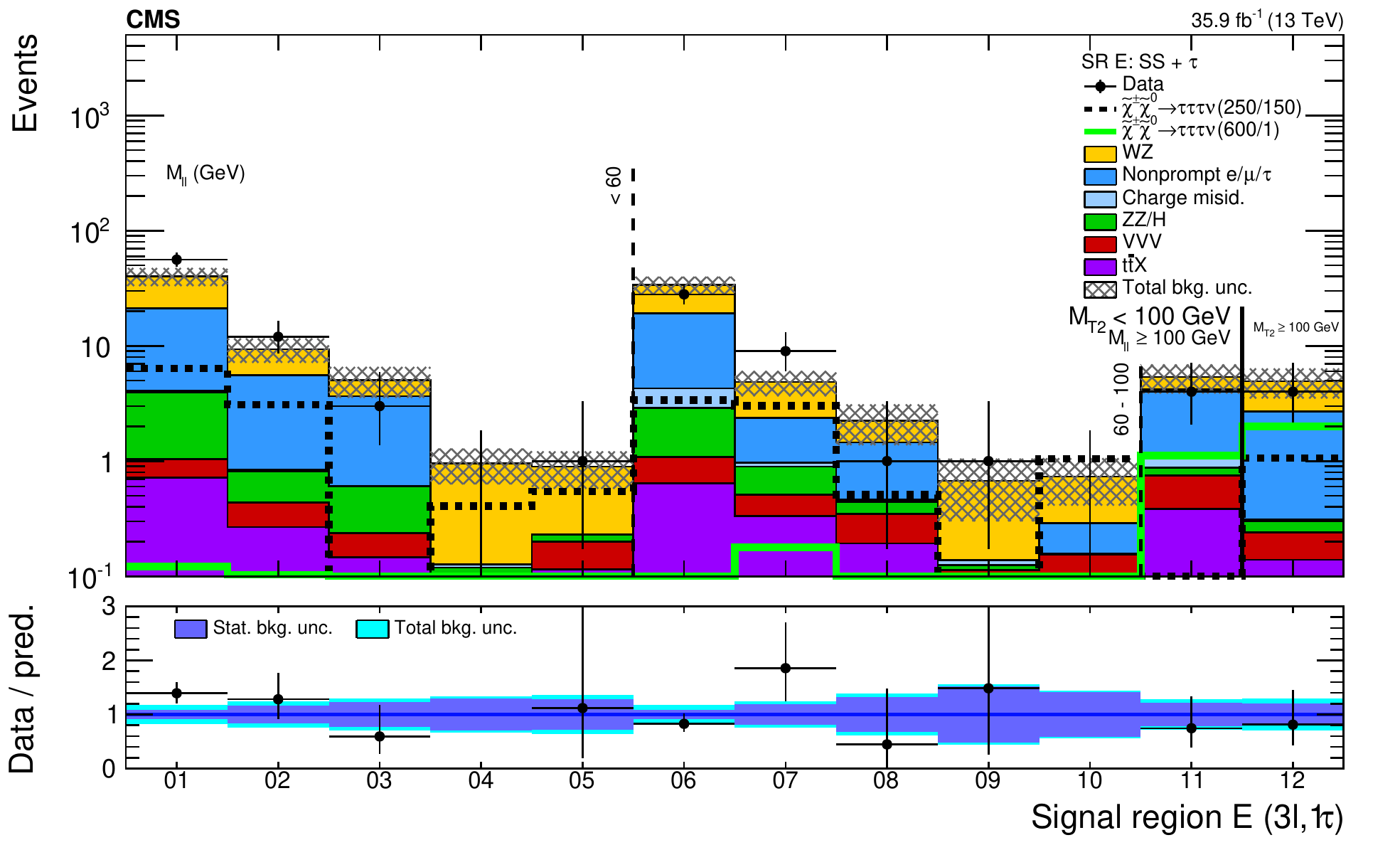}
\includegraphics[width=0.95\textwidth]{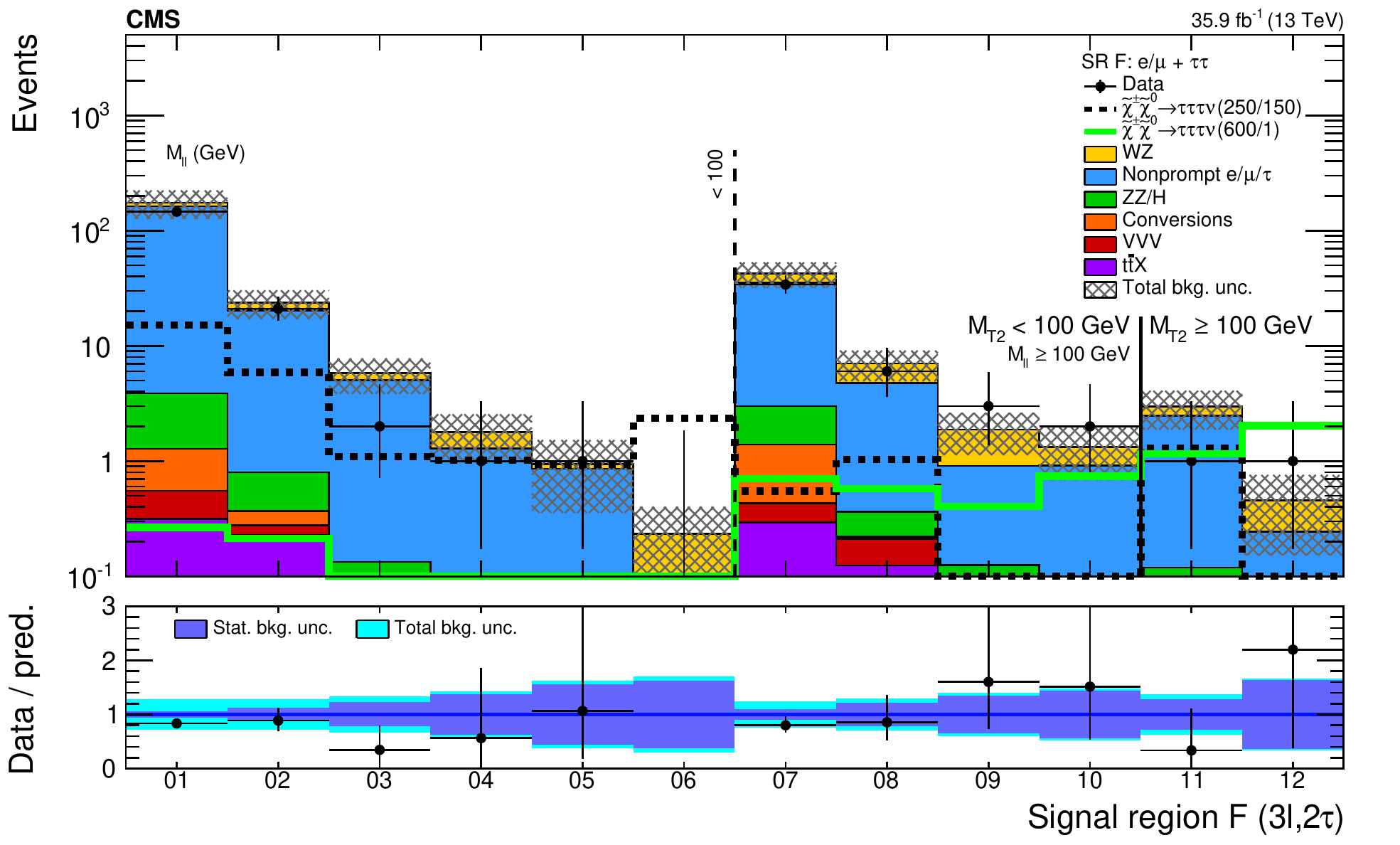}
\caption{Expected and observed yields comparison in events with one $\tauh$: category E (upper);
and in events with two $\tauh$: category F (lower). Two signal mass points in the
$\tau$-dominated model with mass parameter $x=0.5$ are displayed for illustration. The notation
is analogous to that used in Fig.~\ref{fig:yields:2lss:met}.}
\label{fig:yields:SR3lEF}
\end{figure}

\begin{figure}[!hbtp]
\centering
\includegraphics[width=0.95\textwidth]{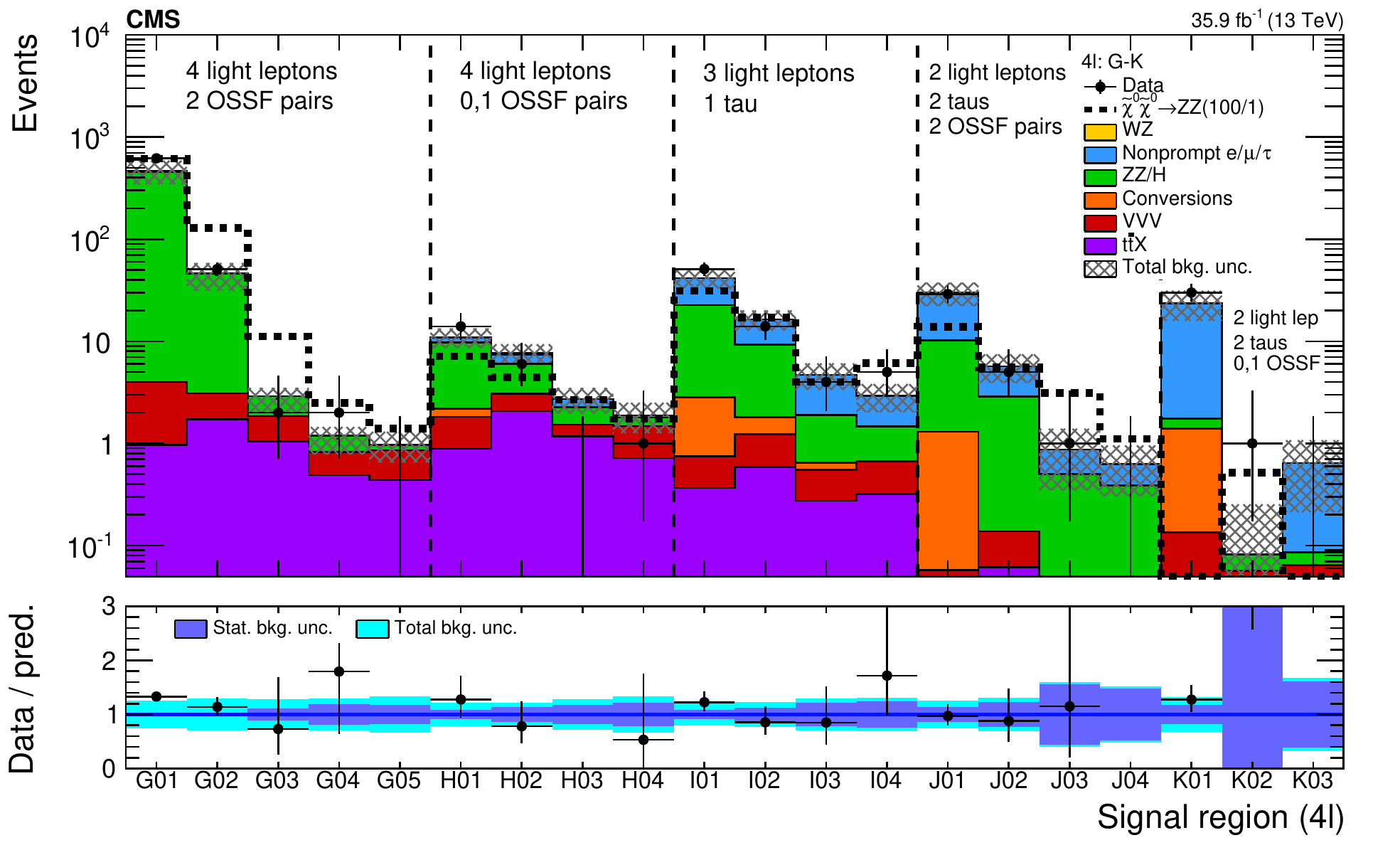}
\caption{Expected and observed yields comparison in signal regions with at least four leptons
(categories G--K). An example mass point in the $\PSGczDo\PSGczDo$ production GMSB model is
displayed for illustration. The notation is analogous to that used
in Fig.~\ref{fig:yields:BRGK:met}.}
\label{fig:yields:SR4l}
\end{figure}

\begin{figure}[h!]
\centering
\includegraphics[width=0.45\textwidth]{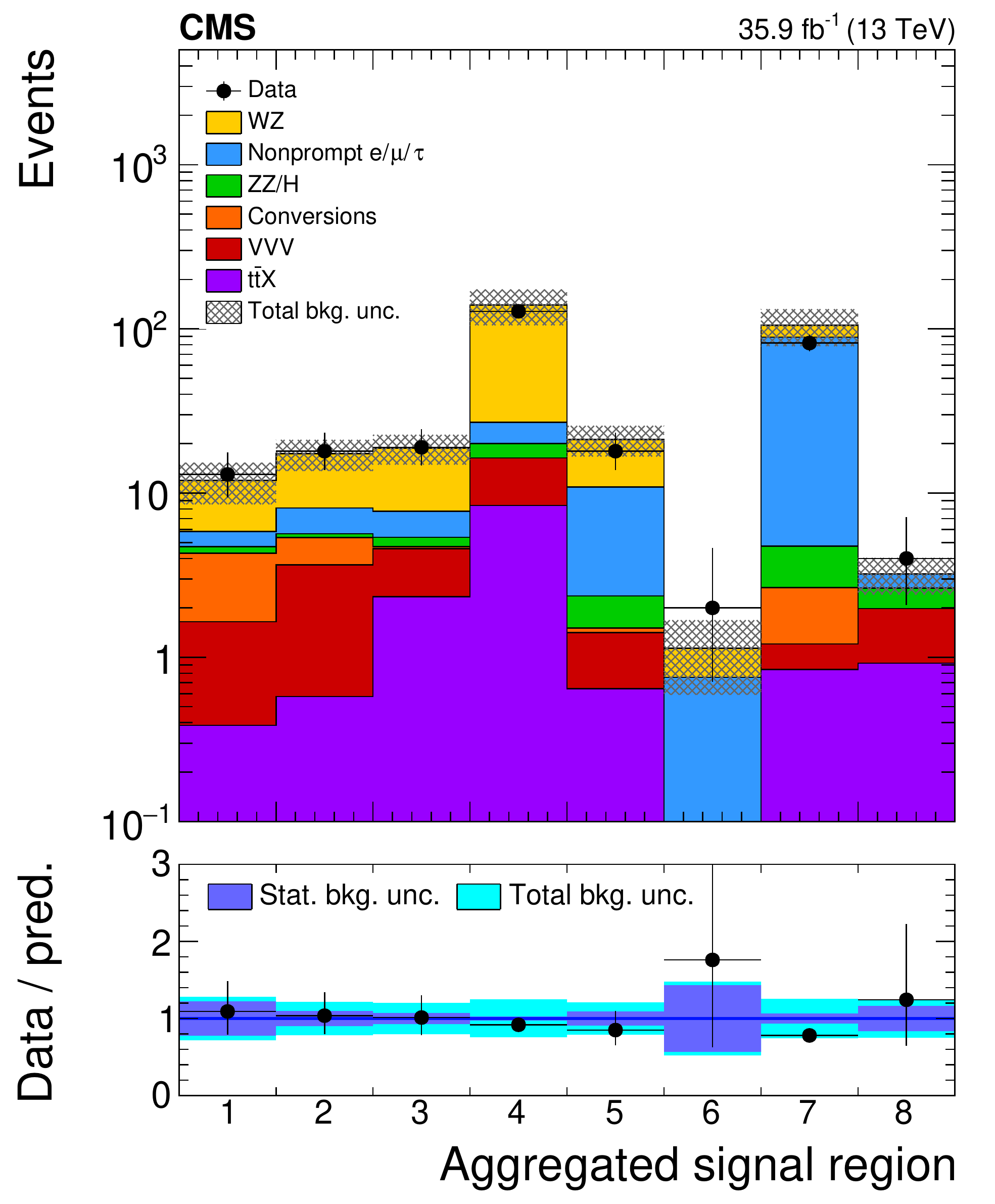}
\caption{Expected and observed yields comparison in the aggregated signal regions. In this plot, the charge misidentification background prediction from control samples in data (that is only relevant in the first two bins due to the SS dilepton final state) are included in the nonprompt background prediction.}
\label{fig:yields:SSR}
\end{figure}

The comparisons between the expected and observed yields are presented in Fig.~\ref{fig:results_SR_ss} for the SS dilepton channel, in Figs.~\ref{fig:yields:SR3lAB}--\ref{fig:yields:SR3lEF} for the trilepton channel, and in Fig.~\ref{fig:yields:SR4l} for signal regions with at least four leptons. An example signal mass point in the $\PSGcpmDo$-$\PSGczDo$ mass plane for which each category has sensitivity, is also shown in these figures. Finally, results for the aggregated signal regions are presented in Fig.~\ref{fig:yields:SSR} and Table~\ref{tab:results:SSR}.

\begin{table}[tbh]
\centering
\topcaption{SS dilepton category: Expected and observed yields in events with two SS light-flavor leptons. For each bin, the first number corresponds to the expected yield (exp.) and its uncertainty, and the second denotes the observed yield (obs.). The uncertainty denotes the total uncertainty in the yield.}
\label{tab:resultsSS}
\resizebox{\textwidth}{!}{
\begin{tabular}{|c|c|c|cc|ccc|cc|cc|}
\hline
 $\Njets$ & 	$\MT$ (\GeVns{}) & 	$\pt^{\ell\ell}$ (\GeVns{})  &  \multicolumn{2}{c|}{$\ptmiss < 100\GeV $} & \multicolumn{3}{c|}{$100 \le \ptmiss < 150\GeV $} & \multicolumn{2}{c|}{$150 \le \ptmiss < 200\GeV$} & \multicolumn{2}{c|}{$\ptmiss \ge 200\GeV$}\\
 & 	 &   &  (exp.) & (obs.) & (exp.) & (obs.) &  & (exp.) & (obs.) & (exp.) & (obs.) \\
\hline\hline
\multirow{6}{*}{0} & \multirow{4}{*}{$<$100 } & \multirow{2}{*}{$<$50 } & \multirow{2}{*}{  1430 $\pm$    180} & \multirow{2}{*}{1193} &     56 $\pm$      9 & 50 & $++$ & \multirow{2}{*}{     5.9 $\pm$      1.2} & \multirow{2}{*}{7} & \multirow{2}{*}{     4.5 $\pm$      3.5} & \multirow{2}{*}{2} \\ \cline{6-8}
& & & & &     36 $\pm$      7 & 25 & \NA & & & & \\ \cline{3-12}

& & \multirow{2}{*}{$>$50 } & \multirow{2}{*}{   163 $\pm$     19} & \multirow{2}{*}{143} &     38 $\pm$      6 & 41 & $++$ & \multirow{2}{*}{    14.4 $\pm$      3.2} & \multirow{2}{*}{11} & \multirow{2}{*}{     6.3 $\pm$      0.9} & \multirow{2}{*}{6} \\ \cline{6-8}
& & & & &     23 $\pm$      4 & 24 & \NA & & & & \\ \cline{2-12}

& \multirow{2}{*}{$>$100 } &  & \multirow{2}{*}{    82 $\pm$     12} & \multirow{2}{*}{67} &     27 $\pm$      4 & 19 & $++$ & \multirow{2}{*}{     5.0 $\pm$      0.8} & \multirow{2}{*}{9} & \multirow{2}{*}{     5.1 $\pm$      2.6} & \multirow{2}{*}{3} \\ \cline{6-8}
& & & & &     18 $\pm$      4 & 18 & \NA & & & & \\ \hline

\multirow{6}{*}{1} & \multirow{4}{*}{$<$100 } & \multirow{2}{*}{$<$50 } &  \multirow{2}{*}{   603 $\pm$     80} & \multirow{2}{*}{591} &     98 $\pm$     14 & 116 & $++$ & \multirow{2}{*}{    33 $\pm$      6} & \multirow{2}{*}{43} & \multirow{2}{*}{    11.4 $\pm$      1.7} & \multirow{2}{*}{13} \\ \cline{6-8}
& & & & &     66 $\pm$     10 & 69 & \NA & & & & \\ \cline{3-12}

& &  \multirow{2}{*}{$>$50 } & \multirow{2}{*}{   264 $\pm$     31} & \multirow{2}{*}{232} &     51 $\pm$      7 & 52 & $++$ & \multirow{2}{*}{    29 $\pm$      5} & \multirow{2}{*}{28} & \multirow{2}{*}{    22.2 $\pm$      3.4} & \multirow{2}{*}{27} \\ \cline{6-8}
& & & & &     31 $\pm$      4 & 35 & \NA & & & & \\ \cline{2-12}

& \multirow{2}{*}{$>$100 } &  & \multirow{2}{*}{    44 $\pm$      7} & \multirow{2}{*}{49} &     16.4 $\pm$      2.9 & 18 & $++$ & \multirow{2}{*}{     6.7 $\pm$      1.1} & \multirow{2}{*}{9} & \multirow{2}{*}{  3.9 $\pm$   0.8} & \multirow{2}{*}{7} \\ \cline{6-8}
& & & & &     10.7 $\pm$      1.9 & 13 & \NA & & & & \\ \hline
\end{tabular}
}

\end{table}

\begin{table}[tbh]
\centering
\topcaption{Category A: Expected and observed yields in events with three $\Pe$ or $\mu$ that form at least one OSSF pair. For each bin, the first number corresponds to the expected yield (exp.) and its uncertainty, and the second denotes the observed yield (obs.). The uncertainty denotes the total uncertainty in the yield.}
\label{tab:resultsA}
\resizebox{0.8\textwidth}{!}{
\begin{tabular}{|c|c|cc|cc|cc|}
\hline
$\MT$ (\GeVns{}) & $\ptmiss$ (\GeVns{}) & \multicolumn{2}{c|}{$\Mll < 75\GeV$} & \multicolumn{2}{c|}{$75 \leq \Mll < 105\GeV $} & \multicolumn{2}{c|}{$\Mll\geq 105\GeV $}\\
 &  & (exp.) & (obs.) & (exp.) & (obs.) & (exp.) & (obs.)\\
\hline\hline
\multirow{7}{*}{$ 0 - 100$}   & $50-100  $ &    185 $\pm$     22  & 186  &   2180 $\pm$    260 & 2278 &    121 $\pm$     14 & 123 \\ \cline{2-8}
                              & $100-150 $ &     35 $\pm$      6  & 34  &    440 $\pm$     70 & 429 &     32 $\pm$      5 & 32 \\ \cline{2-8}
                              & $150-200 $ &      9.3 $\pm$      2.2  & 11  &    129 $\pm$     28 & 123 &     11.6 $\pm$      2.6 & 4 \\ \cline{2-8}
                              & $200-250 $ &      3.3 $\pm$      1.0  & 1  &     48 $\pm$     10 & 37 &   2.9 $\pm$   0.8 & 6 \\ \cline{2-8}
                              & $250-400 $ & \multirow{3}{*}{     4.0 $\pm$      1.0} & \multirow{3}{*}{5} &     42 $\pm$      9 & 38 & \multirow{3}{*}{     3.7 $\pm$      1.0} & \multirow{3}{*}{5} \\ \cline{2-2}\cline{5-6}
                              & $400-550 $ &          &          &      8.5 $\pm$      2.1 & 5 &          & \\ \cline{2-2}\cline{5-6}
                              & $\geq$550  &          &          &   2.6 $\pm$   0.8 & 2 &          & \\ \hline
\multirow{4}{*}{$ 100 - 160$} & $50-100  $ &     50 $\pm$      8  & 60  &    390 $\pm$     50 & 391 &     32 $\pm$      5 & 17 \\ \cline{2-8}
                              & $100-150 $ &     15 $\pm$      4  & 19  &     72 $\pm$     19 & 61 &     9.6 $\pm$      2.4 & 9 \\ \cline{2-8}
                              & $150-200 $ &   1.9 $\pm$   0.6  & 1  &     10 $\pm$      4 & 9 &   2.4 $\pm$   0.7 & 0 \\ \cline{2-8}
                              & $\geq$200  &   0.8 $\pm$   0.4  & 3  &      4.9 $\pm$      1.9 & 8 &   1.0 $\pm$   0.4 & 2 \\ \hline
\multirow{6}{*}{$\geq$160 }   & $50-100  $ &     13.0 $\pm$      2.8 & 16 &     37 $\pm$      9 & 35 &      9.4 $\pm$      2.4 & 9 \\ \cline{2-8}
                              & $100-150 $ &     11.9 $\pm$      3.2 & 17 &     21 $\pm$      8 & 17 &      6.6 $\pm$      2.1 & 3 \\ \cline{2-8}
                              & $150-200 $ &   3.1 $\pm$   1.2 & 4 &      8.9 $\pm$      3.1 & 7 &   3.1 $\pm$   1.0 & 0 \\ \cline{2-8}
                              & $200-250 $ &   2.1 $\pm$   0.8 & 3 &      3.6 $\pm$      1.3 & 5 & \multirow{3}{*}{  2.5 $\pm$   0.8} & \multirow{3}{*}{0} \\ \cline{2-6}
                              & $250-400 $ & \multirow{2}{*}{  0.9 $\pm$   0.4} & \multirow{2}{*}{1} &      4.1 $\pm$      1.6 & 3 &         & \\ \cline{2-2}\cline{5-6}
                              & $\geq$400  &          &          &   1.0 $\pm$   0.5 & 1 &         & \\ \hline
\end{tabular}
}
\end{table}

\begin{table}[tbh]
\centering
\topcaption{Category B: Expected and observed yields in events with three $\Pe$ or $\mu$ that do not form an OSSF pair.
For each bin, the first number corresponds to the expected yield (exp.) and its uncertainty, and the second denotes the observed yield (obs.). The uncertainty denotes the total uncertainty in the yield.}
\label{tab:resultsB}
\resizebox{0.6\textwidth}{!}{
\begin{tabular}{|c|c|cc|cc|}
\hline
$\MT$ (\GeVns{}) & $\ptmiss$ (\GeVns{}) & \multicolumn{2}{c|}{$\Mll < 100\GeV$} & \multicolumn{2}{c|}{$\Mll\geq 100\GeV $}\\
			& 				  & (exp.) & (obs.) & (exp.) & (obs.)\\
\hline\hline
\multirow{2}{*}{$ 0 - 120$} & $50-100  $ &     52 $\pm$     11 & 47 &      4.6 $\pm$      1.6 & 2 \\ \cline{2-6}
                            & $\geq$100  &     23 $\pm$      6 & 19 &   1.8 $\pm$   0.8 & 3 \\ \hline
$\geq$120                   & $\geq 50 $ &     31 $\pm$      7 & 20 &   4.1 $\pm$   1.0 & 6 \\ \hline
\end{tabular}
}
\end{table}

\begin{table}[tbh]
\centering
\topcaption{Category C: Expected and observed yields in events with two $\Pe$ or $\mu$ forming and OSSF pair and one $\tauh$.
For each bin, the first number corresponds to the expected yield and its uncertainty, and the second denotes the observed yield. The uncertainty denotes the total uncertainty in the yield.}
\label{tab:resultsC}
\resizebox{0.8\textwidth}{!}{
\begin{tabular}{|c|cc|c|cc|cc|}
\hline
$\ptmiss$ (\GeVns{}) & \multicolumn{2}{c|}{$75 \leq \Mll < 105\GeV $} & $\MTT(\ell_{1},\ell_{2})$ (\GeVns{}) & \multicolumn{2}{c|}{$\Mll < 75\GeV$} & \multicolumn{2}{c|}{$\Mll\geq 105\GeV $}\\
\hline\hline
$50-100  $ & 3700 $\pm$ 1100 & 3427 & \multirow{7}{*}{$ 0 - 100$} & 440 $\pm$ 130 & 420 & 231 $\pm$ 65 & 223 \\ \cline{1-3}\cline{5-8}
$100-150 $ &   83 $\pm$   14 &   97 &                             &  30 $\pm$   8 & 22  &  41 $\pm$ 11 & 32  \\ \cline{1-3}\cline{5-8}
$150-200 $ &   19.4 $\pm$    3.1 &   18 &                             &   8.5 $\pm$   2.6 & 2   &  11 $\pm$  4 & 6   \\ \cline{1-3}\cline{5-8}
$200-250 $ & \multirow{2}{*}{14.6 $\pm$ 2.6} & \multirow{2}{*}{17} &  & 1.9 $\pm$ 0.8 & 2   &   2.7 $\pm$  1.1 & 1   \\ \cline{1-1}\cline{5-8}
$250-300 $ &                 &      &                             & \multirow{3}{*}{1.1 $\pm$ 0.6} & \multirow{3}{*}{1} & \multirow{3}{*}{2.9 $\pm$ 1.0} & \multirow{3}{*}{3} \\ \cline{1-3}
$300-400 $ &    3.2 $\pm$    0.7 & 0    &                             & & & &  \\ \cline{1-3}
$\geq$400  &  1.5 $\pm$  0.6 & 1    &                             & & & &  \\ \hline
$50-200 $  &                 &      & \multirow{2}{*}{$\geq$100 } & \multicolumn{2}{c}{8.8 $\pm$ 2.0} & \multicolumn{2}{c|}{10} \\ \cline{1-1}\cline{5-8}
$\geq$200  &                 &      &                             & \multicolumn{2}{c}{1.9 $\pm$ 0.7} & \multicolumn{2}{c|}{1} \\ \hline
\end{tabular}
}
\end{table}

\begin{table}[tbh]
\centering
\topcaption{Category D: Expected and observed yields in events with an opposite-sign $\Pe\mu$ pair and one $\tauh$. For each bin, the first number corresponds to the expected yield and its uncertainty, and the second denotes the observed yield. The uncertainty denotes the total uncertainty in the yield.}
\label{tab:resultsD}
\resizebox{0.8\textwidth}{!}{
\begin{tabular}{|c|c|cc|cc|cc|}
\hline
$\MTT(\ell_1,\ell_2)$ (\GeVns{}) & $\ptmiss$ (\GeVns{}) & \multicolumn{2}{c|}{$\Mll < 60\GeV$} & \multicolumn{2}{c|}{$60 \leq \Mll < 100\GeV $} & \multicolumn{2}{c|}{$\Mll\geq 100\GeV $}\\
\hline\hline
\multirow{5}{*}{$ 0 - 100$}   & $50-100  $ &    140 $\pm$     40 & 126 &    117 $\pm$     32 & 102 &     32 $\pm$     10 & 21 \\ \cline{2-8}
                              & $100-150 $ &     41 $\pm$     12 & 37 &     35 $\pm$     10 & 24 &     11 $\pm$      4 & 7 \\ \cline{2-8}
                              & $150-200 $ &      8.7 $\pm$      2.7 & 7 &      7.8 $\pm$      2.5 & 4 &      3.7 $\pm$      1.5 & 7 \\ \cline{2-8}
                              & $200-250 $ &      2.7 $\pm$      1.1 & 2 &      2.9 $\pm$      1.2 & 1 & \multirow{2}{*}{     3.0 $\pm$      1.2} & \multirow{2}{*}{0} \\ \cline{2-6}
                              & $\geq250 $ &   2.3 $\pm$   0.9 & 1 &   1.4 $\pm$   0.6 & 1 & &  \\ \hline
\multirow{2}{*}{$\geq$100 }   & $50-200  $ & \multicolumn{3}{c}{  1.4 $\pm$   0.7} & \multicolumn{3}{c|}{1} \\ \cline{2-8}
                              & $\geq$200  & \multicolumn{3}{c}{0.06 $\pm$ 0.05} & \multicolumn{3}{c|}{0} \\ \hline
\end{tabular}
}
\end{table}

\begin{table}[tbh]
\centering
\topcaption{Category E: Expected and observed yields in events with one SS $\Pe$ or $\mu$ and one $\tauh$. For each bin, the first number corresponds to the expected yield and its uncertainty, and the second denotes the observed yield. The uncertainty  denotes the total uncertainty in the yield.}
\label{tab:resultsE}
\resizebox{0.8\textwidth}{!}{
\begin{tabular}{|c|c|cc|cc|cc|}
\hline
$\MTT(\ell_1,\tau_1)$ (\GeVns{}) & $\ptmiss$ (\GeVns{}) & \multicolumn{2}{c|}{$\Mll < 60\GeV$} & \multicolumn{2}{c|}{$60 \leq \Mll < 100\GeV $} & \multicolumn{2}{c|}{$\Mll\geq 100\GeV $}\\
\hline\hline
\multirow{5}{*}{$ 0 - 100$}   & $50-100  $ &     40 $\pm$      7 & 56 &     34 $\pm$      6 & 28   & \multirow{5}{*}{     5.4 $\pm$      1.5} & \multirow{5}{*}{4} \\ \cline{2-6}
                              & $100-150 $ &      9.4 $\pm$      2.2 & 12 &      4.8 $\pm$      1.2 & 9   &  & \\ \cline{2-6}
                              & $150-200 $ &      5.0 $\pm$      1.5 & 3 &   2.2 $\pm$   0.9 & 1   &  & \\ \cline{2-6}
                              & $200-250 $ &   0.95 $\pm$   0.32 & 0 &   0.7 $\pm$   0.4 & 1   &  & \\ \cline{2-6}
                              & $\geq$250  &   0.89 $\pm$   0.32 & 1 &   0.73 $\pm$   0.32 & 0 &  & \\ \hline
$\geq$100                     & $\geq 50 $ & \multicolumn{3}{c}{     4.9 $\pm$      1.4} & \multicolumn{3}{c|}{4}\\ \hline
\end{tabular}
}
\end{table}

\begin{table}[tbh]
\centering
\topcaption{Category F: Expected and observed yields in events with one $\Pe$ or $\mu$ and two $\tauh$. For each bin, the first number corresponds to the expected yield and its uncertainty, and the second denotes the observed yield. The uncertainty denotes
total uncertainty in the yield.}
\label{tab:resultsF}
\resizebox{0.6\textwidth}{!}{
\begin{tabular}{|c|c|cc|cc|}
\hline
$\MTT(\ell,\tau_1)$ (\GeVns{}) & $\ptmiss$ (\GeVns{}) & \multicolumn{2}{c|}{$\Mll < 100\GeV$} & \multicolumn{2}{c|}{$\Mll\geq 100\GeV $}\\
	 & 	 & (exp.) & (obs.) & (exp.) & (obs.) \\
\hline\hline
\multirow{6}{*}{$ 0 - 100$}   & $50-100  $ &    170 $\pm$     50 & 146 &     42 $\pm$     11 & 34 \\ \cline{2-6}
                              & $100-150 $ &     24 $\pm$      7 & 21 &      7.0 $\pm$      2.1 & 6 \\ \cline{2-6}
                              & $150-200 $ &     5.8 $\pm$      2.0 & 2 &   1.9 $\pm$   0.8 & 3 \\ \cline{2-6}
                              & $200-250 $ &   1.8 $\pm$   0.8 & 1 & \multirow{3}{*}{  1.3 $\pm$   0.6} & \multirow{3}{*}{2} \\ \cline{2-4}
                              & $250-300 $ &   0.9 $\pm$   0.6 & 1 & & \\ \cline{2-4}
                              & $\geq$300  &   0.23 $\pm$   0.17 & 0 & & \\ \hline
						 	
\multirow{2}{*}{$\geq$100 }   & $50-200  $ & \multicolumn{2}{c}{     3.0 $\pm$      1.1} & \multicolumn{2}{c|}{1} \\ \cline{2-6}
                              & $\geq$200  & \multicolumn{2}{c}{  0.45 $\pm$   0.30} & \multicolumn{2}{c|}{1} \\ \hline
\end{tabular}
}
\end{table}

\begin{table}[htp]
\centering
\topcaption{Categories G--K: Expected and observed yields in the $4\ell$ category of the analysis. For each bin, the first number corresponds to the expected yield (exp.) and its uncertainty, and the second denotes the observed yield (obs.). The uncertainty denotes the total uncertainty in the yield.}\label{tab:yields:SR4l}
\resizebox{\textwidth}{!}{
\begin{tabular}{|c|cc|cc|cc|cc|cc|}
\hline
\multirow{2}{*}{$\ptmiss$ (\GeVns{})} & \multicolumn{4}{c|}{$0\tauh$} & \multicolumn{2}{c|}{$1\tauh$} & \multicolumn{4}{c|}{$2\tauh$} \\ \cline{2-11}
                                 & \multicolumn{2}{c|}{$\mathrm{nOSSF}\geq 2$} & \multicolumn{2}{c|}{$\mathrm{nOSSF}<2$} & \multicolumn{2}{c|}{$\mathrm{nOSSF}\geq 0$} & \multicolumn{2}{c|}{$\mathrm{nOSSF}\geq 2$} & \multicolumn{2}{c|}{$\mathrm{nOSSF}<2$} \\ \cline{2-11}
                                 & \multicolumn{2}{c|}{G} & \multicolumn{2}{c|}{H} & \multicolumn{2}{c|}{I} & \multicolumn{2}{c|}{J} & \multicolumn{2}{c|}{K} \\
                                 & (exp.) & (obs.) & (exp.) & (obs.) & (exp.) & (obs.) & (exp.) & (obs.) & (exp.) & (obs.) \\
								
\hline\hline
$0-50    $ &    460 $\pm$    120 & 619 &     10.9 $\pm$      2.2 & 14 &     42 $\pm$      8 & 51 &     30 $\pm$      7 & 29 &     24 $\pm$      8 & 30 \\ \hline
$50-100  $ &     45 $\pm$     14 & 51 &      7.7 $\pm$      1.5 & 6 &     16.4 $\pm$      3.5 & 14 &     5.7 $\pm$      1.7 & 5 &   0.07 $\substack{+0.18 \\ -0.07}$ & 1 \\ \hline
$100-150 $ &   2.7 $\pm$   0.8 & 2 &   2.7 $\pm$   0.6 & 0 &      4.7 $\pm$      1.4 & 4 &   0.9 $\pm$   0.5 & 1 & \multirow{3}{*}{  0.6 $\pm$   0.4} & \multirow{3}{*}{0} \\ \cline{1-9}
$150-200 $ &   1.12 $\pm$   0.33 & 2 & \multirow{2}{*}{  1.9 $\pm$   0.6} & \multirow{2}{*}{1} & \multirow{2}{*}{  2.9 $\pm$   0.9} & \multirow{2}{*}{5} & \multirow{2}{*}{  0.63 $\pm$   0.32} & \multirow{2}{*}{0} & & \\\cline{1-3}
$\geq$200  &   0.97 $\pm$   0.32 & 0 &         &         &         &         &         &         & & \\ \hline
\end{tabular}
}
\end{table}

\clearpage

\begin{table}[h!]
\centering
\topcaption{Expected and observed yields in the aggregated signal regions defined in Section~\ref{sec:regionsSSR}. For each bin, the first number corresponds to the expected yield (exp.) and its uncertainty, and the second denotes the observed yield (obs.). The uncertainty denotes the total uncertainty in the yield.}
\resizebox{0.95\textwidth}{!}{
\begin{tabular}{|c|c|c|cc|}
\hline
 & Final state 	& Definition & \multicolumn{2}{c|}{Event yield} \\
 &  		 	&  			 & (exp.) & (obs.) \\

\hline\hline
1 & 2 SS leptons &  0 jets, $\MT>100\GeV$ and $\ptmiss>140\GeV$                           &     12.5 $\pm$      3.4 & 13 \\ \hline
2 & 2 SS leptons &  1 jets, $\MT<100\GeV$, $\pt^{\ell\ell}<100\GeV$ and $\ptmiss>200\GeV$ &     18 $\pm$      4 & 18 \\ \hline
3 & 3 light leptons & $\MT>120\GeV$ and $\ptmiss>200\GeV$                                        &     19 $\pm$      4 & 19 \\ \hline
4 & 3 light leptons & $\ptmiss>250\GeV$                                                          &    142 $\pm$     34 & 128 \\ \hline
5 & 2 light leptons and 1 $\tauh$ & $\MTT(\ell_1,\tau)>50\GeV$ and $\ptmiss>200\GeV$                 &     22 $\pm$      5 & 18 \\ \hline
6 & 1 light lepton and 2 $\tauh$s  & $\MTT(\ell,\tau_1)>50\GeV$ and $\ptmiss>200\GeV$                 &   1.2 $\pm$   0.6 & 2 \\ \hline
7 & 1 light lepton and 2 $\tauh$s & $\ptmiss>75\GeV$                                                 &    109 $\pm$     28 & 82 \\ \hline
8 & more than 3 leptons & $\ptmiss>200\GeV$                                                      &    3.2 $\pm$     0.8 & 4 \\ \hline
\end{tabular}
}
\label{tab:results:SSR}
\end{table}

This analysis is mostly statistically limited, as can be seen from the ratio plots of Figs.~\ref{fig:yields:2lss:met}--\ref{fig:yields:SR4l}. However, there are search regions for which other sources of uncertainties are also relevant. Such is the case for the on-Z signal regions A, which is dominated by the systematic uncertainty of the \WZ normalization for the low-$\MT$ bins and the \WZ shape uncertainty for the high-$\MT$ bins. In regions that contain at least one $\tauh$ candidate, the systematic uncertainty on the nonprompt background is comparable to the statistical uncertainty. Finally, the uncertainty on the four-lepton signal regions G--K, is dominated by the $\PZ\PZ$ normalization uncertainty.

\clearpage
\section{Interpretations of the searches}
\label{sec:limits}
No evidence of any significant deviation with respect to the SM prediction is observed. The results of this search are interpreted in the context of the simplified models covering the scenarios described in Section~\ref{sec:models}.

We compute 95\% confidence level (CL) upper limits on the new-physics cross sections using the CL$_\mathrm{s}$ method~\cite{Cowan:2010js,Junk:1999kv,Read:2002hq,ATLAS:1379837}, incorporating the uncertainties in the signal efficiency and acceptance and the uncertainties in the expected background described in Section~\ref{sec:systematics}. Lognormal nuisance parameters are used for the signal and background estimate uncertainties. Only the categories with the lepton flavor, multiplicity, and charge requirements corresponding to the topology of the model under consideration are combined to maximize sensitivity to the model in question.

The production cross sections are computed at NLO plus next-to-leading-log (NLL) precision~\cite{Beenakker:1999xh,Fuks:2012qx,Fuks:2013vua,Fuks:2013lya} in a limit of mass-degenerate wino $\PSGczDt$ and $\PSGcpmDo$, light bino $\PSGczDo$, and with all the other sparticles assumed to be heavy and decoupled. Similarly, for the higgsino models, production cross sections are computed in a limit of mass-degenerate higgsino $\PSGczDt$, $\PSGcpmDo$, and $\PSGczDo$ with all the other sparticles assumed to be heavy and decoupled.

\begin{figure}[htbp]
\centering
  \includegraphics[width=0.45\textwidth]{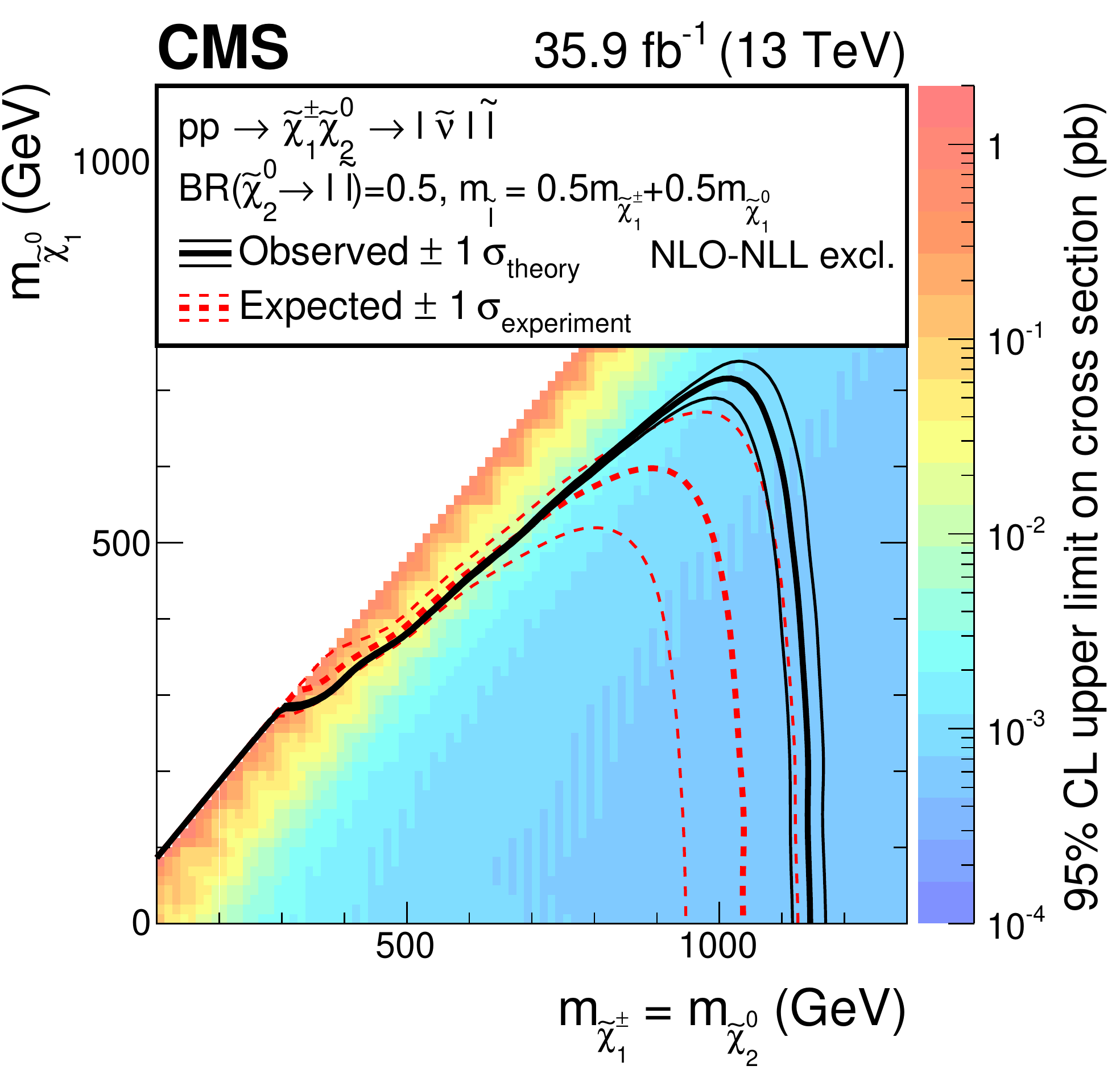}
\caption{Interpretation of the results in the flavor-democratic model with mass parameter $x=0.5$ obtained with events of category A. The shading in the $m_{\PSGczDo}$  versus $m_{\PSGczDt}$ ($=m_{\PSGcpmDo}$) plane indicates the 95\% CL upper limit on the chargino-neutralino production cross section. The contours bound the mass regions excluded at 95\% CL assuming the NLO+NLL cross sections. The observed, ${\pm}1\,\sigma_{\text{theory}}$ ($\pm$1 standard deviation of the theoretical cross section) observed, median expected, and ${\pm}1\,\sigma_{\text{experiment}}$ expected bounds are shown.}
\label{fig:interpr:TChiSlepSnu:FD}
\end{figure}

\begin{figure}[htbp]
\centering
\includegraphics[width=0.45\textwidth]{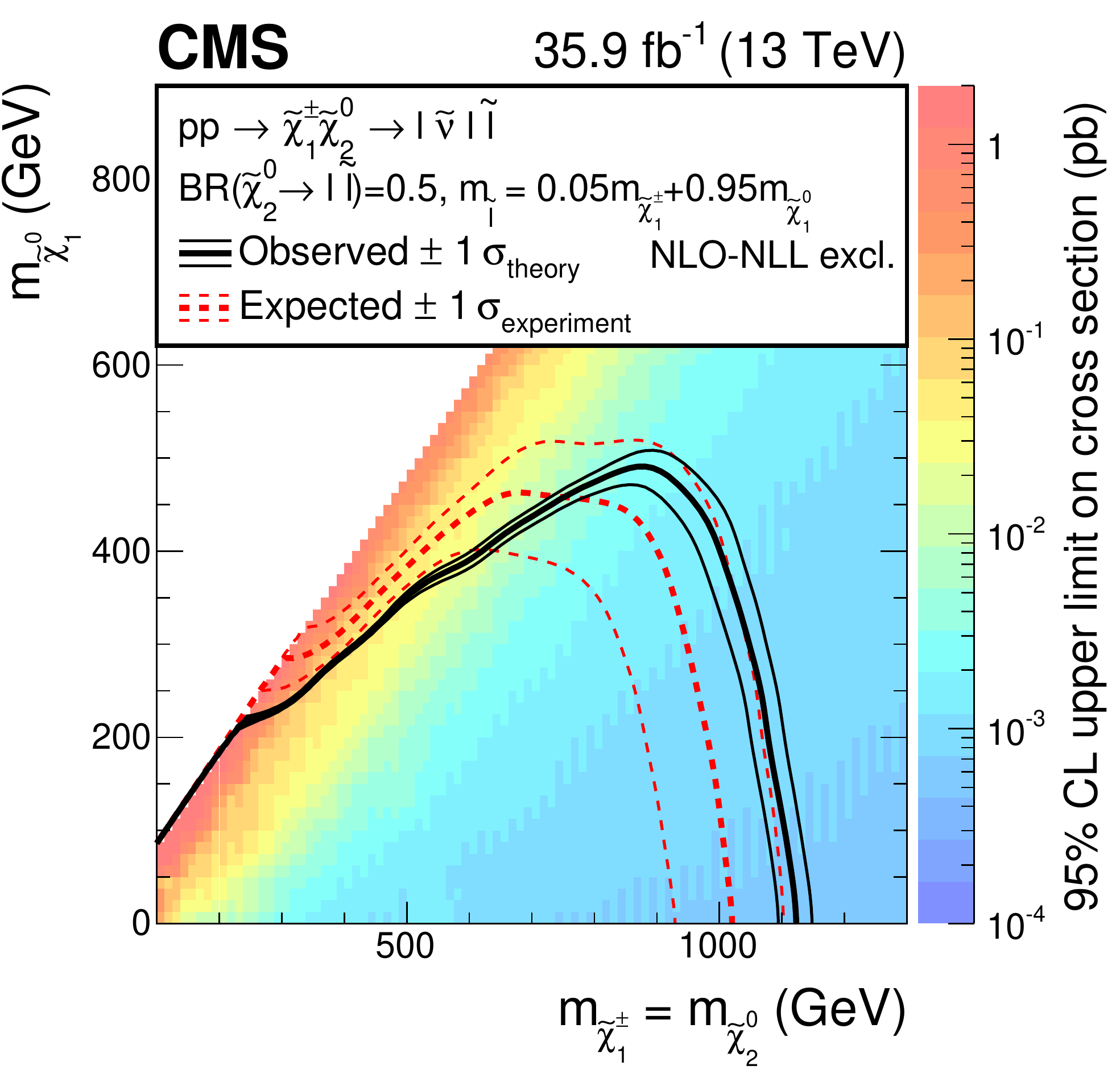}
\includegraphics[width=0.45\textwidth]{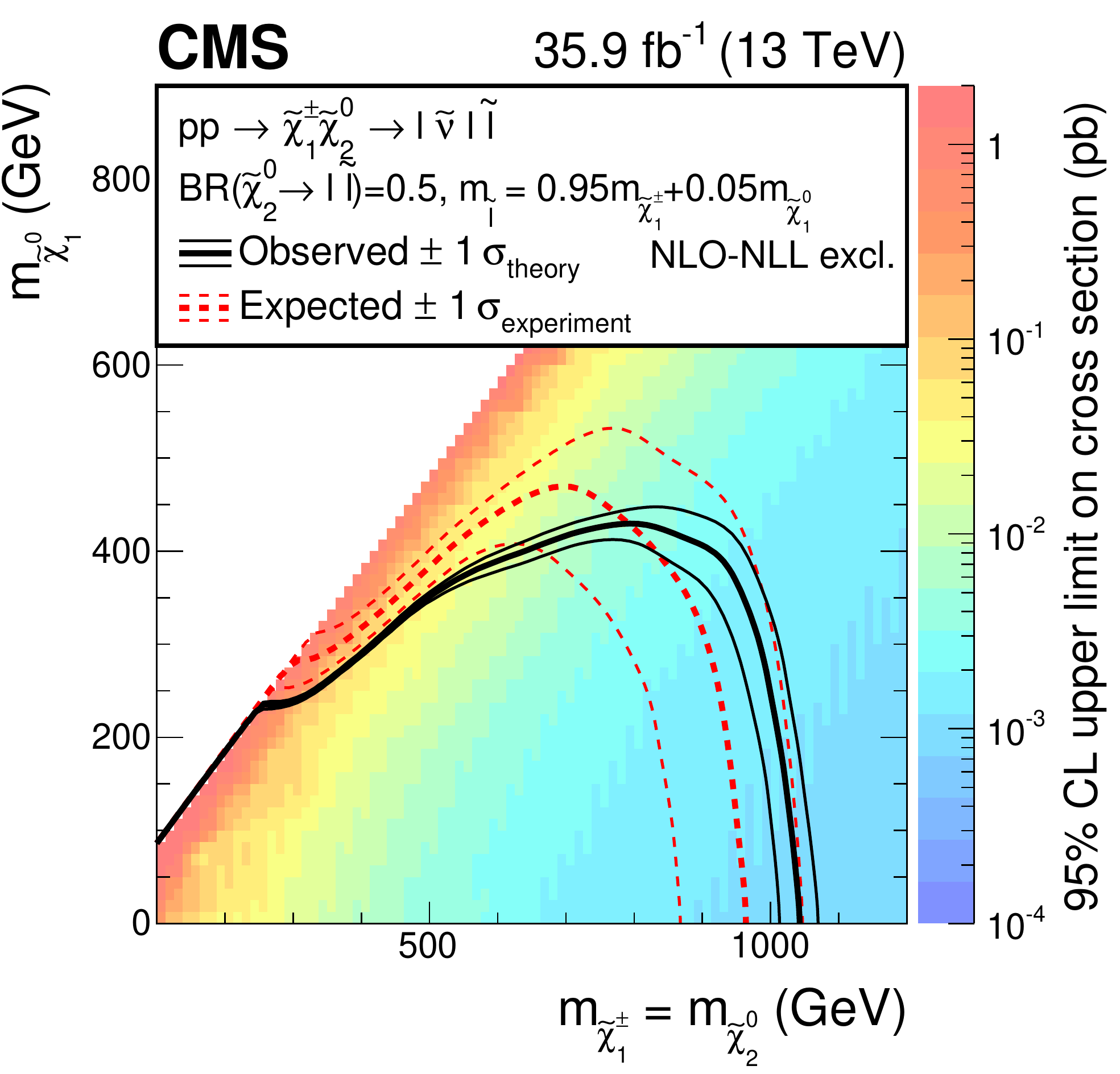}
\caption{Interpretation of the results in the flavor-democratic model with mass parameter $x=0.05$ (left) and $x=0.95$ (right) obtained with the combination of the SS dilepton category and category A. The shading in this figure is as described in Fig.~\ref{fig:interpr:TChiSlepSnu:FD}.}
\label{fig:interpr:TChiSlepSnu:05}
\end{figure}

\begin{figure}[htbp]
\centering
  \includegraphics[width=0.45\textwidth]{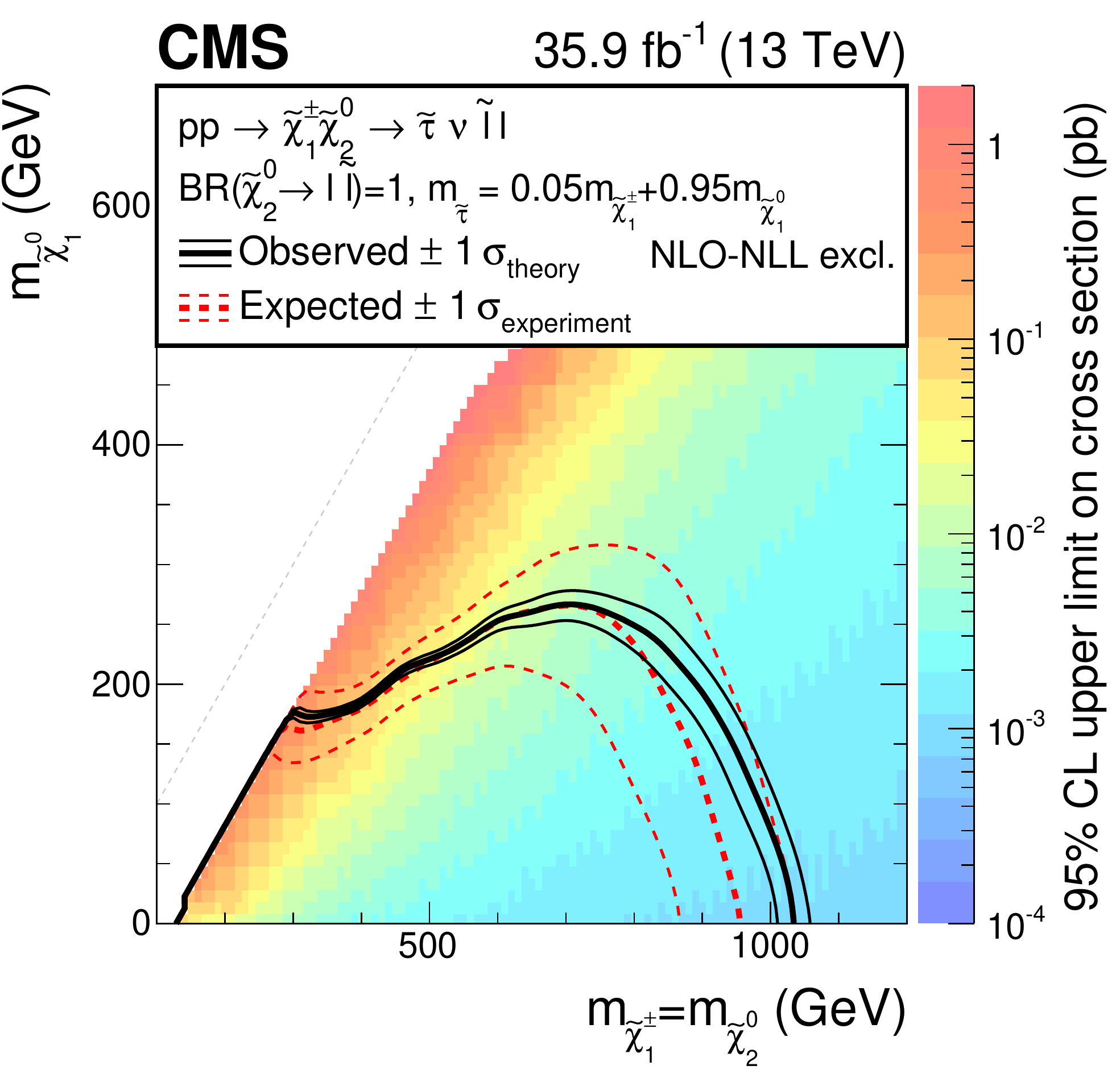}
  \includegraphics[width=0.45\textwidth]{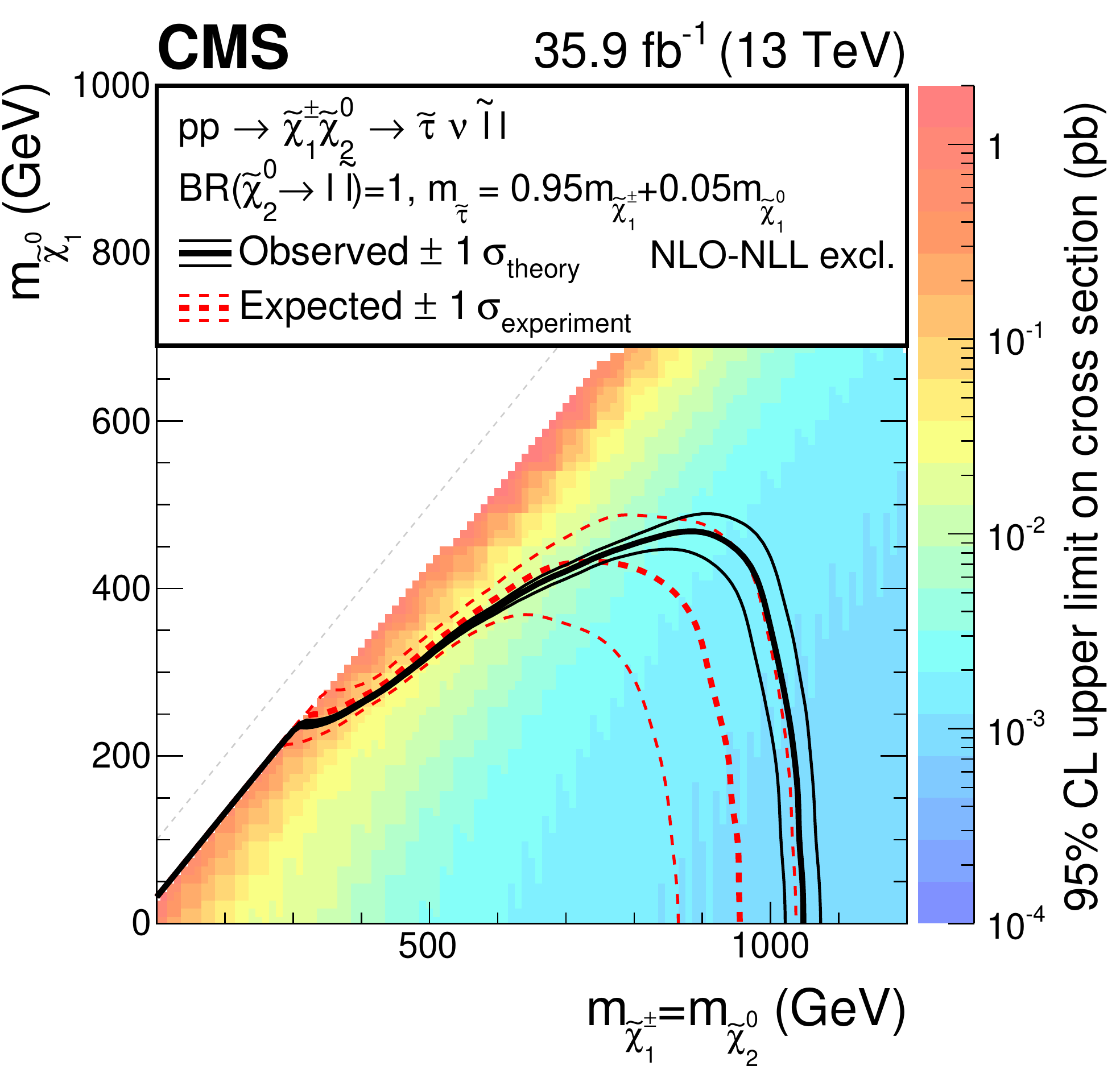}
  \includegraphics[width=0.45\textwidth]{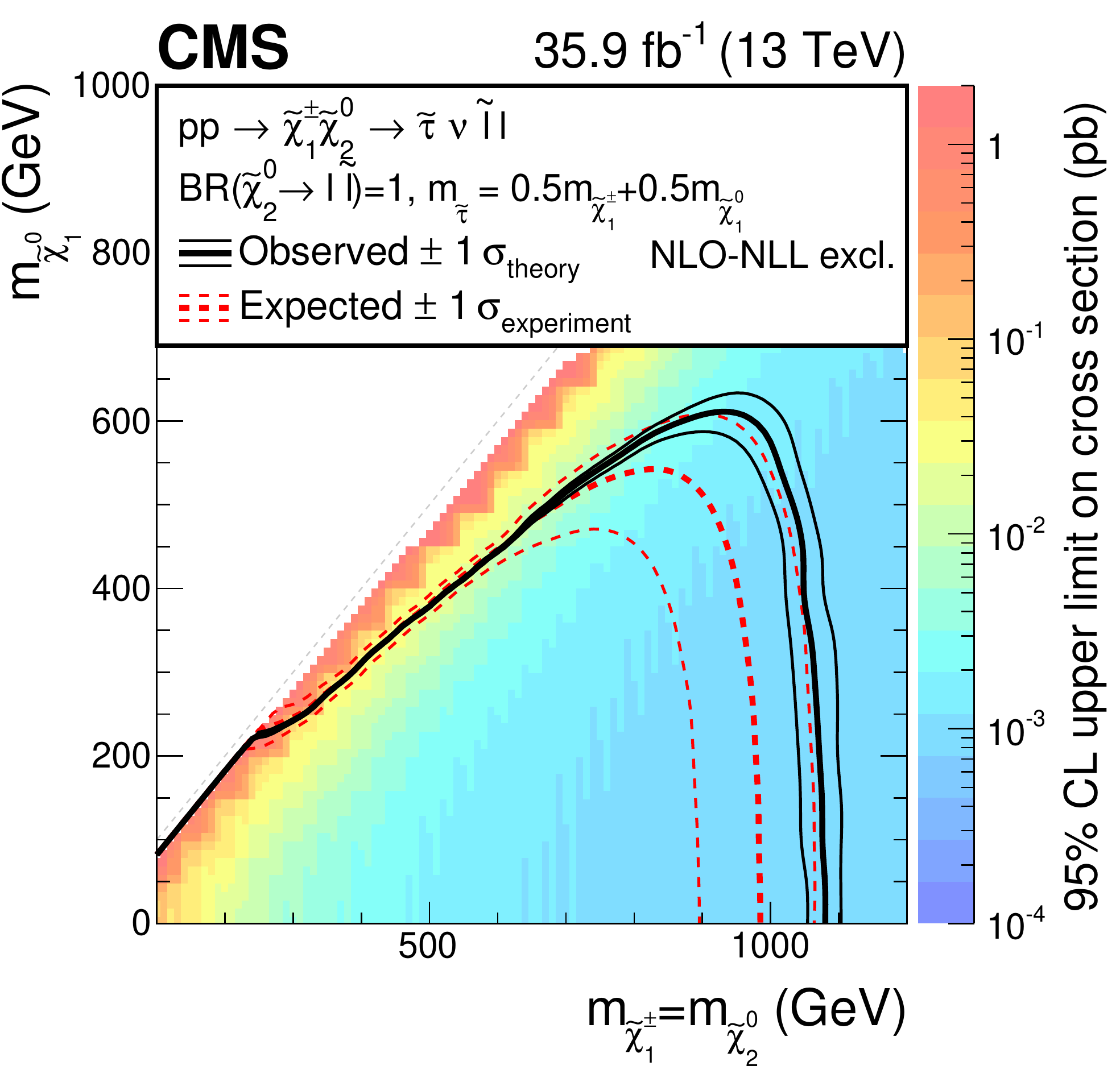}
\caption{
Interpretation of the results in the $\tau$-enriched model with mass parameter $x=0.05$
(upper-left), $x=0.95$ (upper-right) and $x=0.5$ (lower) obtained with events of categories A
and C. The shading in this figure is as described in Fig.~\ref{fig:interpr:TChiSlepSnu:FD}.}
\label{fig:interpr:TChiSlepSnu:TE}
\end{figure}

\begin{figure}[htbp]
\centering
  \includegraphics[width=0.45\textwidth]{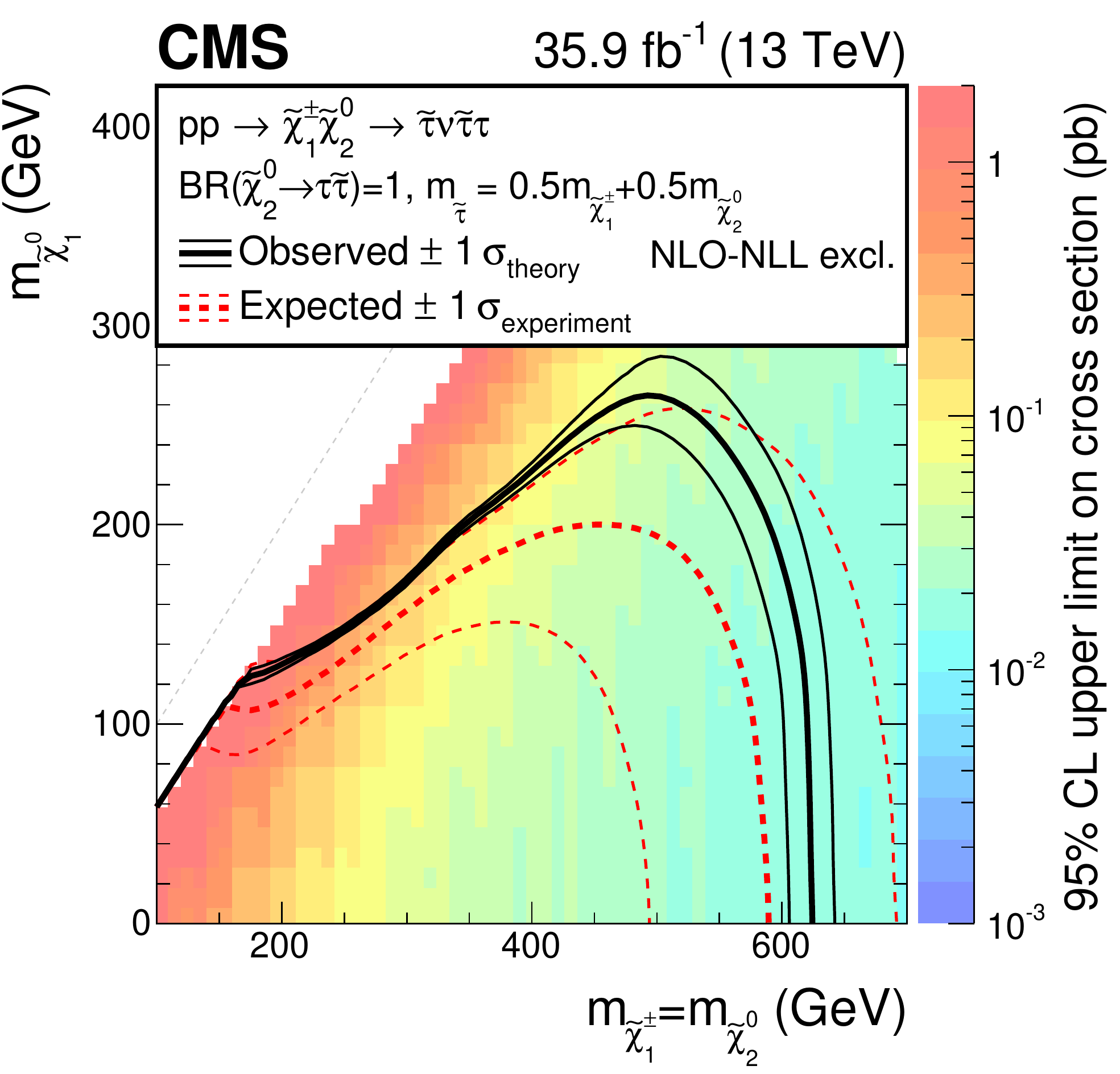}
\caption{
Interpretation of the results in the $\tau$-dominated model with mass parameter
$x=0.5$ obtained with events of category B--F. The shading in this figure is as described in Fig.~\ref{fig:interpr:TChiSlepSnu:FD}.}
\label{fig:interpr:TChiSlepSnu:TD}
\end{figure}

\begin{figure}[htbp]
\centering
\includegraphics[width=0.45\textwidth]{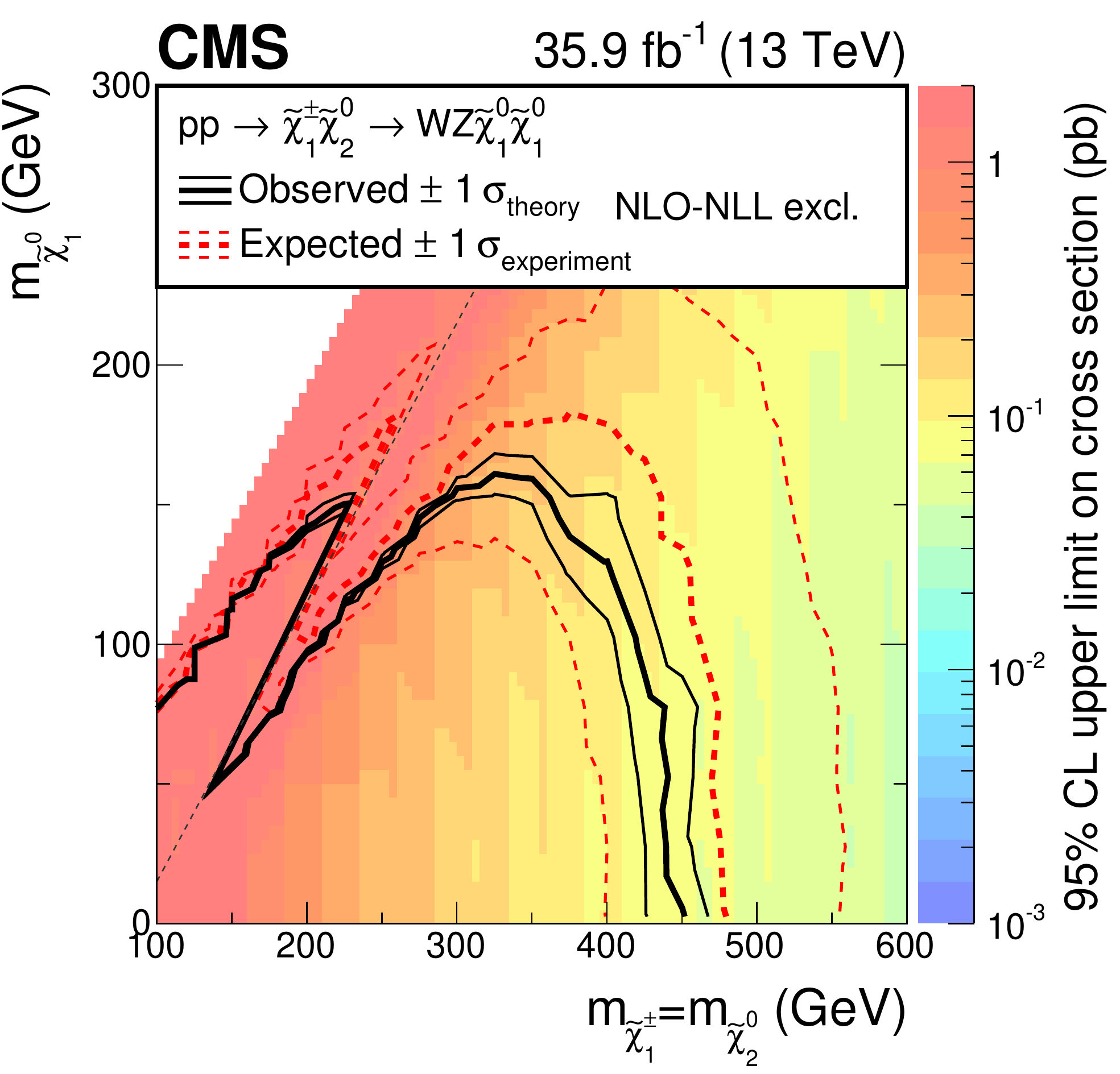} \includegraphics[width=0.45\textwidth]{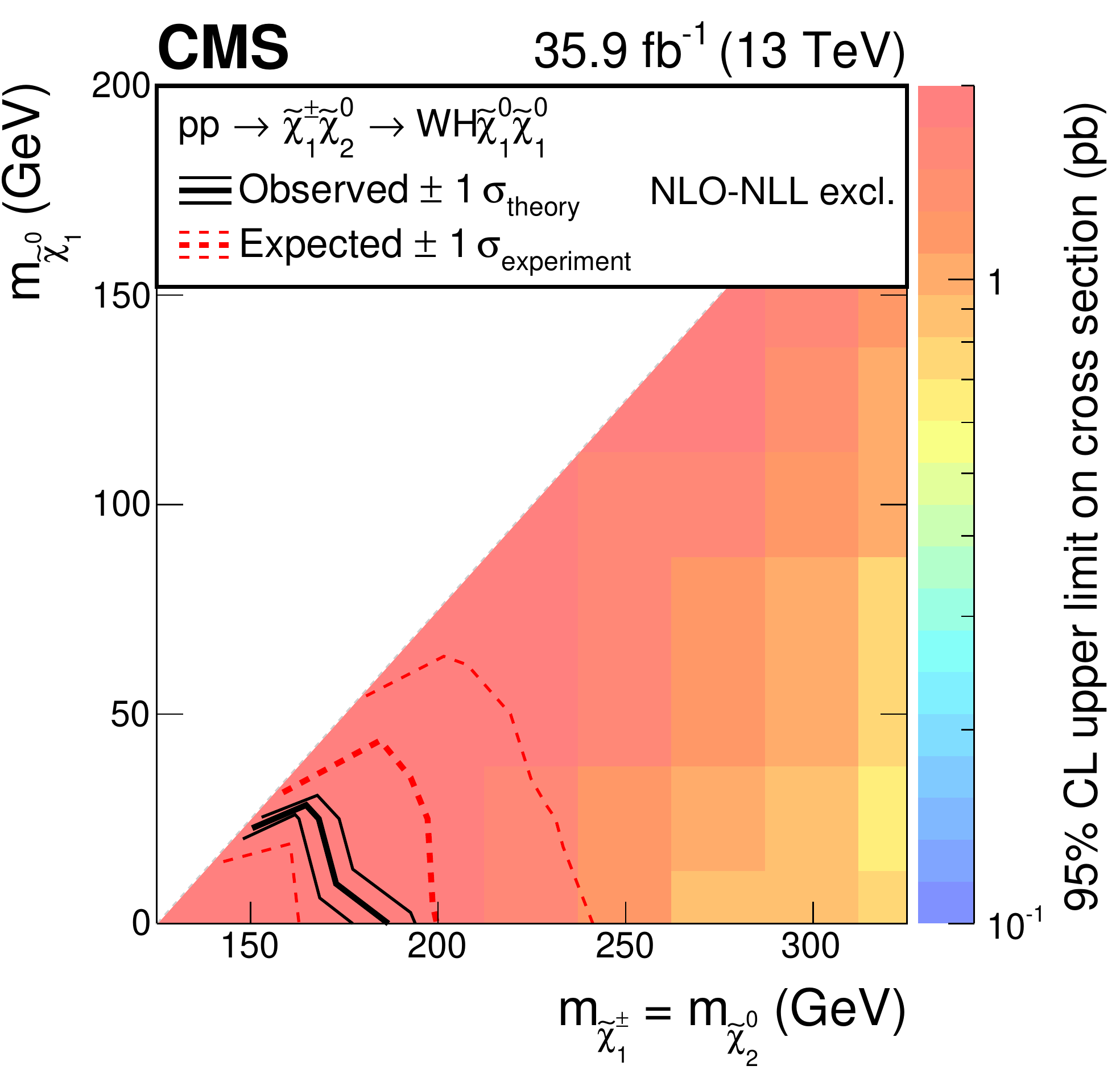}
\caption{
Interpretation of the results in the $\PSGcpmDo\PSGczDt\to\PW\PZ\PSGczDo\PSGczDo$ (left) model obtained with events of category A and the $\PSGcpmDo\PSGczDt\to\PW\PH\PSGczDo\PSGczDo$ (right) model obtained with events of all categories (SS dilepton, trilepton, and four-lepton). The shading in this figure is as described in Fig.~\ref{fig:interpr:TChiSlepSnu:FD}. The dashed grey line on the left plot corresponds to a mass difference between the $\PSGcpmDo$ and $\PSGczDo$ equal to the $\cPZ$ mass.
}
\label{fig:interpr:TChiWZWH}
\end{figure}

\begin{figure}[htbp]
\centering
\includegraphics[width=0.6\textwidth]{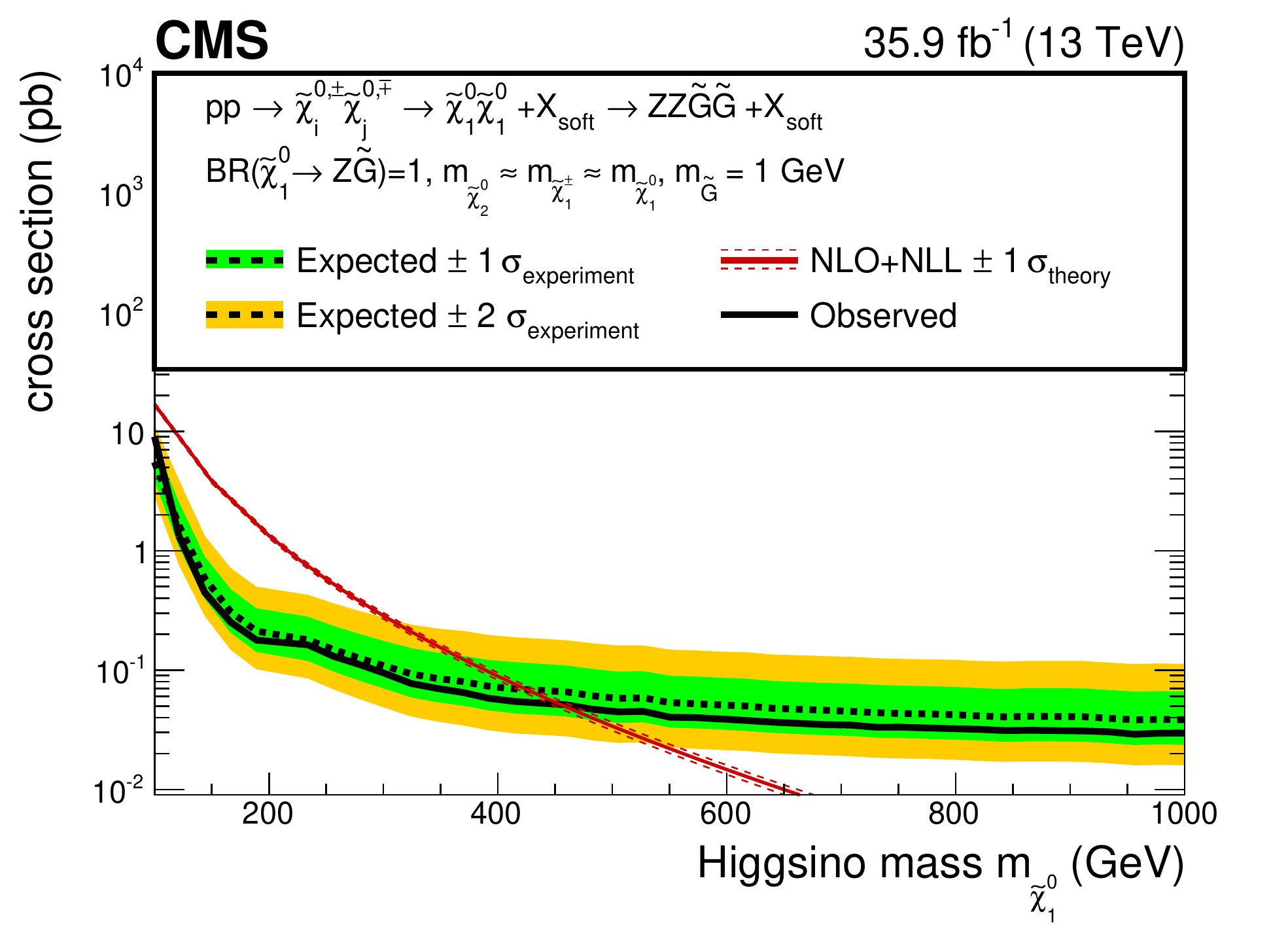} \\
\includegraphics[width=0.6\textwidth]{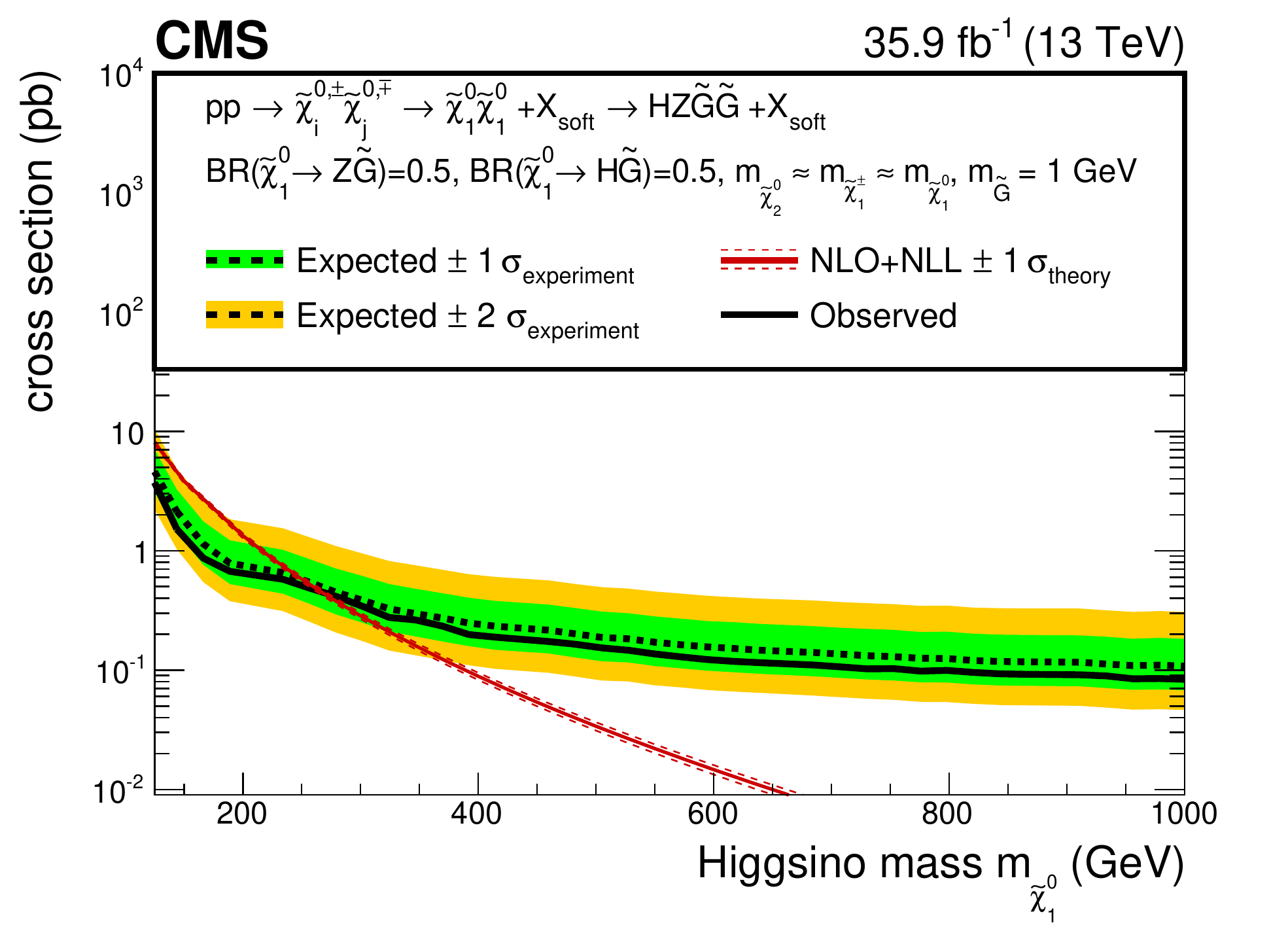} \\
\includegraphics[width=0.6\textwidth]{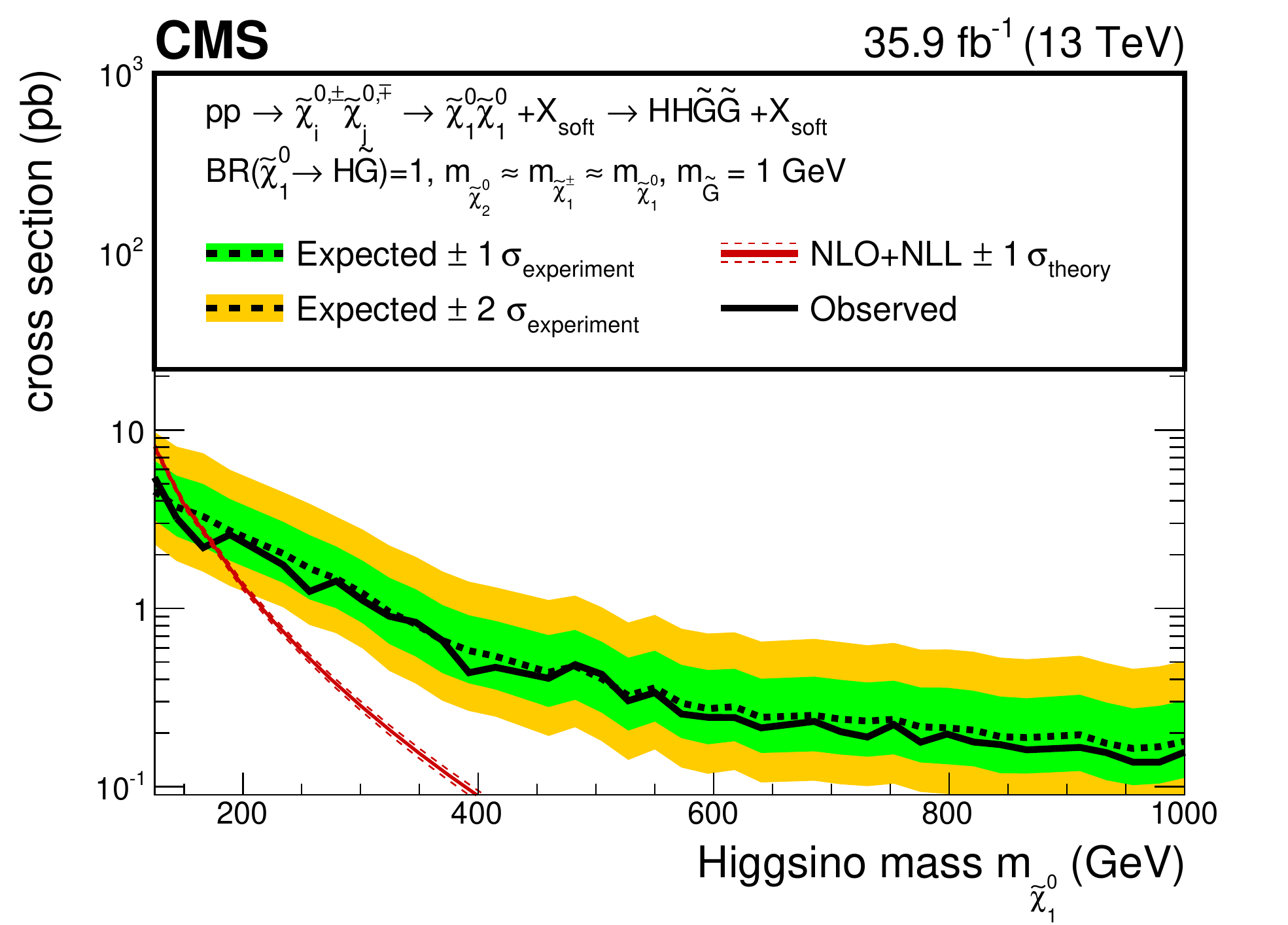}
\caption{
Interpretation of the results in the $\PSGczDo\PSGczDo\to\PZ\PZ\PXXSG\PXXSG$ model (upper), the $\PSGczDo\PSGczDo\to\PH\PZ\PXXSG\PXXSG$ model (middle), and the $\PSGczDo\PSGczDo\to\PH\PH\PXXSG\PXXSG$ model (lower) obtained with events of all trilepton (A--F) and all four-lepton (G--K) categories. The observed, median expected, ${\pm}1\,\sigma_{\text{experiment}}$ expected, and ${\pm}2\,\sigma_{\text{experiment}}$ expected 95\% CL upper limit on the neutralino pair production cross section are compared to the NLO+NLL cross sections ${\pm}1\,\sigma_{\text{theory}}$.
}
\label{fig:interpr:TChiZZHZHH}
\end{figure}

The interpretations of the results are displayed in Figs.~\ref{fig:interpr:TChiSlepSnu:FD}--\ref{fig:interpr:TChiZZHZHH} for all of the models described in Section~\ref{sec:models}. Each plot shows the 95\% CL upper limit on the chargino-neutralino production cross section as a function of the relevant pair of sparticle masses. The  observed, ${\pm}1\,\sigma_{\text{theory}}$ observed, median expected, and ${\pm}1\,\sigma_{\text{experiment}}$ expected contours are also shown.
The assumed BR for each model is displayed in each figure. For each interpretation, a combined global fit is performed using only the signal region categories listed in Table~\ref{tab:interpretations}, in order to consider all the correlations across different bins and to constrain the background estimation and uncertainties. No systematic uncertainty gets significantly constrained by the global fit. The figure displaying each interpretation is also cited in the table.

\begin{table}[tbh]
\centering
\topcaption{Summary of the interpretations of the results using different models.}
\label{tab:interpretations}
{\def\arraystretch{1.25}
\resizebox{\textwidth}{!}{
\begin{tabular}{l|cc}
Model & Categories used & Figure \\
\hline \hline
$\PSGcpmDo\PSGczDt$ production, flavor-democratic, $m_{\tilde{\ell}} = m_{\PSGczDo} + 0.5\cdot (m_{\PSGczDt} - m_{\PSGczDo})$  & A & \ref{fig:interpr:TChiSlepSnu:FD} \\
$\PSGcpmDo\PSGczDt$ production, flavor-democratic, $m_{\tilde{\ell}} = m_{\PSGczDo} + 0.05\cdot (m_{\PSGczDt} - m_{\PSGczDo})$  & SS, A & \ref{fig:interpr:TChiSlepSnu:05} (left) \\
$\PSGcpmDo\PSGczDt$ production, flavor-democratic, $m_{\tilde{\ell}} = m_{\PSGczDo} + 0.95\cdot (m_{\PSGczDt} - m_{\PSGczDo})$  & SS, A & \ref{fig:interpr:TChiSlepSnu:05} (right)\\\hline

$\PSGcpmDo\PSGczDt$ production, $\tau$-enriched, $m_{\tilde{\ell}} = m_{\PSGczDo} + 0.05\cdot (m_{\PSGczDt} - m_{\PSGczDo})$  & A, C & \ref{fig:interpr:TChiSlepSnu:TE} (left) \\
$\PSGcpmDo\PSGczDt$ production, $\tau$-enriched, $m_{\tilde{\ell}} = m_{\PSGczDo} + 0.5\cdot (m_{\PSGczDt} - m_{\PSGczDo})$  & A, C & \ref{fig:interpr:TChiSlepSnu:TE}  (center)\\
$\PSGcpmDo\PSGczDt$ production, $\tau$-enriched, $m_{\tilde{\ell}} = m_{\PSGczDo} + 0.95\cdot (m_{\PSGczDt} - m_{\PSGczDo})$  & A, C & \ref{fig:interpr:TChiSlepSnu:TE} (right)\\\hline

$\PSGcpmDo\PSGczDt$ production, $\tau$-dominated, $m_{\tilde{\tau}} = m_{\PSGczDo} + 0.5\cdot (m_{\PSGczDt} - m_{\PSGczDo})$  & B--F & \ref{fig:interpr:TChiSlepSnu:TD} \\\hline

$\PSGcpmDo\PSGczDt$ production, heavy sleptons, $\PSGcpmDo\PSGczDt\to\PW\PZ$   & A & \ref{fig:interpr:TChiWZWH} (left) \\\hline
$\PSGcpmDo\PSGczDt$ production, heavy sleptons, $\PSGcpmDo\PSGczDt\to\PW\PH$   & SS, A--K & \ref{fig:interpr:TChiWZWH} (right)\\\hline
$\PSGczDo\PSGczDo$ production, $\PSGczDo\PSGczDo\to\PZ\PZ\PXXSG\PXXSG$ & A--K & \ref{fig:interpr:TChiZZHZHH} (upper) \\
$\PSGczDo\PSGczDo$ production, $\PSGczDo\PSGczDo\to\PH\PZ\PXXSG\PXXSG$ & A--K & \ref{fig:interpr:TChiZZHZHH} (middle) \\
$\PSGczDo\PSGczDo$ production, $\PSGczDo\PSGczDo\to\PH\PH\PXXSG\PXXSG$ & A--K & \ref{fig:interpr:TChiZZHZHH} (lower) \\
\hline\hline
\end{tabular}}}
\end{table}

The sensitivity in the flavor-democratic and $\tau$-enriched models, where the mass difference between the $\PSGcpmDo$, $\PSGczDt$ and the \PSGczDo is large,
is driven by the signal regions A (three $\Pe$ or $\mu$ with an OSSF pair) with large $\Mll$, $\MT$ and \ptmiss values (SR A42, SR A43, and SR A44).
As in these signal regions observed counts in data are below the expected SM background, the observed upper limit on the chargino and neutralino  masses with a light \PSGczDo is stronger than the expected one in Figs.~\ref{fig:interpr:TChiSlepSnu:FD}--\ref{fig:interpr:TChiSlepSnu:TE}.

On the other hand, in the compressed scenarios of the flavor-democratic and $\tau$-enriched models with $x=0.05$ or $x=0.95$, where masses of
the heavier $\PSGcpmDo$, $\PSGczDt$ are close to the mass of the \PSGczDo, the analysis sensitivity is dominated by the SS dilepton signal regions with an ISR jet, and large $\MT$ and \ptmiss values. Since in these regions the observed yields in data are slightly above the expected SM background, the
observed upper limit close to the diagonal is weaker than the expected one in Fig.~\ref{fig:interpr:TChiSlepSnu:05}. To enhance the sensitivity to the $\tau$-dominated model (Fig.~\ref{fig:interpr:TChiSlepSnu:TD}), yields from signal regions B to F are used in the interpretation.  Among these regions, the ones contributing
the most to the total result are signal regions F (one $\Pe$ or $\mu$ and two $\tauh$ candidates), E (two $\Pe$ or $\mu$ of same sign and one $\tauh$ candidate),
and D (one $\Pe$ and one $\mu$ of opposite sign, and one $\tauh$ candidate) in order of importance.

In the case in which $\PSGcpmDo$ and $\PSGczDt$ decay via $\PW$ and $\PZ$ bosons, the constraints on the chargino and neutralino masses are weaker (Fig.~\ref{fig:interpr:TChiWZWH} left)
due to the lower branching fraction to a multilepton final state, and higher WZ SM background in the on-Z signal regions A which drive the sensitivity in this model.
In the region where the mass difference between the $\PSGcpmDo$, the $\PSGczDt$, and the \PSGczDo is close to the Z boson mass, the analysis sensitivity drops significantly.
This happens because the $\PSGcpmDo$ and $\PSGczDt$ decay products are produced at rest, and the \PSGczDo does not carry large momentum, leading to low additional \ptmiss
in the detector. Such a final state is very similar to the SM $\PW\PZ$ production, and $\Mll$ variable becomes the only discriminating variable in the signal regions A. In the case of the considered signals, the $\Mll$ has an upper cutoff at $m_{\PSGczDt}-m_{\PSGczDo}$ while the SM $\PW\PZ$ process does not. This reduces the overall analysis sensitivity in this region and leads to a structure in the mass limit in Fig.~\ref{fig:interpr:TChiWZWH} (left).

For the model with $\PSGcpmDo$ and $\PSGczDt$ decaying via $\PW$ and Higgs bosons, all signal regions of the analysis are used. This approach
is motivated by the diverse decay modes of the Higgs boson to leptonic final states (via an intermediate $\PW$ or $\PZ$ boson, or $\tau$ leptons).
The most important signal regions in the order of their sensitivity are B (three $\Pe$ or $\mu$ that do not form an OSSF pair), A,
and to a smaller extent signal regions F, E, D, and SS. The signal regions with more than three leptons (G to K) do not bring any visible improvement
to the sensitivity in this scenario.

Finally, in the considered GMSB scenarios leading to the $\PZ\PZ$, $\PH\PZ$, and $\PH\PH$ bosons in the final states (Fig.~\ref{fig:interpr:TChiZZHZHH})
all trilepton and four-lepton signal regions are used for the interpretation. The sensitivity in the model with two $\PZ$ bosons in the final state is dominated
by the four-lepton signal regions; in the model with two Higgs bosons in the final state by the trilepton signal regions; and in the mixed scenario four-lepton signal
regions are more sensitive when the higgsino masses are low, while the trilepton signal regions contribute the most when higgsinos are heavy.

\clearpage

\section{Summary}
\label{sec:summary}
Results are presented from a search for new physics in same-sign dilepton, trilepton, and four-lepton events containing up to two hadronically decaying $\tau$ leptons in proton-proton collision data at $\sqrt{s} = 13$\TeV, recorded with the CMS detector at the LHC and corresponding to an integrated luminosity of $\ewklumi$. The data are categorized based on the number, charge, and flavor of the leptons, and are subdivided into several kinematic regions to be sensitive to a broad range of supersymmetric particles produced via the electroweak interaction.

No significant deviation from the standard model expectations is observed. The results are used to set limits on various simplified models of supersymmetry (SUSY) that entail the production of superpartners of electroweak gauge or Higgs bosons (charginos and neutralinos). Specifically, we consider chargino-neutralino pair production, the electroweak process that is expected to have the largest cross section, and higgsino pair production in a gauge-mediated SUSY breaking inspired SUSY scenario. The resulting signal topologies depend on the masses of the lepton superpartners.

Models with light left-handed sleptons lead to enhanced branching fractions to final states with three leptons. The results imply limits on the masses of charginos and neutralinos up to 1150\GeV at 95\% confidence level for the flavor-democratic scenario, extending the reach of our previous search~\cite{SUS-13-006} by about 450\GeV. In these models, searches in the same-sign dilepton final state enhance the sensitivity in the experimentally challenging region with small mass difference between the produced gauginos and the lightest supersymmetric particle (LSP) that is inaccessible with the trilepton signature.

Assuming light right-handed sleptons, we consider two scenarios, one in which the chargino decays to $\tau$ leptons while the neutralino decays democratically, and another in which both chargino and neutralino decay to $\tau$ leptons. For the former we exclude masses of charginos and neutralinos up to 1050\GeV, while for the latter masses up to 625\GeV are excluded.

We also consider scenarios that involve the direct decay of gauginos to the LSP via $\PW$ and $\Z$ or Higgs bosons. For the models  with $\PW$ and $\Z$ bosons, chargino masses up to 475\GeV are excluded, improving the previous reach by 200\GeV. In the case of the neutralino decay via a Higgs boson, masses up to 180\GeV are excluded.

In the case of the gauge-mediated SUSY breaking model with four higgsinos and an effectively massless gravitino as the LSP, we exclude higgsino masses up to 450\GeV depending on the assumed next-to-LSP branching fraction to $\Z$ or H boson. Finally, results are presented in a form suitable for alternative theoretical interpretations.

\begin{acknowledgments}
We congratulate our colleagues in the CERN accelerator departments for the excellent performance of the LHC and thank the technical and administrative staffs at CERN and at other CMS institutes for their contributions to the success of the CMS effort. In addition, we gratefully acknowledge the computing centers and personnel of the Worldwide LHC Computing Grid for delivering so effectively the computing infrastructure essential to our analyses. Finally, we acknowledge the enduring support for the construction and operation of the LHC and the CMS detector provided by the following funding agencies: BMWFW and FWF (Austria); FNRS and FWO (Belgium); CNPq, CAPES, FAPERJ, and FAPESP (Brazil); MES (Bulgaria); CERN; CAS, MoST, and NSFC (China); COLCIENCIAS (Colombia); MSES and CSF (Croatia); RPF (Cyprus); SENESCYT (Ecuador); MoER, ERC IUT, and ERDF (Estonia); Academy of Finland, MEC, and HIP (Finland); CEA and CNRS/IN2P3 (France); BMBF, DFG, and HGF (Germany); GSRT (Greece); OTKA and NIH (Hungary); DAE and DST (India); IPM (Iran); SFI (Ireland); INFN (Italy); MSIP and NRF (Republic of Korea); LAS (Lithuania); MOE and UM (Malaysia); BUAP, CINVESTAV, CONACYT, LNS, SEP, and UASLP-FAI (Mexico); MBIE (New Zealand); PAEC (Pakistan); MSHE and NSC (Poland); FCT (Portugal); JINR (Dubna); MON, RosAtom, RAS, RFBR and RAEP (Russia); MESTD (Serbia); SEIDI, CPAN, PCTI and FEDER (Spain); Swiss Funding Agencies (Switzerland); MST (Taipei); ThEPCenter, IPST, STAR, and NSTDA (Thailand); TUBITAK and TAEK (Turkey); NASU and SFFR (Ukraine); STFC (United Kingdom); DOE and NSF (USA).

\hyphenation{Rachada-pisek} Individuals have received support from the Marie-Curie program and the European Research Council and Horizon 2020 Grant, contract No. 675440 (European Union); the Leventis Foundation; the A. P. Sloan Foundation; the Alexander von Humboldt Foundation; the Belgian Federal Science Policy Office; the Fonds pour la Formation \`a la Recherche dans l'Industrie et dans l'Agriculture (FRIA-Belgium); the Agentschap voor Innovatie door Wetenschap en Technologie (IWT-Belgium); the Ministry of Education, Youth and Sports (MEYS) of the Czech Republic; the Council of Science and Industrial Research, India; the HOMING PLUS program of the Foundation for Polish Science, cofinanced from European Union, Regional Development Fund, the Mobility Plus program of the Ministry of Science and Higher Education, the National Science Center (Poland), contracts Harmonia 2014/14/M/ST2/00428, Opus 2014/13/B/ST2/02543, 2014/15/B/ST2/03998, and 2015/19/B/ST2/02861, Sonata-bis 2012/07/E/ST2/01406; the National Priorities Research Program by Qatar National Research Fund; the Programa Clar\'in-COFUND del Principado de Asturias; the Thalis and Aristeia programs cofinanced by EU-ESF and the Greek NSRF; the Rachadapisek Sompot Fund for Postdoctoral Fellowship, Chulalongkorn University and the Chulalongkorn Academic into Its 2nd Century Project Advancement Project (Thailand); the Welch Foundation, contract C-1845; and the Weston Havens Foundation (USA).
\end{acknowledgments}
\bibliography{auto_generated}

\cleardoublepage \appendix\section{The CMS Collaboration \label{app:collab}}\begin{sloppypar}\hyphenpenalty=5000\widowpenalty=500\clubpenalty=5000\textbf{Yerevan Physics Institute,  Yerevan,  Armenia}\\*[0pt]
A.M.~Sirunyan, A.~Tumasyan
\vskip\cmsinstskip
\textbf{Institut f\"{u}r Hochenergiephysik,  Wien,  Austria}\\*[0pt]
W.~Adam, F.~Ambrogi, E.~Asilar, T.~Bergauer, J.~Brandstetter, E.~Brondolin, M.~Dragicevic, J.~Er\"{o}, M.~Flechl, M.~Friedl, R.~Fr\"{u}hwirth\cmsAuthorMark{1}, V.M.~Ghete, J.~Grossmann, J.~Hrubec, M.~Jeitler\cmsAuthorMark{1}, A.~K\"{o}nig, N.~Krammer, I.~Kr\"{a}tschmer, D.~Liko, T.~Madlener, I.~Mikulec, E.~Pree, D.~Rabady, N.~Rad, H.~Rohringer, J.~Schieck\cmsAuthorMark{1}, R.~Sch\"{o}fbeck, M.~Spanring, D.~Spitzbart, J.~Strauss, W.~Waltenberger, J.~Wittmann, C.-E.~Wulz\cmsAuthorMark{1}, M.~Zarucki
% ...
\vskip\cmsinstskip
% ...
1:~~Also at Vienna University of Technology, Vienna, Austria\\
% ...

\end{sloppypar}
\end{document}